\documentclass[11pt, a4paper]{article}
\pdfoutput=1 

\usepackage[height=8.85in,width=6.4in]{geometry}

\usepackage{graphicx,rotating}     
\usepackage[bookmarksopen,colorlinks=true,linkcolor=light_blue,
citecolor=light_pink,urlcolor=dark_red,linktoc=all]{hyperref}

\usepackage{amsmath}
\usepackage{mathtools}
\usepackage{amsfonts}
\usepackage{amssymb}
\usepackage{slashed}
\usepackage{braket}
\usepackage{bm}
\usepackage{cite}
\usepackage{comment}
\usepackage{xcolor}
\usepackage[utf8]{inputenc}
\usepackage[footnotesize]{caption}
\usepackage{bbold}

\usepackage{fourier}

\newcommand{\vev}[1]{ \left\langle {#1} \right\rangle }
\newcommand{\dd}{\mathrm{d}}

\newcommand{\abs}[1]{\left\vert {#1} \right\vert}

\makeatletter
\@addtoreset{equation}{section}

\makeatother

\usepackage{multicol}
\definecolor{dark_red}{rgb}{0.7, 0., 0.}
\definecolor{light_pink}{rgb}{1,0.4,0.4}
\definecolor{light_blue}{rgb}{0.284602,0.317763,0.963947}
\definecolor{cred}{RGB}{180,50,40} 
\definecolor{darkgreen}{RGB}{0, 100, 0}
\definecolor{desy_blue}{HTML}{009EE2}
\definecolor{desy_orange}{HTML}{FD8800}
\definecolor{forestgreen}{HTML}{228B22}
\definecolor{ochre}{HTML}{CCAA2B}

\begin{document}

\hypersetup{pageanchor=false}
\begin{titlepage}

\begin{center}

\hfill DESY 19-166\\

\vskip 1.in

{\Huge \bfseries 
Chiral Anomaly, Schwinger Effect,\\ \vspace*{3mm} Euler-Heisenberg Lagrangian,} \\ \vspace*{4mm} \textit{\huge and application to axion inflation}\\
\vskip .8in

{\Large Valerie Domcke, Yohei Ema, Kyohei Mukaida}

\vskip .3in
\begin{tabular}{ll}
\emph{DESY, Notkestra{\ss}e 85, D-22607 Hamburg, Germany}
\end{tabular}

\end{center}
\vskip .6in

\begin{abstract}
\noindent
Particle production in strong electromagnetic fields is a recurring theme in solid state physics, heavy ion collisions, early universe cosmology and formal quantum field theory. In this paper we discuss the Dirac equation in a background of parallel electric and magnetic fields. We review the Schwinger particle production rate, clarify the emergence of the chiral anomaly equation and compute the induced current of charged fermions. We distinguish the contributions from non-perturbative particle production, from the running of the gauge coupling constant and from non-linearities in the effective QED Lagrangian, and clarify how these contributions arise within a single framework. We apply these results to axion inflation. A Chern-Simons coupling between the pseudoscalar particle driving cosmic inflaton and an abelian gauge group induces a dual production of gauge fields and charged fermions. We show that the resulting scalar and gravitational wave power spectra strongly depend on the fermion mass.
\end{abstract}

\end{titlepage}

\tableofcontents
\thispagestyle{empty}
\renewcommand{\thepage}{\arabic{page}}
\renewcommand{\thefootnote}{$\natural$\arabic{footnote}}
\setcounter{footnote}{0}
\newpage
\hypersetup{pageanchor=true}

\section{Introduction}
\label{sec:intro}

The production and motion of charged fermions in (strong) electric and magnetic fields is a recurring phenomena in many different areas of physics. Schwinger production describes the pair creation of particles and antiparticles in strong electric fields~\cite{Heisenberg:1935qt,Schwinger:1951nm}, exponentially suppressed by the particle mass. Adding a magnetic field, the particle mass is replaced by the effective magnetic mass, quantized in Landau levels~\cite{Nikishov:1969tt,Bunkin:1969if} (a phenomena well known in solid state physics in the context of the Quantum Hall Effect~\cite{vonKlitzing:1980pdk}). In the limit of massless fermions, the lowest (gap-less) energy level induces asymmetric particle production. This in turn is a beautiful realization of the chiral anomaly in quantum field theory (QFT) (Adler Bell Jackiw anomaly~\cite{Adler:1969gk,Bell:1969ts}) in the presence of helical electric and magnetic background fields~\cite{Warringa:2012bq,Domcke:2018eki,Copinger:2018ftr}, discussed first in~\cite{Nielsen:1983rb} in the context of Weyl fermions in a crystal. These arguments have been extended to non-abelian gauge fields in~\cite{Domcke:2018gfr}.

Going beyond solid state physics and formal QFT,
particle production and transport phenomena associated with the chiral anomaly
also play an important role in the study of the quark gluon plasma (QGP) and in cosmology.
In heavy ion collision experiments, intense magnetic fields and local $P$-/$CP$-violation are expected~\cite{Kharzeev:1998kz}.
In the presence of a magnetic field on top of the chiral imbalance, a separation of electric charge is induced by the chiral magnetic effect~\cite{Vilenkin:1980fu,Alekseev:1998ds,Fukushima:2008xe,Son:2012wh,Son:2012bg,Zyuzin:2012tv}, which is a signature of the local $P$-/$CP$-violation~\cite{Kharzeev:2007tn,Kharzeev:2007jp,Kharzeev:2013ffa}.
Turning to cosmology, strong helical gauge fields appear in models of axion inflation~\cite{Turner:1987vd,Garretson:1992vt,Anber:2006xt},  
are postulated to permeate intergalactic voids~\cite{Durrer:2013pga}, may play a crucial role in baryogenesis~\cite{Joyce:1997uy,Bamba:2006km,Kamada:2016eeb,Kamada:2016cnb,Anber:2015yca,Jimenez:2017cdr,Domcke:2019mnd} and also appear in some implementations of the relaxion model~\cite{Hook:2016mqo,Choi:2016kke,Tangarife:2017vnd,Tangarife:2017rgl,Fonseca:2018xzp}, constructed to address the hierarchy problem in the Standard Model (SM) of particle physics.

All these examples share the same underlying physical principles. The gauge field background leads to a time-dependent dispersion relation for the fermions, inducing particle production. {In the presence of a magnetic field, the dispersion relation features discrete Landau levels.} The produced fermions are accelerated along the field lines of the background gauge field, leading to an induced current~\cite{Abramchuk:2016afc,Bavarsad:2017oyv}. In general, the final expression for the induced current receives contributions from fermionic excitations, from the vacuum (fermion loops) as well as from non-linearities in the gauge field induced by virtual charged particles {(described by the Euler-Heisenberg action~\cite{Heisenberg:1935qt})}. If the gauge fields are not external fields but treated dynamically (as in realistic cosmological models), this induced current will inhibit the gauge production.

Various aspects of this picture have been studied in the literature. The production rate of (massive) fermions in parallel, static electric and magnetic fields was derived in~\cite{Nikishov:1969tt,Bunkin:1969if} (see also~\cite{Dunne:2004nc}). The induced current resulting from this particle production was discussed \textit{e.g.}, in~\cite{Tanji:2008ku}. {Our work is moreover closely related to Ref.~\cite{Warringa:2012bq}, which discusses the contribution to the induced current sourced by particle production in the context of heavy ion collisions, using a methodology similar to ours.} 
For a discussion of the Euler-Heisenberg terms see~\cite{Dunne:2004nc}. 
For a recent analysis of the induced current in scalar quantum electrodynamics (QED) in de Sitter space without magnetic fields, clarifying nicely all the contributions to the induced current, see~\cite{Banyeres:2018aax}.

In this work, we compute the full expressions for fermion production and the induced current in a background of constant, helical {electric and magnetic} gauge fields. 
We keep the fermion mass as a free parameter and work in conformal coordinates, ensuring that our results can be applied both to flat space time and de Sitter space time. We derive and solve the equation of motion for the fermions (and equivalently for the Bogoliubov coefficients), accounting for the non-perturbative result for fermion production. 
We clarify how the chiral anomaly equation arises~\cite{Warringa:2012bq,Copinger:2018ftr}, connecting smoothly to the results obtained in the case of massless fermions~\cite{Nielsen:1983rb,Domcke:2018eki}. Based on these results, we compute all contributions to the induced current:
(i) the current from particle production,
(ii) the running of the gauge coupling, and
(iii) the $1/m$ suppressed terms of the Euler Heisenberg action.
The framework presented here, based on solving the Dirac equation of motion in a background of helical electric and magnetic fields, captures all these different aspects of fermions in abelian gauge theories in a simple and unified description. 

The resulting expression for the induced current allows us to tackle applications in cosmology. As an example, we study the case of axion inflation. Here, the rolling pseudo-scalar inflaton field induces a tachyonic instability in one of the gauge field helicities. The naive exponential gauge field production is significantly inhibited by the induced current for fermion masses smaller than ${\cal O}(100)~H$, with $H$ denoting the Hubble rate. Consequently, the resulting scalar and gravitational wave power spectrum at small scales become highly sensitive to the fermion mass, providing a unique opportunity to test the microphysical implementation of these inflation models.

The remainder of this paper is organized as follows. In Section~\ref{sec:eom} we discuss the fermion equation of motion, introducing the notion of discretized Landau levels in the dispersion relation and clarifying some subtleties about the definition of particles and antiparticles in a gauge field background. This enables us to study particle production in Section~\ref{sec:PP}, including the derivation of the anomaly equation and the derivation of all the contributions to the induced current. In Section~\ref{sec:applications} we turn to axion inflation as an example for the relevance of these computations in early Universe cosmology. We conclude in Section~\ref{sec:conclusions}. Several important but rather technical steps are relegated to the appendices: Appendix~\ref{app:notation} clarifies our notation and conventions, App.~\ref{app:eom} provides some detail on the computations of Sec.~\ref{sec:eom} whereas App.~\ref{app:PP} contains some of the necessary computations for Sec.~\ref{sec:PP}. Finally, App.~\ref{sec:WFE} contains the weak field expansion necessary to understand the vacuum contributions to the induced current.

\section{Solving the fermion equation of motion}
\label{sec:eom}

This section is dedicated to the analysis of the Dirac equation in a background of classical (anti-)parallel electric and magnetic fields. In particular, we identify the eigenvalues (energy levels) and eigenstates (particles and anti particles), both in the absence and presence of an electric field. This sets the stage for computing particle production and the different contributions to the induced current in Sec.~\ref{sec:PP}.

\subsection{Preliminaries}

\paragraph{The fermion equation of motion.} A Dirac fermion $\psi$ of mass $m$ which is charged under an Abelian gauge group obeys the equation of motion
\begin{align}
 0 = ( i \slashed D - m a ) \psi \,,
 \label{eq:eom}
\end{align}
with $\slashed D = (\partial_\mu + i g Q A_\mu)\gamma^\mu$, $Q$ denoting the charge of the fermion $\psi$ under the abelian gauge group with vector potential $A_\mu$ and gauge coupling $g$, and $a$ denoting the scale factor of the FRW metric. For more details on our notation and conventions, see App.~\ref{app:notation}. 

It will prove easier to solve the second order differential equation obtained by acting on Eq.~\eqref{eq:eom} with the differential operator $( i \slashed D + m a)$.\footnote{This procedure is completely equivalent to solving the equation of motion for the auxiliary field $\Phi = ( i \slashed D + m a) \psi$, as performed \textit{e.g.}, in Ref.~\cite{Domcke:2018eki}.} After some algebra (see App.~\ref{app:eom}), we obtain
\begin{align}
 0 & = ( i \slashed D + m a) ( i \slashed D -  m a) \psi  \\
   & = \left[ \left(- \partial_0^2 + (\bm \nabla - i g Q \bm A)^2  - m^2 a^2 \right)  + g Q \bm \sigma \cdot \begin{pmatrix} 
                                                                  \bm B + i \bm E & 0 \\
                                                                  0 & \bm B - i \bm E
                                                                 \end{pmatrix}
        - i m \gamma^0 a' \right] \psi \,.
        \label{eq:eom3}
\end{align} 
Here $\bm E$ and $\bm B$ denote co-moving `electric' and `magnetic' fields associated with the vector potential $\bm A$, \textit{i.e.,} $\bm E = - \partial_0 \bm A$ and $\bm B = \nabla \times \bm A$, $\bm \sigma$ contains the Pauli matrices and we denote the derivative with respect to conformal time with a prime.

From this point on we will ignore the expansion of the Universe, $a' = 0$, assuming it to be slow compared to all relevant microphysical processes (see Sec.~\ref{subsec:EBbound} for details).
Moreover, motivated by the applications discussed in Sec.~\ref{sec:applications}, we will consider constant (anti-)parallel electric and magnetic fields. This in particular implies a non-vanishing Chern-Simons term, $\bm E \cdot \bm B \neq 0$.
Without loss of generality, one may take the $z$-axis to match with the direction of the electric field:
\begin{align}
A^\mu = (0, 0, B x , A_z ) \,,
\label{eq:A}
\end{align}
with $E = - \dot A_z > 0$. Here the sign of the magnetic field, $B \gtrless 0$, encodes the two possible configurations (parallel vs antiparallel), related by a $CP$ transformation.
Note that, throughout this paper, we take temporal gauge for convenience.
Inserting Eq.~\eqref{eq:A} and performing a Fourier transform in the $y$- and $z$-direction, 
\begin{align}
 \psi(t, \bm x) =  \int \frac{\dd k_y \dd k_z}{(2 \pi)^2} e^{i( k_y y + k_z z)} \widetilde \psi(t, x, k_y, k_z) \,,
 \label{eq:FT}
\end{align}
we can simplify Eq.~\eqref{eq:eom3} to
\begin{align}
 \left[\partial_x^2 - (k_y - g Q \lambda B x)^2 - \partial_t^2 - (k_z - g Q A_z)^2 - m^2 a^2 + g Q  \begin{pmatrix}
         (B + i E)\sigma_z & 0 \\
         0 & (B - i E)\sigma_z                                                                                                            \end{pmatrix}
   \right] \widetilde \psi(t, x) = 0 \,.
   \label{eq:eom_2nd}
\end{align}
Here and hereafter, we mostly drop the $k_y$ and $k_z$ arguments in $\tilde \psi$ for brevity.

\paragraph{Mode expansion.} 
Let us take a spinor basis which diagonalizes the last term in Eq.~\eqref{eq:eom_2nd}:
\begin{align}
 \chi^\text{L}_+ & = \begin{pmatrix}
             1 & 0 & 0 & 0
            \end{pmatrix}^T , \quad 
 \chi^\text{L}_- = \begin{pmatrix}
             0 & 1 & 0 & 0
            \end{pmatrix}^T , \nonumber \\ 
   \chi^\text{R}_+ & = \begin{pmatrix}
             0 & 0 & 1 & 0
            \end{pmatrix}^T , \quad 
   \chi^\text{R}_- = \begin{pmatrix}
             0 & 0 & 0 & 1
            \end{pmatrix}^T ,    
\end{align}
where
\begin{align}
 \begin{pmatrix} \sigma_z & 0 \\ 0 & \sigma_z \end{pmatrix} \chi^H_\sigma = \sigma \chi^H_\sigma  \,. 
\end{align}
with $\sigma = \pm$ and $H = \text{L}, \text{R}$ describes left- and right-handed particles, respectively.
Anticipating that the differential equation \eqref{eq:eom_2nd} is separable, we can expand $\tilde \psi$ as follows~\cite{Nielsen:1983rb,Bavarsad:2017oyv,Domcke:2018eki}:
\begin{align}
 \widetilde \psi(x,t) = \sum_{n, \sigma, H} h_n(x) g_{n,\sigma}^H(t) \chi_\sigma^H \,.
 \label{eq:mode}
\end{align}
The $x$-dependent part of Eq.~\eqref{eq:eom_2nd} describes a re-scaled harmonic oscillator with shifted zero-point, 
\begin{align}
 \left[\partial_x^2 - (k_y - g Q B x)^2 \right] h_n(x) 
= s g Q B (\partial_{x_{s}}^2 - x_s^2 ) h_n(x)
 = - sgQB\, (2 n + 1)\,h_n (x) \,,
  \label{eq:oscillator}
\end{align}
where we have defined 
\begin{align}
	s \equiv \operatorname{sgn} (Q B)\,, \quad
	x_s \equiv \sqrt{s gQB} x - s \frac{k_y}{\sqrt{s gQB}}\,.
\end{align}
The solutions of Eq.~\eqref{eq:oscillator} can be expressed in terms of Hermite polynomials $H_n$,
\begin{align}
	h_n (x) \equiv \left( \frac{s g Q B}{\pi} \right)^{1/4} \left( \frac{1}{2^n n ! } \right)^{1/2}  e^{- x_{s}^2 / 2} H_n (x_{s}) \,,
\label{eq:hnsolution}
	\end{align}
where have normalized $h_n(x)$ such that $\int \dd x \,  h_n(x)    h_{n'}(x)  = \delta_{nn'}$ and the introduction of the discrete parameter $s$ ensures that all minus signs are kept track off in the manipulations of Eq.~\eqref{eq:oscillator}. 
After inserting Eq.~\eqref{eq:mode} and \eqref{eq:hnsolution} into \eqref{eq:eom_2nd},
the $t$-dependent part of~\eqref{eq:mode} is given by the solution of
\begin{align}
\left\{ \partial_t^2 + \Pi_z^2 + m^2 a^2 + s g Q B \left[ (2 n + 1) - s \sigma \right] \mp i \sigma g Q  E \right\} g_{n, \sigma}^\text{L/R}(t) = 0 \,,
 \label{eq:eom_g}
\end{align}
where 
\begin{align}
	\Pi_z \equiv k_z - g Q A_z \,.
\end{align}
We note the appearance of the discrete energy levels $(2 n + 1) s g Q B$ labeling the solutions of the harmonic oscillator, see Eq.~\eqref{eq:oscillator}. The magnetic field confines the trajectories of the charged particles projected onto the $x$-$y$ plane to circular orbits, resulting in quantized energy levels, commonly referred to as Landau levels. 

In the following Sec.~\ref{sec:ll}, we first turn off the electric field and study the dispersion relation in the presence of the magnetic field only.
For a vanishing electric field, \textit{i.e.,} for constant $A_z$, one may easily solve Eq.~\eqref{eq:eom_g} by $g^H_{n, \sigma} \propto e^{\mp i \Omega_{n,\sigma} t}$, where the dispersion relation is given by
\begin{align}
	\Omega_{n, \sigma} = \sqrt{ \Pi_z^2 + m^2 a^2 + s g Q B \left[ (2 n + 1) - s \sigma \right] }.
\end{align}
The lowest energy solution is obtained from $n = 0$ and $\sigma = s$, which is referred to as the \textit{lowest Landau level}. For the higher energy levels, referred to as \textit{higher Landau levels}, we note the degeneracy
$\Omega_{n, s} = \Omega_{n - 1, - s}$, for $n \geq 1$.
Taking into account left-/right-handed particles, $H = \text{L}, \text{R}$, and positive/negative energies, one might think that there are four independent solutions for the lowest Landau level and eight for each higher Landau level.
However, only a half of the solutions for the second order differential equation \eqref{eq:eom3} actually solve the equation of motion \eqref{eq:eom} which is first order, as we will see in Sec.~\ref{sec:ll}.
Hence, the number of true independent solutions is two for the lowest Landau level and four for each higher Landau level.
In the next Sec.~\ref{sec:ll} we first discuss the lowest Landau level and then move on to the higher Landau levels.


\subsection{Landau levels}
\label{sec:ll}

\paragraph{Lowest Landau Level.}

In the absence of an electric field, \textit{i.e.}, for constant $A_z$ in Eq.~\eqref{eq:A}, the operator on the left-hand side of Eq.~\eqref{eq:eom_g} becomes independent of the index $H$. As we have mentioned, the terms proportional to the $B$-field cancel  and the energy is minimized for $n = 0$ and $\sigma = s$:
\begin{align}
0 = 
 \left[ \partial_t^2 + \Pi_z^2 + m^2 a^2 \right]  g^H_{0,s} \,,
\end{align}
whose solution is 
\begin{align}
	g_0^{H\,(\pm)}(t) = \exp(\mp i \Omega_0 t) \,, \qquad 
	\Omega_0 \equiv \sqrt{\Pi_z^2 + m^2 a^2}\,.
	\label{eq:g0}
\end{align}
Although apparently we have four independent solutions, 
$h_0 g_0^{H \, (\pm)} \chi^H_s$ for $H = \text{L}, \text{R}$,
only two particular linear combinations solve the original Dirac equation \eqref{eq:eom}, which provide positive and negative frequency modes, respectively (see App.~\ref{app:eom}):
\begin{align}
	u_0 &= \frac{e^{- i \Omega_0 t}}{\sqrt{2 \Omega_0}} h_0 \left[ \sqrt{\Omega_0 - s \Pi_z}\, \chi_s^\text{L} + \sqrt{\Omega_0 + s \Pi_z}\, \chi_s^\text{R} \right]\,, \label{eq:u0} \\
	v_0 &= \frac{e^{i \Omega_0 t}}{\sqrt{2 \Omega_0}} h_0 \left[ \sqrt{\Omega_0 + s \Pi_z}\, \chi_s^\text{L} - \sqrt{\Omega_0 - s \Pi_z}\, \chi_s^\text{R} \right]\,. \label{eq:v0}
\end{align}
These solutions span an orthonormal basis for the wave functions in the lowest Landau level, \textit{i.e.},
$\int \dd x\, u_0^\dag u_0 = \int \dd x \, v_0^\dag v_0 = 1$ and
$\int \dd x\, u_0^\dag v_0 = 0$.

\begin{figure}
\centering
 \includegraphics[width = 0.4\textwidth]{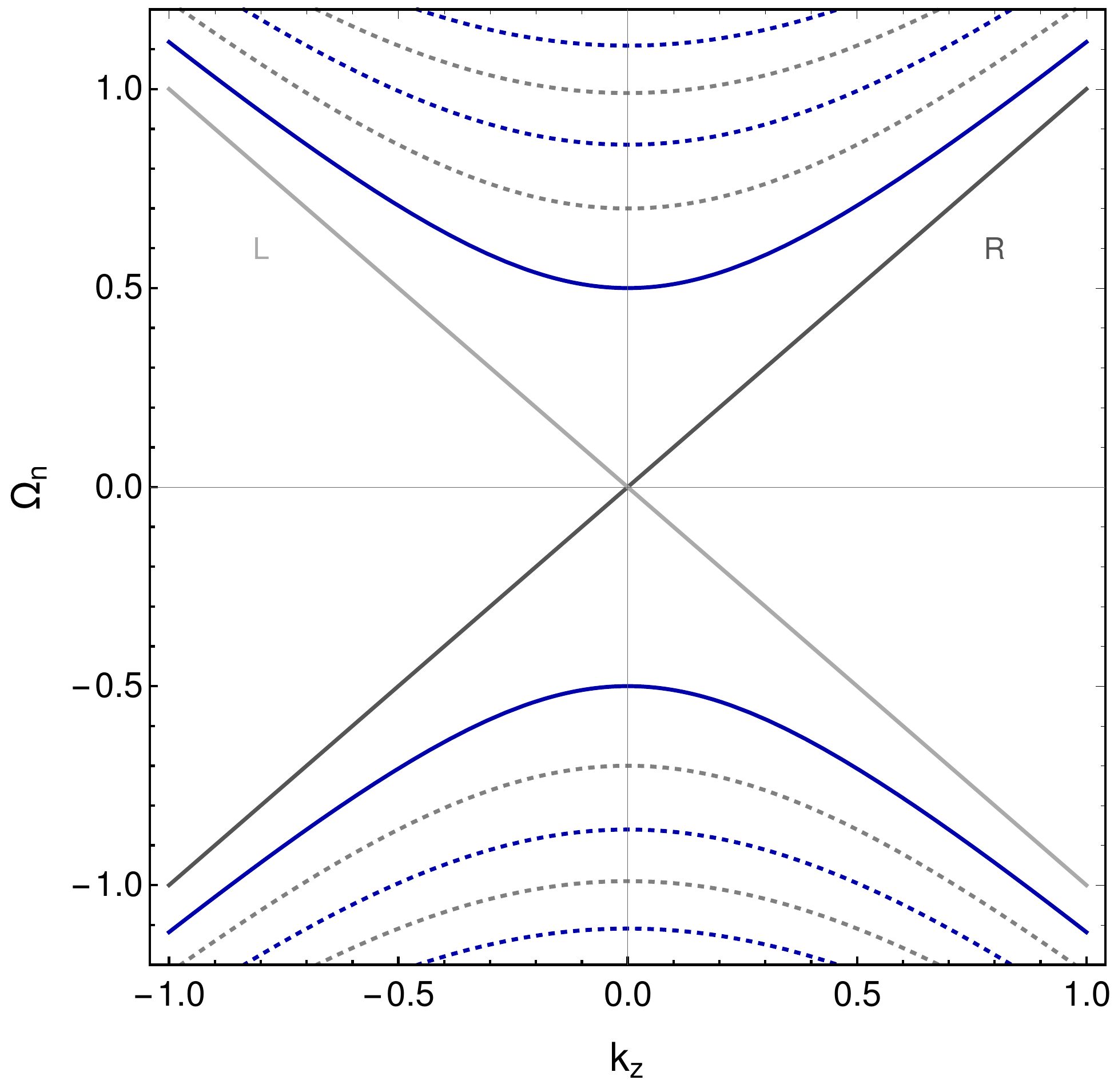} 
 \hspace{10mm}
 \includegraphics[width = 0.4\textwidth]{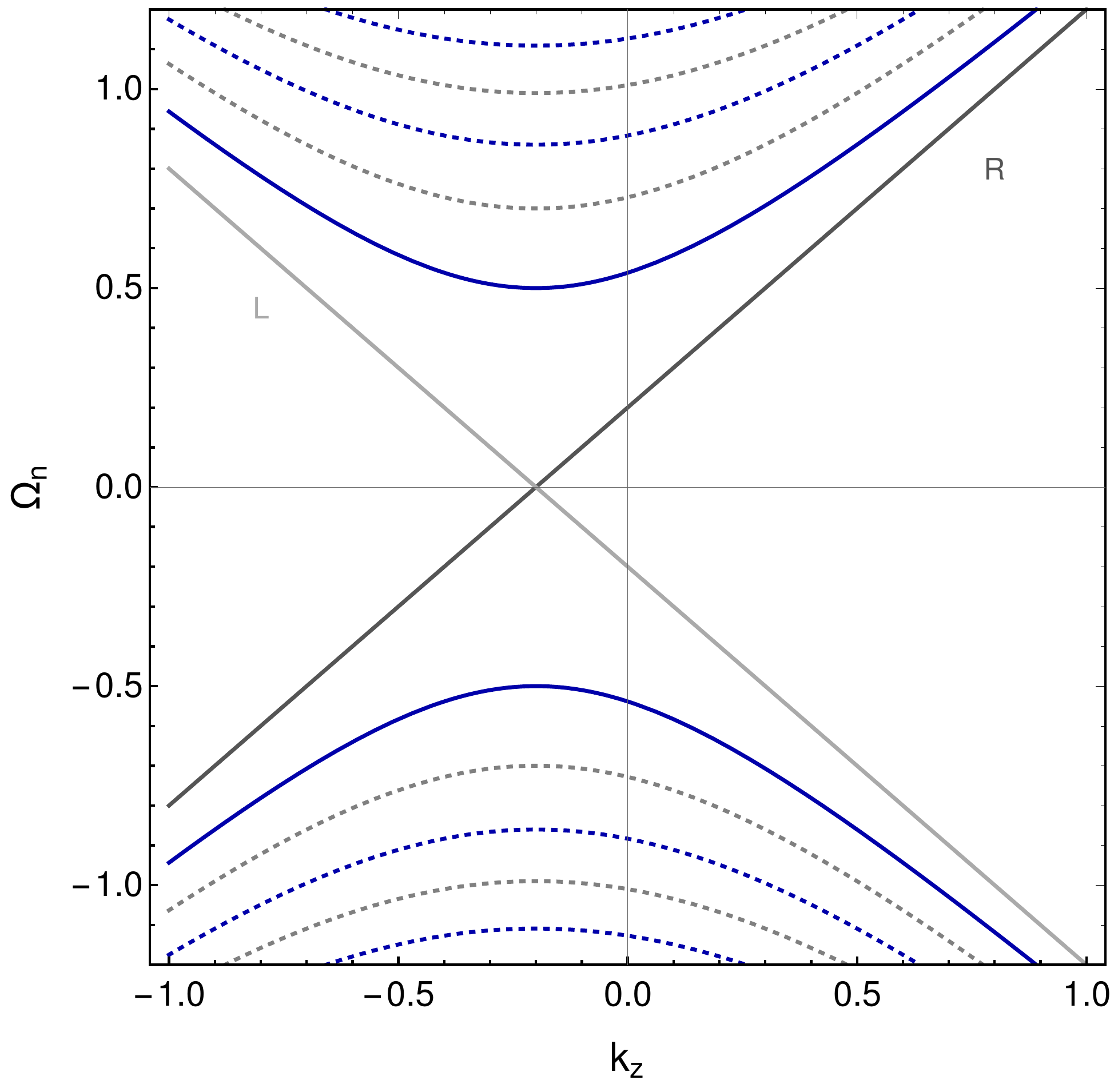}
 \caption{Energy levels for fermions in the presence of background gauge fields. The solid blue lines denote the lowest Landau level $n = 0$ for a massive fermion, the dotted lines correspond to higher Landau levels $n \geq 1$. The gray lines indicate the corresponding energy levels for massless fermions, for which in the lowest Landau level is asymmetric between left-handed (L) and right-handed (R) fermions. Note that all higher Landau levels (dotted curves) have a two-fold degeneracy, \textit{i.e.}, the dispersion relation is identical for left- and right-handed particles. For the purpose of these plots, we use units where $g |Q| E = 1$. In these units, we set $m a = 0.5$, $2 g |Q B| = 0.7$  and $ t = 0$ ($t = 0.2$) 
 in the left (right) panel. }
 \label{fig:energylevel0}
\end{figure}

It is instructive to consider the limit of chiral fermions, \textit{i.e.} taking the limit $m \rightarrow 0$ in Eqs.~\eqref{eq:u0} and \eqref{eq:v0}:
\begin{align}
	u_0 \to e^{- i \abs{\Pi_z}t} \left[ \theta (- s \Pi_z) \chi^\text{L}_s + \theta (s \Pi_z) \chi^\text{R}_s \right]\,, \quad 
	v_0 \to  e^{ i \abs{\Pi_z}t} \left[ \theta (s \Pi_z) \chi^\text{L}_s - \theta ( -s \Pi_z) \chi^\text{R}_s \right]\,,
\end{align}
where $\theta$ denotes the Heaviside function.
From this equation, one can see that the dispersion relation for the left-/right-handed fermions (proportional to $\chi^\text{L}_s, \chi^\text{R}_s$ respectively) is given by $\omega_{0}^\text{L/R} = \mp s \Pi_z$, reproducing the well known result for the lowest Landau level~\cite{Nielsen:1983rb,Domcke:2018eki}. This is also illustrated in Fig.~\ref{fig:energylevel0}. 

\paragraph{Higher Landau levels.}
For a constant $A_z$, the operator on the left-hand side of Eq.~\eqref{eq:eom_g} becomes independent of the index $H$,
and both $g_{n,s}^H$ and $g^H_{n-1,-s}$ solve the same equation:
\begin{align}
	0 = \left[ \partial_t^2 + \Pi_z^2 + m^2 a^2 + 2 n s g Q B \right] g_{n, s}^H 
	= \left[ \partial_t^2 + \Pi_z^2 + m^2 a^2 + 2 n s g Q B \right] g_{n-1, -s}^H \,.
\end{align}
The solution can be easily obtained as
\begin{align}
	g_{n,s}^{H\, (\pm)} = g_{n - 1, -s}^{H\, (\pm)} = \exp (\mp i \Omega_n t)\,, \qquad
	\Omega_n \equiv \sqrt{\Pi_z^2 + m_T^2 a^2}\,, \qquad
	m^2_T a^2 \equiv 2 n s g Q B + m^2 a^2\,.
	\label{eq:mT}
\end{align}
Here we have defined the transverse mass $m_T$ for later convenience.
We get eight independent solutions for each $n$, $h_n g^H_{n,s} \chi^H_s$ and $h_{n-1} g^H_{n-1, -s} \chi^H_{-s}$, which solve the second order differential equation \eqref{eq:eom_2nd}.
However, only four particular linear combinations actually solve the Dirac equation \eqref{eq:eom}:
\begin{align}
	u^{(1)}_n & = \frac{e^{- i \Omega_n t}}{2 \sqrt{\Omega_n}}
	\left[ i h_n \sqrt{1 - \frac{m}{m_T}}
	\left( - \sqrt{\Omega_n - s \Pi_z} \chi_s^\text{L} + \sqrt{\Omega_n + s \Pi_z} \chi_s^\text{R} \right) \right. \nonumber \\ & \qquad \qquad \qquad \qquad
	+ \left. h_{n-1} \sqrt{1 + \frac{m}{m_T}} 
	\left( \sqrt{\Omega_n + s \Pi_z} \chi_{-s}^\text{L} + \sqrt{\Omega_n - s \Pi_z} \chi^\text{R}_{-s} \right)
	\right] \,, \label{eq:u1} \\[.5em]
	u^{(2)}_n & = \frac{e^{- i \Omega_n t}}{2 \sqrt{\Omega_n}}
	\left[ i h_{n-1} \sqrt{1 - \frac{m}{m_T}}
	\left( - \sqrt{\Omega_n - s \Pi_z} \chi_{-s}^\text{R} + \sqrt{\Omega_n + s \Pi_z} \chi_{-s}^\text{L} \right) \right. \nonumber \\ & \qquad \qquad \qquad \qquad
	+ \left. h_{n} \sqrt{1 + \frac{m}{m_T}} 
	\left( \sqrt{\Omega_n + s \Pi_z} \chi_{s}^\text{R} + \sqrt{\Omega_n - s \Pi_z} \chi^\text{L}_{s} \right)
	\right] \,, \label{eq:u2}
\end{align}
for positive frequency modes; and
\begin{align}
	v^{(1)}_n & = \frac{e^{ i \Omega_n t}}{2 \sqrt{\Omega_n}}
	\left[ - i h_n \sqrt{1 - \frac{m}{m_T}}
	\left( \sqrt{\Omega_n + s \Pi_z} \chi_s^\text{L} + \sqrt{\Omega_n - s \Pi_z} \chi_s^\text{R} \right) \right. \nonumber \\ & \qquad \qquad \qquad \qquad
	+ \left. h_{n-1} \sqrt{1 + \frac{m}{m_T}} 
	\left( - \sqrt{\Omega_n - s \Pi_z} \chi_{-s}^\text{L} + \sqrt{\Omega_n + s \Pi_z} \chi^\text{R}_{-s} \right)
	\right] \,, \label{eq:v1} \\[.5em]
	v^{(2)}_n & = \frac{e^{ i \Omega_n t}}{2 \sqrt{\Omega_n}}
	\left[ - i h_{n-1} \sqrt{1 - \frac{m}{m_T}}
	\left( \sqrt{\Omega_n + s \Pi_z} \chi_{-s}^\text{R} + \sqrt{\Omega_n - s \Pi_z} \chi_{-s}^\text{L} \right) \right. \nonumber \\ & \qquad \qquad \qquad \qquad
	+ \left. h_{n} \sqrt{1 + \frac{m}{m_T}} 
	\left( - \sqrt{\Omega_n - s \Pi_z} \chi_{s}^\text{R} + \sqrt{\Omega_n + s \Pi_z} \chi^\text{L}_{s} \right)
	\right] \,, \label{eq:v2}
\end{align}
for negative frequency modes.
{Note that due to the degeneracy in $\Omega_n$, any linear combination of $u_n^{(1)}$ and $u_n^{(2)}$ (and correspondingly of  $v_n^{(1)}$ and $v_n^{(2)}$) solves the Dirac equation. The basis above is chosen such that}
i) for each higher Landau level $n$, these solutions span an orthonormal basis for the wave functions, \textit{i.e.}, 
$\int \dd x\, u_n^{(r)\, \dag} u_{\bar n}^{(\bar r)} = \int \dd x\, v_n^{(r)\, \dag} v_{\bar n}^{(\bar r)} = \delta_{n \bar n} \delta_{r \bar r}$ and $\int \dd x\, u_n^{(r) \, \dag} v_{\bar n}^{(\bar r)} = 0$ and ii) the $n = 0$ solutions~\eqref{eq:u0} and \eqref{eq:v0} are obtained from the expressions \eqref{eq:u1} to \eqref{eq:v2} by choosing $u_0^{(1)} = v_0^{(1)} = 0$, $u_0^{(2)} = u_0$, and $v_0^{(2)} = v_0$.

\subsection{Particles and anti-particles}
\label{sec:pandantip}

We here clarify the definition of positive and negative energy states without assuming a vanishing electric field.
This discussion is helpful, in particular, when we turn on the electric field and discuss particle production in Sec.~\ref{sec:PP}.
To define the positive and negative energy states unambiguously, we need to look at the Hamiltonian for the Dirac fermion, which is given by
\begin{align}
	H_\psi = \int \dd^3 x\,\, \overline \psi \left( - i \bm{\nabla} \cdot \bm{\gamma} - g Q \bm{A} \cdot \bm{\gamma} + m a\right) \psi\,.
\end{align}
Inserting the mode expansion given in Eq.~\eqref{eq:mode} into this equation, one may express it in the following matrix form:
\begin{align}
	H_\psi = & \int \frac{\dd k_y \dd k_z}{(2 \pi)^2}\,
	\Bigg[
	\begin{pmatrix}
		g^{\text{L}\, \ast}_{0,s} &\!\!\!\!\!\! g^{\text{R}\, \ast}_{0,s}
	\end{pmatrix}
		\begin{pmatrix}
		-s \Pi_z(t) & ma \\
		ma & s \Pi_z (t)
	\end{pmatrix}
		\begin{pmatrix}
		g^\text{L}_{0,s}\\ g^\text{R}_{0,s}
	\end{pmatrix} \\
	&+ \sum_{n=1}
		\begin{pmatrix}
		g^{\text{L}\, \ast}_{n,s} &\!\!\!\!\!\! g^{\text{L}\, \ast}_{n-1,-s} &\!\!\!\!\!\! g^{\text{R}\, \ast}_{n,s} &\!\!\!\!\!\! g^{\text{R}\, \ast}_{n-1,-s}
	\end{pmatrix}
		\begin{pmatrix}
		- s \Pi_z (t) & - i m_{B} a & ma & 0 \\
		i m_{B} a & s \Pi_z (t) & 0 & ma \\
		ma & 0 & s \Pi_z (t) & i m_{B} a \\
		0 & ma & - i m_{B} a & - s \Pi_z (t)
	\end{pmatrix}
		\begin{pmatrix}
		g^\text{L}_{n,s}\\ g^\text{L}_{n-1,-s} \\g^\text{R}_{n,s} \\ g^\text{R}_{n-1,-s}
	\end{pmatrix}
	\Bigg]\,,
\end{align}
where $m_{B} a \equiv \sqrt{2 n s g Q B}$.
We write the time argument explicitly in $\Pi_z$ to emphasize that this expression is applicable for $\dot A_z \neq 0$.
From this equation, it is clear that the $n = 0$ and $n \geq 1$ solutions never mix, and hence we can study  them separately.

Let us first consider the first two-by-two matrix.
Its eigenvalues are $\pm \Omega_0$ with
\begin{align}
	\Omega_0 (t) = \sqrt{\Pi_z^2 (t) + m^2 a^2}\,.
\end{align}
The positive and negative eigenvalues correspond to particles and anti-particles, respectively.
The corresponding eigenvectors are given by
\begin{align}
	\bm{g}_0^{(+)} \equiv
	\begin{pmatrix}
		\sqrt{\frac{\Omega_0 (t) - s \Pi_z (t)}{2 \Omega_0 (t)}} \\ \sqrt{\frac{\Omega_0 (t) + s \Pi_z (t)}{2 \Omega_0 (t)}}
	\end{pmatrix}\,, \quad
	\bm{g}_0^{(-)}\equiv
	\begin{pmatrix}
		\sqrt{\frac{\Omega_0 (t) + s \Pi_z (t)}{2 \Omega_0 (t)}} \\ - \sqrt{\frac{\Omega_0 (t) - s \Pi_z (t)}{2 \Omega_0 (t)}}
	\end{pmatrix}\,,
\end{align}
for the positive and negative eigenvalues respectively.
These eigenvectors form the basis for the wave functions for the positive and negative energy states:
\begin{align}
	U_0 &= e^{-i \int^t \dd t' \Omega_0 (t')} h_0\,
	\left[ \bm{g}_0^{(+)} \right]^T
	\begin{pmatrix}
		\chi_s^\text{L} \\ \chi_s^\text{R}
	\end{pmatrix} \,,
	\quad
	V_0  =  e^{i \int^t \dd t' \Omega_0 (t')} h_0\,
	\left[ \bm{g}_0^{(-)} \right]^T
	\begin{pmatrix}
		\chi_s^\text{L} \\ \chi_s^\text{R}
	\end{pmatrix}\,.
	\label{eq:UV0}
\end{align}

One can easily see that they coincide with the solutions for the equation of motion for a vanishing electric field as expected, \textit{i.e.},
$U_0 = u_0$ and $V_0 = v_0$ for $\dot A_z = 0$.
This means that the positive and negative energy modes provide distinct solutions for the equation of motion for $\dot A_z = 0$, \textit{i.e.}, they do not mix with each other.
However, this is no longer true once we turn on the electric field.
There the positive and negative frequencies get mixed during the course of evolution due to the time-dependence in $\Pi_z(t)$, resulting in the particle production as we will see in Sec.~\ref{sec:PP}.

Next, we move on to the four-by-four matrix. Its eigenvalues are degenerate, \textit{i.e.}, we find two positive and two negative energy states, $\pm \Omega_n$ with
\begin{align}
	\Omega_n (t) = \sqrt{ \Pi_z^2 (t) + m_T^2 a^2} \,, 
	\qquad m_T^2 a^2 \equiv 2 n s g Q B + m^2 a^2\,.
\end{align}
The corresponding eigenvectors for the positive energy state are
\begin{align}
	\bm{g}_n^{1, (+)} \equiv \frac{1}{\sqrt{2}}
	\begin{pmatrix}
		- i \sqrt{1 - \frac{m}{m_T}} \sqrt{\frac{\Omega_n(t) - s\Pi_z(t)}{2\Omega_n (t)}} \\
		\sqrt{1 + \frac{m}{m_T}} \sqrt{\frac{\Omega_n(t) + s\Pi_z (t)}{2\Omega_n (t)}} \\
		i \sqrt{1 - \frac{m}{m_T}} \sqrt{\frac{\Omega_n (t) + s\Pi_z (t)}{2\Omega_n (t)}} \\
		\sqrt{1 + \frac{m}{m_T}} \sqrt{\frac{\Omega_n (t) - s\Pi_z (t)}{2\Omega_n (t)}}
	\end{pmatrix}\,,
	\qquad
	\bm{g}_n^{2, (+)} \equiv \frac{1}{\sqrt{2}}
	\begin{pmatrix}
		\sqrt{1 + \frac{m}{m_T}} \sqrt{\frac{\Omega_n (t) - s\Pi_z (t)}{2\Omega_n (t)}} \\
		i \sqrt{1 - \frac{m}{m_T}} \sqrt{\frac{\Omega_n (t) + s\Pi_z (t)}{2\Omega_n (t)}} \\
		\sqrt{1 + \frac{m}{m_T}} \sqrt{\frac{\Omega_n (t) + s\Pi_z (t)}{2\Omega_n (t)}} \\
		-i\sqrt{1 - \frac{m}{m_T}} \sqrt{\frac{\Omega_n (t) - s\Pi_z (t)}{2\Omega_n (t)}}
	\end{pmatrix}\,.
	\label{eq:gn1}
\end{align}
Those for the negative energy state are
\begin{align}
		\bm{g}_n^{1, (-)} \equiv \frac{1}{\sqrt{2}}
	\begin{pmatrix}
	- i \sqrt{1 - \frac{m}{m_T}} \sqrt{\frac{\Omega_n (t) + s\Pi_z (t)}{2\Omega_n (t)}} \\
		- \sqrt{1 + \frac{m}{m_T}} \sqrt{\frac{\Omega_n(t) - s\Pi_z (t)}{2\Omega_n (t)}} \\
		i \sqrt{1 - \frac{m}{m_T}} \sqrt{\frac{\Omega_n(t) - s\Pi_z(t)}{2\Omega_n (t)}} \\
		\sqrt{1 + \frac{m}{m_T}} \sqrt{\frac{\Omega_n (t) + s\Pi_z (t)}{2\Omega_n (t)}}
	\end{pmatrix}\,,
	\qquad
	\bm{g}_n^{2, (-)} \equiv \frac{1}{\sqrt{2}}
	\begin{pmatrix}
		 \sqrt{1 + \frac{m}{m_T}} \sqrt{\frac{\Omega_n(t) + s\Pi_z (t)}{2\Omega_n (t)}} \\
	- i \sqrt{1 - \frac{m}{m_T}} \sqrt{\frac{\Omega_n (t) - s\Pi_z (t)}{2\Omega_n (t)}} \\
	- \sqrt{1 + \frac{m}{m_T}} \sqrt{\frac{\Omega_n (t) - s\Pi_z (t)}{2\Omega_n (t)}}
\\
		- i \sqrt{1 - \frac{m}{m_T}} \sqrt{\frac{\Omega_n(t) + s\Pi_z(t)}{2\Omega_n (t)}} 
	\end{pmatrix}\,.
	\label{eq:gn2}
\end{align}
From these eigenvectors, we obtain the following wave functions for the positive and negative energy states:
\begin{align}
	U_n^{(r)} = e^{- i \int^t \dd t' \Omega_n (t')} 
	\left[ \bm{g}^{r, (+)} \right]^T
	\begin{pmatrix}
		h_n \chi_s^\text{L} \\
		h_{n-1} \chi_{-s}^\text{L} \\
		h_n \chi_s^\text{R} \\
		h_{n-1} \chi_{-s}^\text{R}
	\end{pmatrix}\,,
	\qquad
	V_n^{(r)} = e^{- i \int^t \dd t' \Omega_n (t')} 
	\left[ \bm{g}^{r, (-)} \right]^T
	\begin{pmatrix}
		h_n \chi_s^\text{L} \\
		h_{n-1} \chi_{-s}^\text{L} \\
		h_n \chi_s^\text{R} \\
		h_{n-1} \chi_{-s}^\text{R}
	\end{pmatrix}\,.
	\label{eq:UV}
\end{align}
Again, they coincide with the solutions for the equation of motion [Eqs.~\eqref{eq:u1} to \eqref{eq:v2}] for a vanishing electric field, \textit{i.e.}, $U_n^{(r)} = u_n^{(r)}$ and $V_n^{(r)} = v_n^{(r)}$ for $\dot A_z = 0$.
Once we turn on the electric field, Eq.~\eqref{eq:UV} no longer solve the equation of motion, implying the particle production.

Now that we know the wave functions for the positive and negative energy states in the presence of the electric field,
we can for any time $t$ define creation (and annihilation) operators for particles and anti-particles, $\hat B_n^{(r)}$ and $\hat D_n^{(r)}$, respectively.
With these, we can expand the fermion field at any given time $t$ as follows:
\begin{align}
	\psi = \int \frac{\dd k_y \dd k_z}{(2 \pi)^2} e^{i ( k_y y + k_z z )} 
	\sum_{n,r} \left[ U_n^{(r)} \hat B_n^{(r)} + V_n^{(r)} \hat D_n^{(r)\, \dag} \right].
	\label{eq:expand_t}
\end{align}
Note that the following rules are implicit to get this formal expression: $U_0^{(1)} = V_0^{(1)} = 0$, $U_0^{(2)} = U_0$, and $V_0^{(2)} = V_0$.
This also implies $\hat B_0^{(1)} = 0$, $\hat D_0^{(1)} = 0$, $\hat B_0^{(2)} = \hat B_0$, and $\hat D_0^{(2)} = \hat D_0$. 
Here creation and annihilation operators fulfill the following commutation relations:
\begin{align}
	\{ \hat B_n^{(r)}, \hat B_{\bar n}^{(\bar r)\,\dag} \} &= (2 \pi)^2 \delta (p_y - \bar p_y) \delta (p_z - \bar p_z) \delta_{n, \bar n} \delta_{r, \bar r}\,, \\
	\{ \hat D_n^{(r)}, \hat D_{\bar n}^{(\bar r)\, \dag} \} &= (2 \pi)^2 \delta (p_y - \bar p_y) \delta (p_z - \bar p_z) \delta_{n, \bar n} \delta_{r, \bar r}\,, \\
	\{ \text{otherwise} \} &= 0\,.
\end{align}

\section{Particle production}
\label{sec:PP}

\begin{figure}[t]
\centering
\includegraphics[width = 0.49\linewidth]{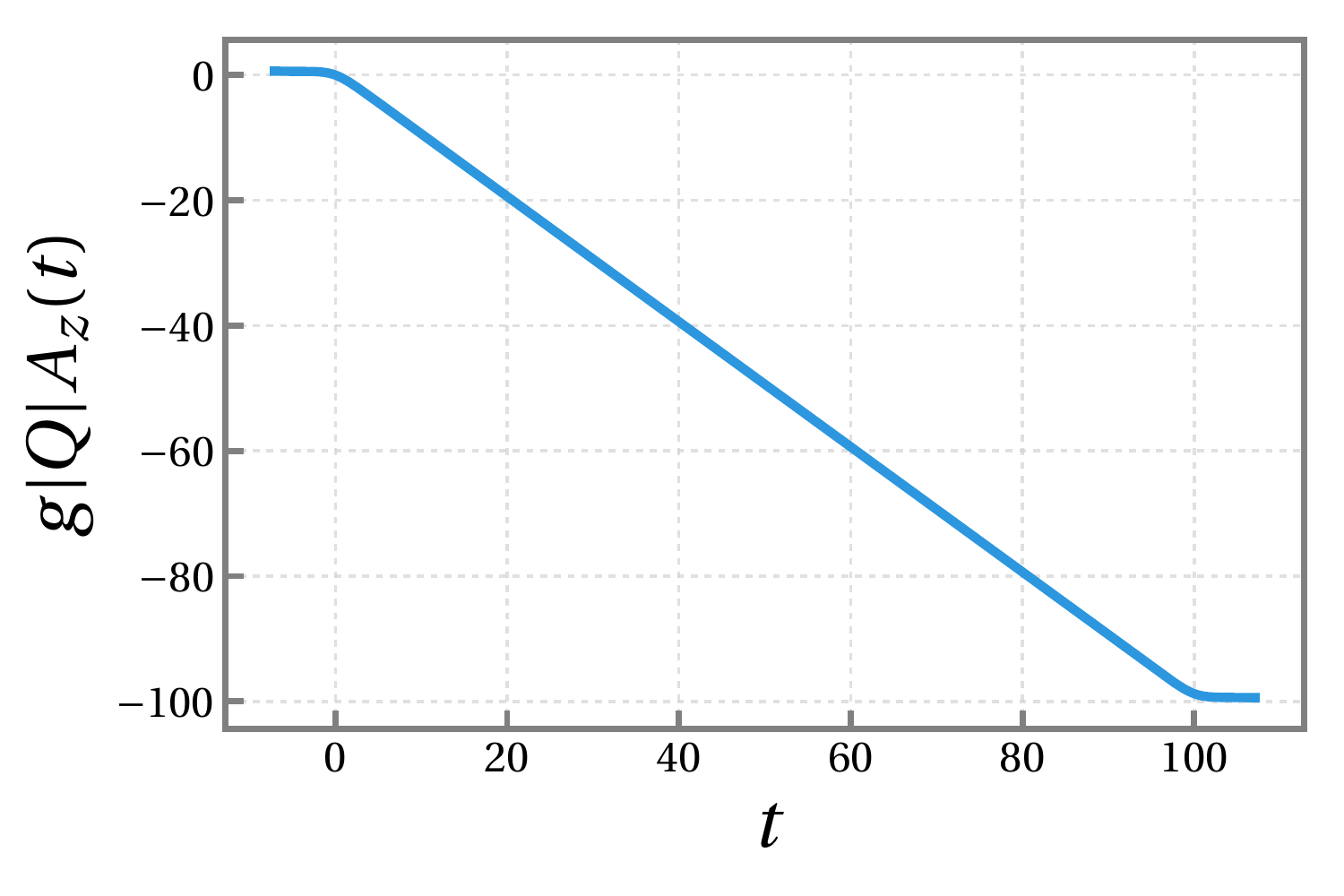}
\includegraphics[width = 0.49\linewidth]{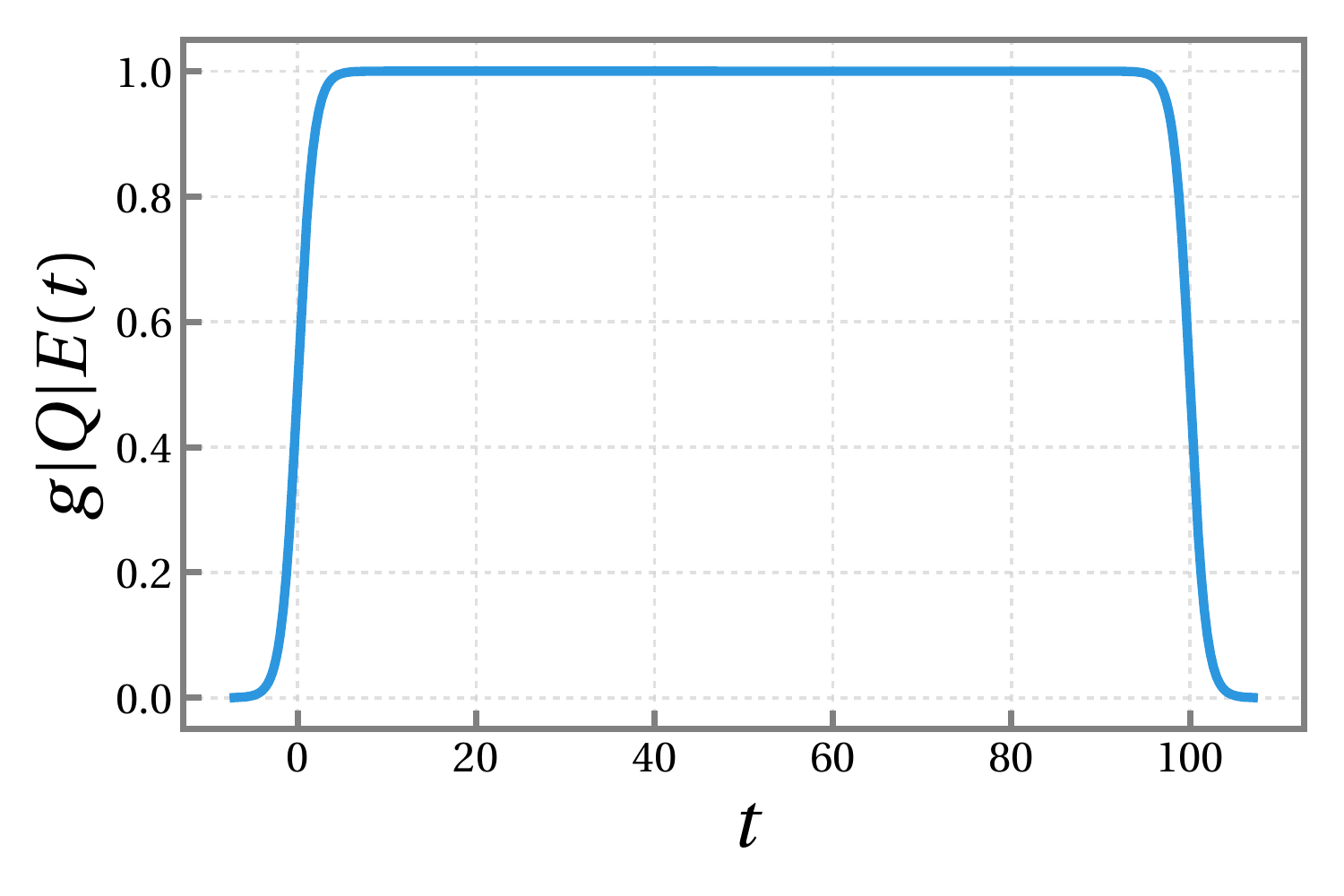}
\caption{
A schematic plot of {the time evolution of the electric field, see} Eq.~\eqref{eq:pumpup}.
Here we work in units of $\sqrt{g |Q| E} = 1$ with $E$ being a constant value of the electric field for $t= 0$ to $\tau$.
In this plot, we have set $\tau = 100$.
}
\label{fig:gauge}
\end{figure}

To study the fermion production induced by the electric field, let us turn on the electric field adiabatically for a finite amount of time. This allows us to define eigenstates with positive/negative frequency at early and at late times, and 
avoid spurious behavior associated with sudden turn on/off of the electric field.
More concretely, we turn on the electric field from $t = 0$ to $t = \tau$.
\begin{align}
 A_z(t) = \begin{cases}
           0 & t \ll 0 \\
           - E t & 0 < t < \tau \\
           - E \tau & \tau \ll t
          \end{cases} \qquad \Rightarrow \qquad 
         E(t) = \begin{cases}
           0 & t \ll 0 \\
           E & 0 < t < \tau \\
           0  & \tau \ll t
          \end{cases} \,.
          \label{eq:pumpup}
\end{align}
Note here that, to avoid spurious singularities at $t=0$ and $t=\tau$, we interpolate these discontinuities smoothly as shown in Fig.~\ref{fig:gauge}.

\subsection{Bogoliubov coefficients}

\paragraph{Definition.}
We consider solutions $\tilde u_n^{(r)}, \, \tilde v_n^{(r)}$ for the equation of motion which fulfill the following boundary condition:
$\tilde u_n^{(r)} \to u_n^{(r)}|_{A_z = 0}$ and $\tilde v_n^{(r)} \to v_n^{(r)} |_{A_z = 0}$ at $t \ll 0$.
Note here that we have used the same convention for the lowest Landau level as Eq.~\eqref{eq:expand_t}, \textit{i.e.}, $u_0^{(1)}=v_0^{(1)} = 0$, $u_0^{(2)} = u_0$, and $v_0^{(2)} = v_0$.
Initially, there is no electric field and hence one may expand the field by means of wave functions defined in Eqs.~\eqref{eq:u0}, \eqref{eq:v0} and \eqref{eq:u1} to \eqref{eq:v2} at $A_z = 0$ together with the creation operators $\hat b_n^{(r)}$, $\hat d_n^{(r)}$ defined at $t \ll 0$ and the corresponding annihilation operators $(\hat b_n^{(r)})^\dagger$, $(\hat d_n^{(r)})^\dagger$ .
This implies
\begin{align}
	\psi (t, \bm{x}) &= \int \frac{\dd k_y \dd k_z}{(2 \pi)^2} e^{i (k_y y + k_z z)} \sum_{n,r} \left[ \tilde u_n^{(r)} \hat b_n^{(r)} + \tilde v_n^{(r)} \hat d_n^{(r)\, \dag} \right]\,,
	\label{eq:expand_ini} \\
	& \to 
	\int \frac{\dd k_y \dd k_z}{(2 \pi)^2} e^{i (k_y y + k_z z)} \sum_{n,r} \left[  u_n^{(r)}|_{A_z = 0}\, \hat b_n^{(r)} +  v_n^{(r)}|_{A_z = 0}\, \hat d_n^{(r)\, \dag} \right]
	\qquad \text{for} \quad t \ll 0\,,
\end{align}
where the annihilation operators erase the initial vacuum state, \textit{i.e.}, $b_n^{(r)} \ket{0} = d_n^{(r)} \ket{0} = 0$.

Once we turn on the electric field, the wave functions which diagonalize the Hamiltonian, $U_n^{(r)}$ and $V_n^{(r)}$, no longer coincide with the solutions for the equation of motion.
Hence, the solutions, $\tilde u_n^{(r)}$ and $\tilde v_n^{(r)}$, must be expressed as a linear combination:
\begin{align}
	\left. \tilde u_n^{(r)} \right|_{t > 0} = \alpha_n^{(r)} U_n^{(r)} + \beta_n^{(r)} V_n^{(r)}\,, \quad
	\left. \tilde v_n^{(r)} \right|_{t > 0} = - \beta_n^{(r) \ast} U_n^{(r)} + \alpha_n^{(r) \ast} V_n^{(r)}\,.
	\label{eq:linear}
\end{align}
Here we have defined the Bogoliubov coefficients:
\begin{align}
	\alpha_n^{(r)} = \left( U_n^{(r)}, \tilde u_n^{(r)} \right)\,, \quad
	\beta^{(r)} = \left( V_n^{(r)}, \tilde u_n^{(r)} \right) \,, \qquad \text{with} \quad  
	\left( f, g \right) \equiv \int \dd x\, f^\dag g\,.
	\label{eq:bogo_def}
\end{align}
Intuitively, the $\alpha$-coefficient describes the probability of a particle-to-particle transition, whereas the $\beta$ coefficient encodes a particle-to-antiparticle transition. Complex conjugation exchanges particle and antiparticle states on both sides of these transitions.
Note here again that, by convention, for the lowest Landau level, we only have $r = 2$, \textit{i.e.}, $\alpha_0^{(2)} = \alpha_0$ and $\beta_0^{(2)} = \beta_0$.
Inserting Eq.~\eqref{eq:linear} into Eq.~\eqref{eq:expand_ini} and comparing it to Eq.~\eqref{eq:expand_t}, one finds the following relations between the initial creation/annihilation operators, $\hat b_n^{(r)}$ and $\hat d_n^{(r)}$, and those at a later time ($t > 0$), $\hat B_n^{(r)}$ and $\hat D_n^{(r)}$:
\begin{align}
	\hat B_n^{(r)} = \alpha_n^{(r)} \hat b_n^{(r)} - \beta_n^{(r) \ast} \hat d_n^{(r) \dag} \,, \quad
	\hat D_n^{(r) \dag} = \beta_n^{(r)} \hat b_n^{(r)} + \alpha_n^{(r) \ast} \hat d_n^{(r) \dag}\,.
\end{align}
Note that the Bogoliubov coefficients satisfy $|\alpha_n^{(r)}|^2 + |\beta_n^{(r)}|^2 = 1$.
From the discussion above, it is clear that the annihilation operators at a later time $t > 0$ do not erase the initial vacuum state.
More explicitly, the number densities become non-vanishing for $|\beta_n^{(r)}| > 0$ implying the particle production:
$\langle\hat B_n^{(r) \dag} B_n^{(r)} \rangle = \langle\hat D_n^{(r) \dag} D_n^{(r)} \rangle = \text{vol}\, (\mathbb{R}^2) |\beta_n^{(r)}|^2$.

\begin{figure}[t]
\centering
\includegraphics[width=0.5\linewidth]{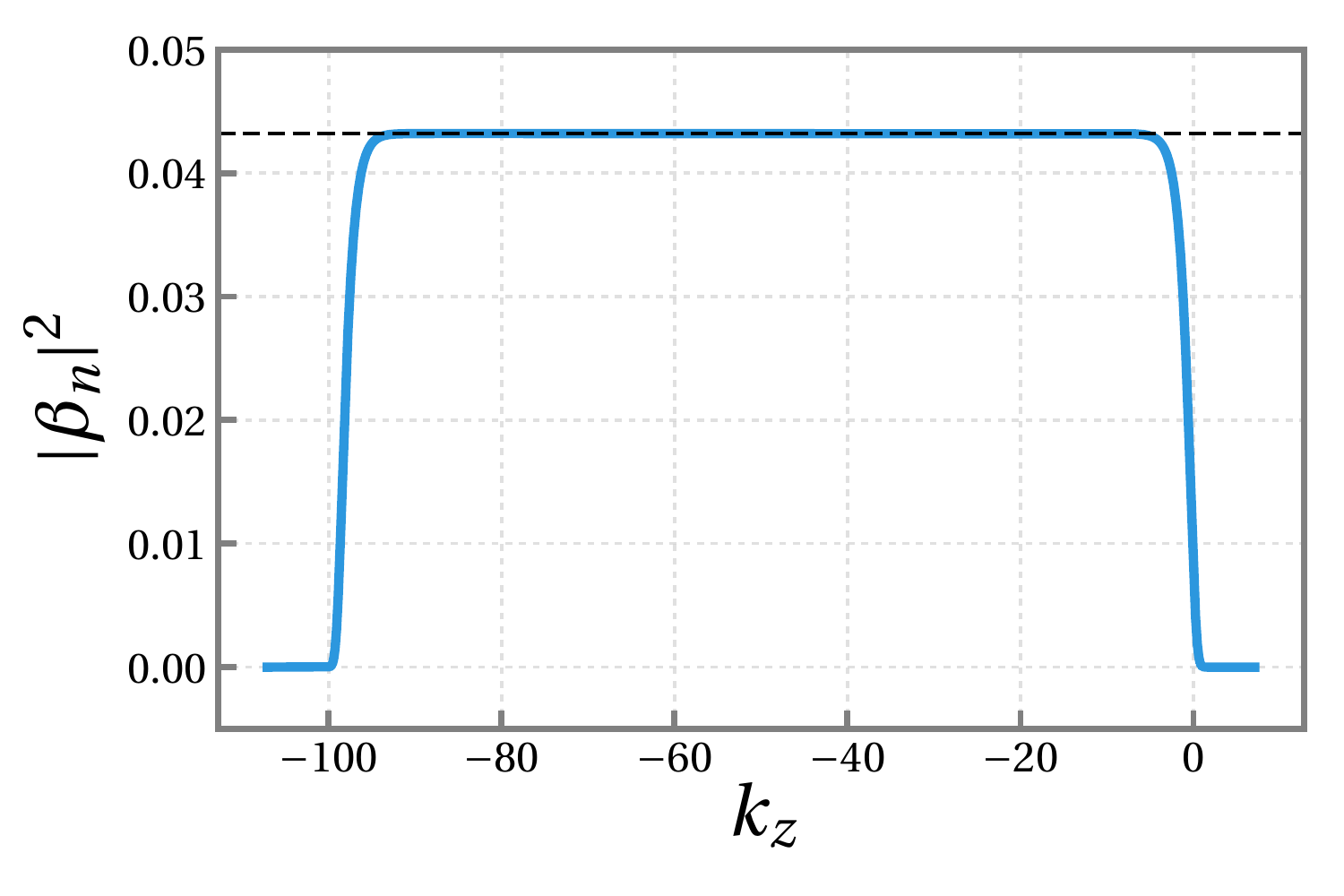}
\caption{Particle production encoded in the Bogoliubov coefficient $|\beta_n|^2$ for $t \gg \tau$ as a function of the momentum $k_z$, obtained by solving Eq.~\eqref{eq:bogo_evo} numerically.
Here we have adopted units of $\sqrt{g |Q| E} = 1$, taken $\tau = 100$ and $m_T a = 1$ in these units and have chosen $Q > 0$.
For comparison, the horizontal black dashed line indicates the value $e^{-\pi m_T^2 a^2 / g |Q| E}$, demonstrating excellent agreement with the analytical estimate \eqref{eq:asymptotic}.}
\label{fig:b2}
\end{figure}

\paragraph{Evolution of Bogoliubov coefficients.}
Instead of obtaining $\tilde u_n^{(r)}$ and $\tilde v_n^{(r)}$ by solving the equation of motion directly, it is convenient to derive an equivalent evolution equation for the Bogoliubov coefficients.
Taking a time derivative of Eqs.~\eqref{eq:bogo_def}, we obtain the evolution equation for the Bogoliubov coefficients (see App.~\ref{app:PP}):
\begin{align}
	\partial_t \alpha_n^{(r)} = - \beta_n^{(r)} \times  \frac{s g Q E}{2 \Omega_n^2}  e^{2 i \int^t \dd t' \Omega_n (t')} m_T a \,, 
	\qquad
	\partial_t \beta_n^{(r)} = \alpha_n^{(r)} \times  \frac{s g Q E}{2 \Omega_n^2}  e^{- 2 i \int^t \dd t' \Omega_n (t')} m_T a\,,
	\label{eq:bogo_evo_temp}
\end{align}
where the initial condition at $t \ll 0$ is $\alpha_n^{(r)} = 1$ and $\beta_n^{(r)} = 0$.
One can see that the structure of the equations is the same as those for the ordinary Schwinger effect without $B$-field by just replacing $m_B a$  with $\sqrt{k_x^2 + k_y^2}$~\cite{Cohen:2008wz}. 
For later convenience, we rewrite Eq.~\eqref{eq:bogo_evo_temp} as follows:
\begin{align}
	\partial_t \abs{\beta_n^{(r)}}^2 &= \frac{s g Q E}{2 \Omega_n^2} 
	\, 2\Re \left( 
		\alpha_n^{(r) \ast} \beta_n^{(r)} e^{2 i \int^t \dd t' \Omega_n} 
	 \right)\,, \nonumber \\
	 \partial_t \left( \alpha_n^{(r) \ast} \beta_n^{(r)} \right)
	 &= \frac{s g Q E}{2 \Omega_n^2} m_T a \, e^{- 2 i \int^t \dd t' \Omega_n}
	 \left( 1 - 2 \abs{\beta_n^{(r)}}^2 \right)\,.
	\label{eq:bogo_evo}
\end{align}
From a comparison with the ordinary Schwinger effect~\cite{Cohen:2008wz}, we obtain the resulting asymptotic,
 non-perturbative behavior of $| \beta_n^{(r)} |^2$ at $t \gg \tau$ for $|k_z| \gg m a$ and $|\Pi_z| \gg m a$ as
\begin{align}
	\abs{ \beta_n^{(r)} }^2 \simeq \theta (- p k_z) \theta (p(k_z + g Q E \tau) )
	\exp \left( - \frac{2 \pi n  \abs{B}}{E} \right) \exp \left(- \frac{\pi m^2 a^2}{g \abs{Q} E} \right) \,,
	\label{eq:asymptotic}
\end{align}
where we have introduced $p = \operatorname{sgn} (Q)$.
To avoid this notational complication, we will sometimes take $Q > 0$ and $B > 0$ in the following.
Be careful that, in this case, all the quantities in the exponent should be regarded as their absolute values if one would like to recover general signs for them.
Note that the inherently non-perturbative solution~\eqref{eq:asymptotic} can never be obtained by a perturbative expansion in $E / m^2a^2$. Vice versa, the full solutions to Eq.~\eqref{eq:bogo_evo} will contain perturbative corrections of order $E / m^2a^2$ to the asymptotic solution~\eqref{eq:asymptotic}. We will return to these in Sec.~\ref{subsec:Jind} and App.~\ref{sec:WFE}. 
Fig.~\ref{fig:b2} shows $| \beta_n^{(r)} |^2$ as a function of $k_z$ as a result of a numerical evaluation of Eq.~\eqref{eq:bogo_evo}.

\subsection{Chiral anomaly and chiral asymmetry}
\label{subsec:anomaly}

\paragraph{Deriving the chiral anomaly.}

In the presence of a chiral anomaly, a chiral fermion rotation relates the Chern-Simons term and the chiral fermion current~\cite{Fujikawa:1979ay,Fujikawa:1980eg}. For a massive fermion, this chiral fermion rotation also transforms the fermion mass term,
\begin{align}
 \partial_\mu  J_5^\mu = - \frac{g^2 Q^2}{8 \pi^2} F_{\mu \nu} \tilde F^{\mu  \nu} + 2 i m a\, \overline \psi \gamma_5 \psi \,,
 \qquad J_5^\mu =  \overline \psi \gamma^\mu \gamma_5 \psi \,.
 \label{eq:anomaly}
\end{align}
In the setup at hand, the (anti-)parallel electric and magnetic fields induce a non-vanishing Chern-Simons term, $F \tilde F = - 4 \bm{E} \cdot \bm{B}$. Consequently, the fermion production discussed in the previous section must obey this anomaly equation. In the following we will verify this explicitly. 
{This is not only an important and non-trivial cross-check of the consistency of our treatment, but also clarifies how the anomaly equation for a massive fermion arises (see also Ref.~\cite{Warringa:2012bq} as well as Refs.~\cite{Nielsen:1983rb,Domcke:2018eki,Domcke:2018gfr} for a derivation for massless fermions).} 
As we shall see in the following, only the lowest Landau level contributes to the anomaly equation.

Let us begin by evaluating the left-hand side of Eq.~\eqref{eq:anomaly}. Taking the spatial average and using the translational invariance of the initial vacuum state, we are left with only the $\mu = 0$ component $J_5^0$, which we can identify as the chiral charge
\begin{align}
	q_5 \equiv \frac{1}{\operatorname{vol} (\mathbb{R}^3)} \int \dd^3 x\, \vev{J_5^0}
	= \frac{1}{\operatorname{vol} (\mathbb{R}^3)} \int \dd^3 x\, \frac{1}{2}\vev{ \left[ \psi^\dag, \gamma_5 \psi \right] } \,,
	\label{eq:chiral_charge}
\end{align}
measuring the number density of right-handed minus left-handed particles. 
Here $\langle \dots \rangle$ indicates the vacuum expectation value with respect to the initial vacuum at $t \ll 0$.\footnote{
Note that the spatial average is redundant when assuming the initial vacuum state has translational invariance, and hence the chiral charge is essentially $q_5 = \vev{J_5^0}$.
Nevertheless we mostly keep it to avoid confusions.}
\footnote{We have here utilized the antisymmetrized chiral current. This is always possible by adding a total derivative to the action. This expression is useful because it makes a $CP$-invariant regularization explicit and allows in general a more transparent treatment of the $\eta$-invariant~\cite{Atiyah:1963zz,Atiyah:1968mp} (see Ref.~\cite{Domcke:2018gfr} for details). 
In the case at hand the $\eta$-invariant vanishes and hence this is not strictly necessary as we will see, but we will stick with the antisymmetrized version anyway.}

Inserting Eqs.~\eqref{eq:expand_ini} and \eqref{eq:linear} into \eqref{eq:chiral_charge}, we obtain (see App.~\ref{app:PP} for an explicit derivation and the reason why one may drop the regularization):
\begin{align}
	q_5 (t) = \frac{gQB}{2 \pi} \int \frac{\dd k_z}{2 \pi} \left[  
		\frac{\Pi_z}{\Omega_0}\, 2 \abs{\beta_0}^2 
		+ s \frac{m a}{\Omega_0} \left( e^{2 i \int^t \dd t' \Omega_0 (t')} \alpha_0^\ast \beta_0 + \text{H.c.} \right)
	\right]\,.
	\label{eq:chiral_charge_bogo}
\end{align}
Since contributions from the higher Landau levels cancel out between the two modes, $r = 1 ,2$, only the lowest Landau level contributes to the chiral charge.
In the next subsection, we will estimate $q_5$ for $t \gg \tau$ by inserting the solution of the evolution equation for the Bogoliubov coefficients.
Here we leave this expression as it is and proceed to the proof of the anomaly equation \eqref{eq:anomaly}.

To evaluate the left-hand-side of the anomaly equation \eqref{eq:anomaly}, let us take a time derivative of $q_5$ given in Eq.~\eqref{eq:chiral_charge_bogo}.
By using the equation of motion for the Bogoliubov coefficients in Eq.~\eqref{eq:bogo_evo}, we obtain
\begin{align}
	\dot q_5 = \frac{g Q B}{2 \pi} \int \frac{\dd k_z}{2 \pi} 
	\left[
		\frac{m^2a^2}{\Omega_0^3} gQE
		+ 2 s m a \left( i \alpha_0^\ast \beta_0 e^{2 i \int^t \dd t' \Omega_0 (t')} + \text{H.c.} \right)
	\right]\,.
	\label{eq:q5dot_temp}
\end{align}
Recalling the definition of the dispersion relation, $\Omega_0 = \sqrt{\Pi_z^2 + m^2 a^2}$, and changing the integration variable $k_z \to \Pi_z/ma \equiv u$, one may integrate the first term in Eq.~\eqref{eq:q5dot_temp} analytically, which provides the Chern-Simons term:
\begin{align}
	\frac{g Q B}{2 \pi} \int \frac{\dd k_z}{2 \pi} \frac{m^2a^2}{\Omega_0^3} gQE = \frac{g^2 Q^2}{4 \pi^2}  E B \int^{\infty}_{-\infty} \dd u \,\frac{1}{(u^2 + 1)^{3/2}} 
	= \frac{g^2 Q^2}{2 \pi^2} EB = - \frac{g^2 Q^2}{8 \pi^2} F_{\mu\nu} \tilde F^{\mu\nu}\,.
	\label{eq:cs_proof}
\end{align}
The last step of the proof is to show the second term in Eq~\eqref{eq:q5dot_temp} coincides with the expectation value of the pseudo scalar operator $2 m a\vev{\overline \psi i\gamma_5 \psi}$.
Again, inserting Eqs.~\eqref{eq:expand_ini} and \eqref{eq:linear} into $2 m a\vev{\overline \psi i\gamma_5 \psi}$, we get
\begin{align}
	\frac{1}{\operatorname{vol} (\mathbb{R}^3)} \int \dd^3 x\,
	 2 m a \, \vev{\overline \psi i\gamma_5 \psi }
	 = \frac{g Q B}{2 \pi} \int \frac{\dd k_z}{2 \pi}\, 2 s ma 
	 \left( i \alpha_0^\ast \beta_0 e^{2 i \int^t \dd t' \Omega_0 (t')} + \text{H.c.} \right)\,,
	 \label{eq:mass_proof}
\end{align}
which corresponds precisely to the second term in Eq.~\eqref{eq:q5dot_temp}.
Combining Eqs.~\eqref{eq:q5dot_temp}, \eqref{eq:cs_proof}, and \eqref{eq:mass_proof}, we finally arrive at
\begin{align}
	\dot q_5 = - \frac{g^2 Q^2}{8 \pi^2} F_{\mu\nu} \tilde F^{\mu\nu} + 2 m a \, \vev{\overline \psi i\gamma_5 \psi}\,,
	\label{eq:anomaly_2}
\end{align}
which is nothing but the anomaly equation \eqref{eq:anomaly}.
Here, for notational brevity, we have dropped the spatial average on $2 m a\vev{\overline \psi i\gamma_5 \psi}$.

This derivation clarifies that, if one studies the particle production by solving Eqs.~\eqref{eq:bogo_evo} and expands the fermion field as in Eqs.~\eqref{eq:expand_ini} and \eqref{eq:linear}, the chiral anomaly equation \eqref{eq:anomaly} is automatically fulfilled.

\begin{figure}[t]
\centering
\includegraphics[width=0.5\linewidth]{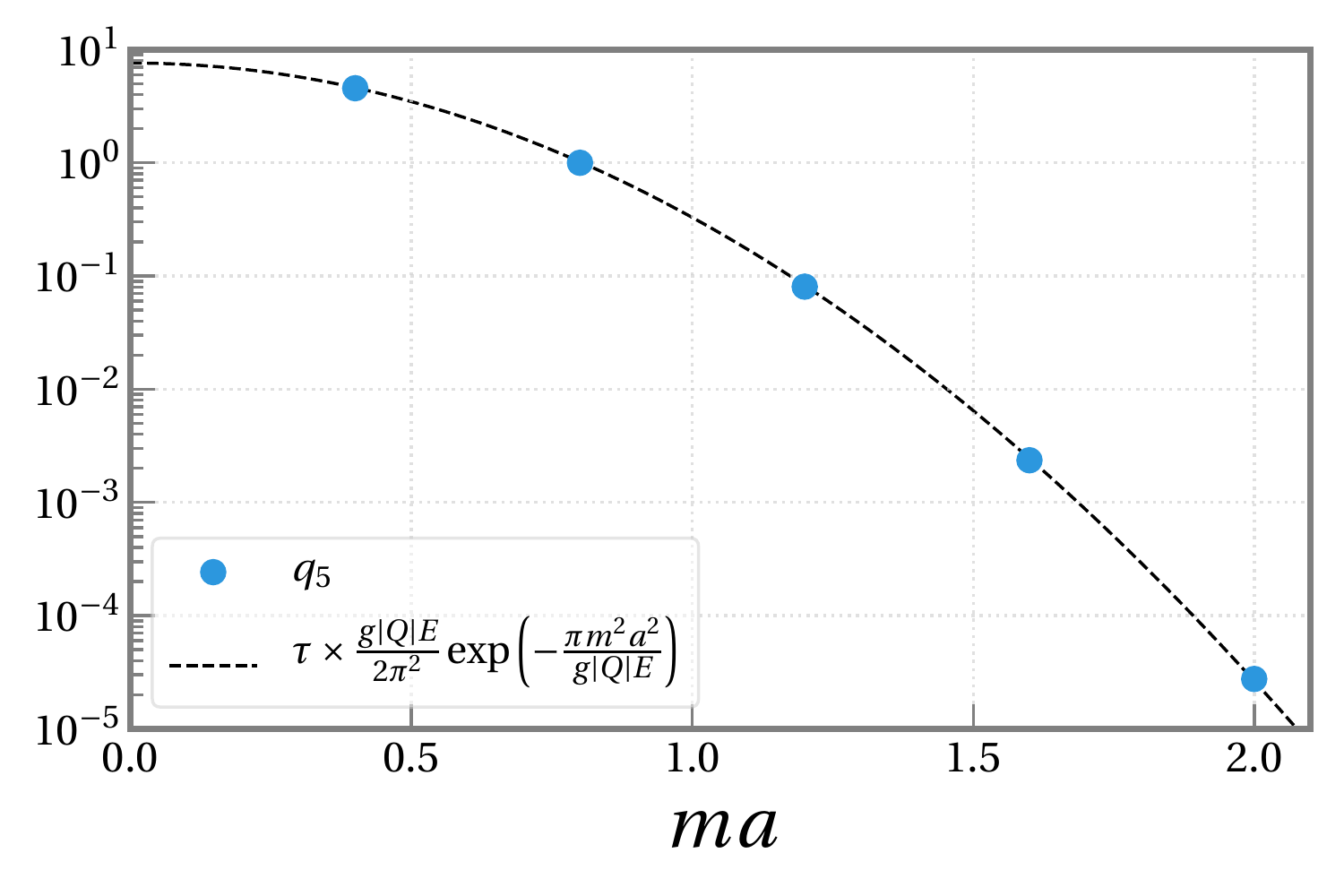}
\caption[Blue points show a numerical evaluation of {the chiral charge} $q_5$ as a function of {the fermion mass} $m$. ]{Blue points show a numerical evaluation of {the chiral charge} $q_5$ as a function of {the fermion mass} $m$. 
Here we have adopted a units $\sqrt{g |Q| E} = 1$, and taken $\tau = 150$.\footnotemark \,
For comparison, we also show the right-hand side of Eq.~\eqref{eq:q5_t_int} as a black dashed line, demonstrating excellent agreement with this analytical estimate.
}
\label{fig:mvsq5}
\end{figure}

\paragraph{Chiral asymmetry.}
We have shown that the chiral anomaly equation is guaranteed for the solutions of Eq.~\eqref{eq:bogo_evo}.
Here we explicitly evaluate the resulting chiral charge after pumping up an electric field from $t = 0$ to $t = \tau$ as in Eq.~\eqref{eq:pumpup}.
In particular, we will clarify how each term in the anomaly equation contributes to the final asymmetry in the chiral charge with the help of a numerical evaluation of Eq.~\eqref{eq:bogo_evo}.
We take $Q > 0$ for notational brevity of the following discussion.

Let us start with the chiral charge, Eq.~\eqref{eq:chiral_charge_bogo}.
To get the final chiral asymmetry, we need to evaluate the chiral charge at a late time, \textit{i.e.}, $q_5 (t \gg \tau)$.
For $t \gg \tau$, the electric field is already turned off and
hence the Bogoliubov coefficients are no longer evolving.
Contrary to the first term in Eq.~\eqref{eq:chiral_charge_bogo}, one can show that the second term, $\int \dd k_z (ma / \Omega_0)\Re(\alpha_0^\ast \beta_0 e^{2 i \int \dd t' \Omega_0})$,  does not contain a term which grows with $\tau$. We have verified this numerically.
Hence using the asymptotic non-perturbative behavior for $|\beta^2|$ given in Eq.~\eqref{eq:asymptotic}, we can estimate the resulting chiral charge as
\begin{align}
	q_5 (t \gg \tau) \simeq \tau \times \frac{g^2 Q^2}{2 \pi^2} E B \, \exp \left( - \frac{\pi m^2 a^2}{g Q E} \right)\,,
	\label{eq:q5_t_int}
\end{align}
which is consistent with the argument given in Ref.~\cite{Fukushima:2010vw,Warringa:2012bq}.
We also check this estimation by solving Eq.~\eqref{eq:bogo_evo} numerically, as depicted in Fig.~\ref{fig:mvsq5}.
Taking the massless limit $m \to 0$, one  recovers the well-known result for chiral fermions in Ref.~\cite{Nielsen:1983rb}.\footnotetext{Note that this $E$ appearing in $(g^2Q^2/2\pi^2) EB e^{-\pi m^2 a^2/g |Q| E}$ is its constant value for $t=0$ to $\tau$ in Eq.~\eqref{eq:pumpup}, while the electric field in $F \tilde F$ is the time-dependent one $E(t)$.}

\begin{figure}[t]
\centering
\includegraphics[width=0.5\linewidth]{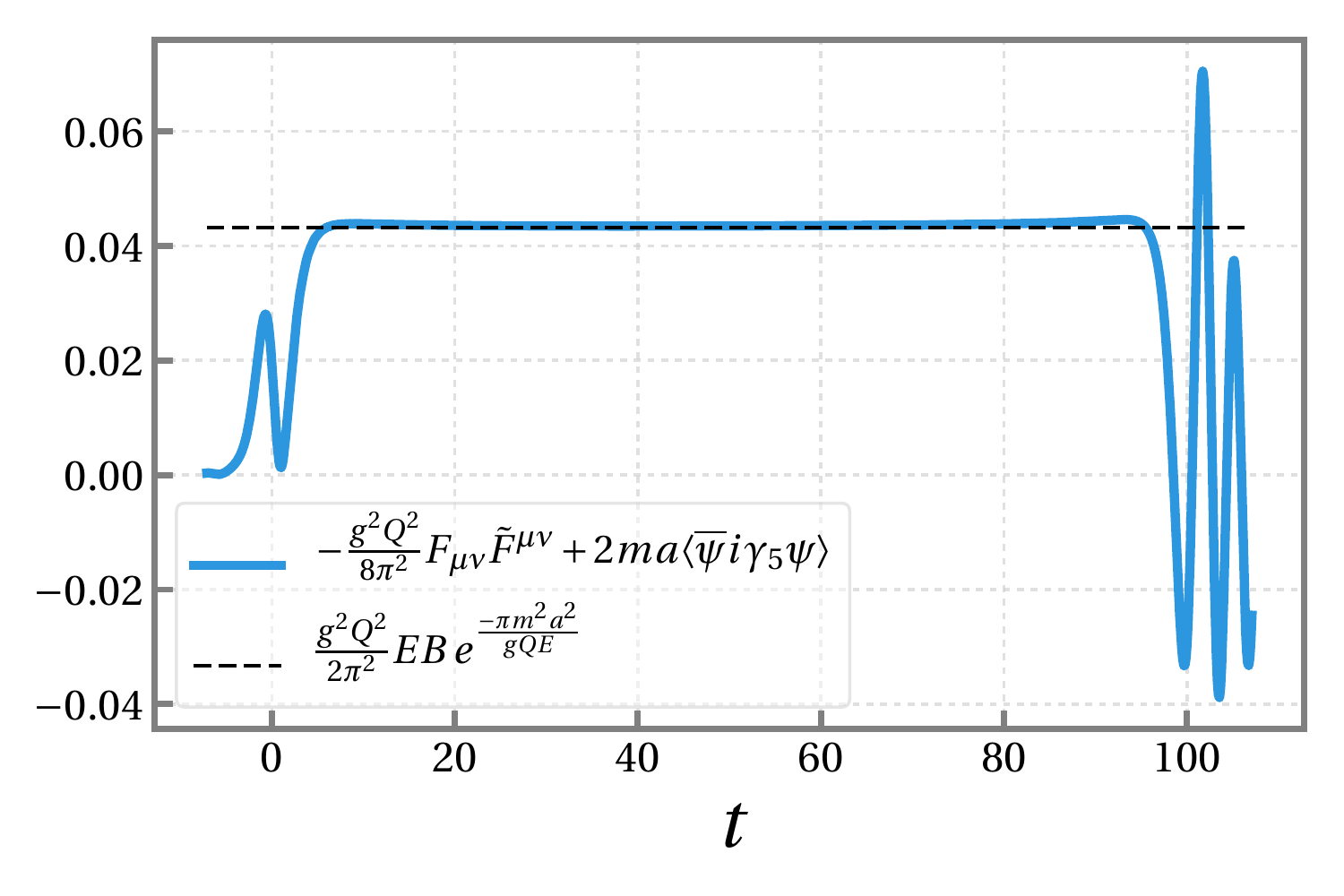}
\caption[Contributions to the chiral anomaly equation.]{{Contributions to the chiral anomaly equation. }
The blue solid line shows a numerical evaluation of {the chiral charge $q_5$, determined by evaluating} the right-hand-side of Eq.~\eqref{eq:anomaly_2} [using Eqs.~\eqref{eq:mass_proof} and \eqref{eq:bogo_evo}] as a function of $t$ for the gauge field given in Eq.~\eqref{eq:pumpup}. Here we have adopted units of $\sqrt{g |Q| E} = 1$, and taken $\tau = 100$, $ma = 1$ and $\sqrt{g \vert Q \vert B} = 1$.
For comparison, the black dashed line indicates $(g^2Q^2/2\pi^2) EB e^{-\pi m^2 a^2/g |Q| E}$.
If we integrate the blue solid line over time, the integrand is dominated by the plateau, which gives $\simeq \tau \times (g^2Q^2/2\pi^2) EB e^{-\pi m^2 a^2/g |Q| E}$, confirming Eq.~\eqref{eq:mass_t_int}.
}
\label{fig:anomaly_rhs}
\end{figure}

It is instructive to see how each term in the right-hand-side of Eq.~\eqref{eq:anomaly_2} contributes to this resulting chiral asymmetry.
The Chern-Simons term obviously yields
\begin{align}
	- \frac{g^2 Q^2}{8 \pi^2} \int \dd t' \, F_{\mu \nu} \tilde F^{\mu\nu} \simeq \tau \times \frac{g^2 Q^2}{2 \pi^2} E B\,.
	\label{eq:CS_t_int}
\end{align}
To estimate the contribution arising from $2 m a \vev{\overline \psi i \gamma_5 \psi}$, we need to evaluate the imaginary part of $\alpha_0^\ast \beta_0 e^{2 i \int \dd t' \Omega_0}$ as Eq.~\eqref{eq:mass_proof}.
As can be seen from Fig.~\ref{fig:anomaly_rhs},
for a long enough $\tau$ such that $|g Q E \tau| \gg ma$, the time integration of the pseudo-scalar operator, $2 m a \vev{\overline \psi i \gamma_5 \psi}$ is dominated by the plateau from $t = 0 $ to $t = \tau$ and hence we get
\begin{align}
	\int \dd t' \, 2 m a \vev{ \overline \psi i \gamma_5 \psi }
	\simeq \tau \times \frac{g^2 Q^2}{2 \pi^2} E B
	\left[ \exp \left( - \frac{\pi m^2 a^2}{g Q E} \right) - 1 \right]\,, \label{eq:mass_t_int}
\end{align}
which coincides with the result in Ref.~\cite{Warringa:2012bq,Copinger:2018ftr}.
In the massless limit, this term vanishes as expected.
Now we see that the estimation for the final chiral charge in Eq.~\eqref{eq:q5_t_int} is consistent with the summation of Eqs.~\eqref{eq:CS_t_int} and \eqref{eq:mass_t_int}.
We note in particular the non-trivial ``-1'' contribution in Eq.~\eqref{eq:mass_t_int}, which cancels the $F \tilde F$ contribution to the chiral charge, ensuring the overall exponential suppression.
Since due to the operator nature of the anomaly equation this cancellation must hold exactly,
one can also see how this term emerges explicitly by perturbative computations omitting the non-perturbative exponential behavior in the limit of $m^2 a^2 \gg E$.
As shown in App.~\ref{sec:WFE}, the leading order perturbative contribution is
\begin{align}
	\alpha_0^\ast \beta_0 e^{2 i \int^t \dd t' \Omega_0}
	= i s m a \frac{g Q E}{4 \Omega_0^3} + \cdots \,.
	\label{eq:cross_lo}
\end{align}
Inserting Eq.~\eqref{eq:cross_lo} into Eq.~\eqref{eq:mass_proof}, we get
\begin{align}
	\lim_{m^2 a^2 \gg E}2 m a \vev{ \overline \psi i \gamma_5 \psi }
	& \simeq - \frac{g^2 Q^2}{4 \pi^2} E B \int^\infty_{- \infty} \dd u\, \frac{1}{(u^2 + 1)^{3/2}} 
	= -  \frac{g^2 Q^2}{2 \pi^2} EB\,.
	\label{eq:mass_estimation}
\end{align}
It shows that the perturbative contribution of the mass term
indeed cancels the $F\tilde{F}$ term at the leading order,
which is essential for the anomaly equation as discussed above.

\subsection{Induced current and Euler-Heisenberg action}
\label{subsec:Jind}
Once the fermions get generated, they are accelerated in the background gauge field and tend to erase it.
On top of this, even if the fermions are so heavy that their production is not efficient, the coupling to the fermions changes the effective action for the gauge field via the running of the gauge coupling and via higher dimensional operators suppressed by the fermion mass.
These contributions correspond to the Euler-Heisenberg action~\cite{Heisenberg:1935qt}, which describes the non-linear dynamics of QED induced by one-loop effects.

In this subsection we explicitly show that the expectation value of the current contains all the effects mentioned above.
More concretely, we will evaluate
\begin{align}
	\vev{J_z} &= \frac{1}{\operatorname{vol} (\mathbb{R}^3)}\int \dd^3 x\, \frac{1}{2} \left[ \vev{\overline \psi, \gamma^3 \psi} \right] \label{eq:current_def} \\[.5em]
	& = \frac{g Q B}{2 \pi}  \int \frac{\dd \Pi_z}{2 \pi}\, \sum_{n,r} \left[ \frac{s \Pi_z}{\Omega_n} 2 \abs{\beta_n^{(r)}}^2 + \frac{m_T a}{\Omega_n} \left( \alpha_n^{(r) \ast} \beta_n^{(r)} e^{2 i \int^t \dd t' \Omega_n} + \text{H.c.} \right) \right]\,.
	\label{eq:current}
\end{align}
To get the expression in the second line, we have inserted Eqs.~\eqref{eq:expand_ini} and \eqref{eq:linear}.

\paragraph{Induced current.}
Let us first discuss the contribution from the particle production.
The effect of particle production is imprinted in the non-perturbative part proportional to $e^{- \pi m^2a^2 / g Q E}$.
As discussed above Eq.~\eqref{eq:q5_t_int}, $\int \dd \Pi_z (m_T a / \Omega_n)\Re( \alpha^\ast \beta e^{2i \int \dd t' \Omega})$ does not contain any term which grows with $\tau$.
As a result, we arrive at the following expression of the current induced for a large enough $\tau$
by the particle production after pumping up the electric field as Eq.~\eqref{eq:pumpup}:
\begin{align}
	g Q \vev{J_z}_\text{ind} &\equiv  \frac{g^2 Q^2 B}{2 \pi^2} \int \dd \Pi_z\, \frac{s \Pi_z}{\Omega_n} \abs{\beta_n^{(r)}}^2_\text{NP} 
	 \simeq \tau \times \frac{(g Q)^3}{2 \pi^2} s E B \, \sum_{n,r} e^{- \frac{\pi}{g \abs{Q} E} ( m^2 a^2 + 2 n s g Q B )} \nonumber \\
	& = \tau \times \frac{(g \abs{Q})^3}{2 \pi^2} E \abs{B}\,
	\coth \left( \frac{\pi \abs{B}}{E} \right)\,
	e^{- \frac{\pi m^2 a^2}{g \abs{Q} E}}\,,
	\label{eq:induced}
\end{align}
reproducing ~\cite{Warringa:2012bq}.
The subscript ``ind'' means the current induced by the particle production.
We have multiplied a factor $g Q$ in order to make the structure which couples to the gauge field explicit. Note that contrary to the computation of the chiral charge, all Landau levels contribute to the induced current.
This result is also consistent with Refs.~\cite{Bunkin:1969if,Nikishov:1969tt,Dunne:2004nc} obtained from evaluating the imaginary part of the effective action after integrating out fermions. See also Ref.~\cite{Cohen:2008wz} to see how to relate their results to ours.

\paragraph{Vacuum contribution.}
As discussed above, the contribution from the particle production is exponentially suppressed for heavy fermions, \textit{i.e.}, $e^{- \pi m^2 a^2 / g \abs{Q} E}$.
However, this does not mean that those heavy fermions don't contribute to $\vev{J_z}$.
One may anticipate their effect because the low-energy effective action for the gauge field after integrating out heavy fermions should involve the threshold correction to
the running of the gauge coupling as well as higher dimensional operators suppressed by the heavy fermion mass.
In the following we explicitly compute these effects by means of a perturbative expansion in $E / m^2 a^2$ and $B / m^2 a^2$. 
{Here we ignore terms that contain more than one time derivatives in $E$.
These terms generate higher derivative corrections to the Euler-Heisenberg terms
in the low-energy effective action.}
See also Ref.~\cite{Banyeres:2018aax} for the case of a vanishing magnetic field.

As discussed in appendix~\ref{sec:WFE}, in addition to the non-perturbative term from the particle production, we also have perturbative contributions in $|\beta|^2$ and $\alpha^\ast \beta e^{2 i \int \dd t' \Omega}$.
By utilizing integration by parts repeatedly, one may systematically perform perturbative expansions in $E / m^2 a^2$ and $B / m^2 a^2$.
After some algebra, we obtain
\begin{align}
	2 \Re \left( \alpha_n^{(r)\ast} \beta_n^{(r)} e^{2 i \int^t \dd t' \Omega_n} \right)_\text{P} &= 
	s \frac{m_T a}{4 \Omega_n} \left[ 
		\frac{g Q \dot E}{\Omega_n^3} - \frac{(g Q)^3}{8 \Omega_n^9} E^2 \dot E \left( 3 m_T^2 a^2 - 38 \Omega_n^2 + 248 \Pi_z^2 \right) 
	\right] + \text{(Odd in $\Pi_z$)} + \cdots \,, \label{eq:cross_pert}\\
	\abs{\beta_n^{(r)}}^2_\text{P} & = 
	\frac{7}{32} m_T^2 a^2 \frac{(g Q)^3}{\Omega_n^{10}} E^2 \dot E \, \Pi_z + \text{(Even in $\Pi_z$)} + \cdots\,,
	\label{eq:beta2_pert}
\end{align}
where the subscript ``P'' indicates perturbative contributions and the ellipsis represents the higher order terms in $E /m^2 a^2$ and $B / m^2 a^2$.
We have dropped terms which do not survive after integration over $\Pi_z$, \textit{i.e.}, terms odd in $\Pi_z$ for Eq.~\eqref{eq:cross_pert} and even in $\Pi_z$ for Eq.~\eqref{eq:beta2_pert}.
See appendix~\ref{sec:WFE} for the derivation.
Inserting Eqs.~\eqref{eq:cross_pert} and \eqref{eq:beta2_pert} into Eq.~\eqref{eq:current}, we get
\begin{align}
	g Q \vev{J_z}_\text{vac} &=
	\frac{g Q \dot E}{16 \pi^2} \sum_{n,r} \frac{s g Q B}{m_T^2 a^2} 
	\int \dd u \, \left[ \frac{1}{(u^2 + 1)^{5/2}}
	- \frac{(g Q E)^2}{8 m_T^6 a^6} \frac{1}{(u^2 + 1)^{11/2}} \left( 234 u^2 - 38 (u^2 + 1) + 3 \right)
	\right] + \cdots \nonumber \\
	&= s \frac{g Q \dot E}{\pi^2} \frac{s g Q B}{m^2 a^2} 
	\sum_{n,r} \left[ \frac{1}{12} \frac{1}{\left(\frac{2 s g QB}{m^2 a^2}\right) n + 1} + \frac{1}{15} \left( \frac{g Q E}{m^2 a^2} \right)^2 \left(\frac{1}{\left(\frac{2 s g QB}{m^2 a^2}\right) n + 1}\right)^3 \right] + \cdots\,,
	\label{eq:Jvac_div}
\end{align}
with $u = \Pi_z / ma$.
The summation over the first term in the square brackets contains a divergence as we will see.
Let us make use of the zeta function regularization.
We can cope with this summation with the help of the Hurwitz zeta function $\zeta (p, q)$:
\begin{align}
	\sum_{n,r} \frac{1}{( b n + 1)^p} = \frac{2}{b^p}\, \zeta (p, 1/b) -  1 \,,
	\qquad \zeta (p ,q) \equiv \sum_{n = 0} \frac{1}{(n + q)^p}\,, 
	\qquad b = \frac{2 s g Q B}{m^2 a^2}\,,
\end{align}
where to evaluate the two terms in Eq.~\eqref{eq:Jvac_div} we will be interested in $p = 1$ and $p = 3$.
Here, in the first equation, we have utilized the fact that higher Landau levels ($n \geq 1$) have two modes ($r = 1, 2$) for each $n$ while the lowest Landau level ($n = 0$) has only one mode.
The last ``-1'' stems from this mismatch.
It is known that the Hurwitz zeta function can be extended to a complex value of $p$ via the analytic continuation, which contains a pole at $p = 1$. By expanding it around $p = 1$ and also Taylor-expanding in $b$, we get
\begin{align}
	\lim_{p \to 1} \, \sum_{n,r} \frac{1}{(b n + 1)^p} = 2\left[ \frac{1}{b (p - 1)} + \frac{1}{2} + \frac{b}{12} + \cdots  \right] - 1
	= \frac{2}{b (p - 1)} + \frac{b}{6} + \cdots\,,
	\label{eq:zeta1}
\end{align}
where the ellipses involve higher order terms in $B/m^2$ and terms which vanish at $p = 1$.
We also need the behavior at $p = 3$, which is given by
\begin{align}
	\sum_{n,r} \frac{1}{(b n + 1)^3} = 2 \left( \frac{1}{2 b} + \frac{1}{2} + \cdots \right) - 1 = \frac{1}{b} + \cdots\,,
	\label{eq:zeta3}
\end{align}
where the ellipses imply higher order terms in $B/m^2$.
Using the relations~\eqref{eq:zeta1} and \eqref{eq:zeta3}, we finally obtain the following form of the current which contains the divergence and finite terms at the leading order in $E / m^2 a^2$ and $B / m^2 a^2$:
\begin{align}
	g Q \vev{J_z}_\text{vac} = g^2 \frac{Q^2}{12 \pi^2} \ln \frac{\hat \Lambda^2}{m^2} \times \dot E + \frac{(g Q)^4}{36 \pi^2} \frac{B^2 \dot E }{m^4 a^4} + \frac{(g Q)^4}{30 \pi^2} \frac{E^2 \dot E}{m^4 a^4}+ \cdots\,.
	\label{eq:current_vac}
\end{align}
Recalling that the level, $n$, represents the transverse momentum squared and the summation is asymptotically $\sum^N 1/ n \sim \ln N$ for $N \gg 1$, we replace the divergence at $p = 1$ with $\ln \hat \Lambda^2 / m^2$ with $\hat \Lambda$ being a UV-cutoff.

Now we are ready to discuss the physical origin of Eq.~\eqref{eq:current_vac}.
For this purpose, it is convenient to see how this current enters the equation of motion for the gauge field once we make the gauge field dynamical.
Since our expression for the current Eq.~\eqref{eq:current_vac} was based on the vector potential \eqref{eq:A}, the Maxwell equation reduces to
\begin{align}
	0 = \dot E_{\hat\Lambda} + g_{\hat\Lambda} Q \vev{J_z}_\text{vac}  \rightarrow 
	0 = \left[  
		\left( \frac{1}{g_{\hat\Lambda}^2} + \frac{Q^2}{12 \pi^2} \ln \frac{\hat \Lambda^2}{m^2} \right) 
		+ \frac{Q^4}{36 \pi^2} \frac{(g_{\hat\Lambda} B_{\hat\Lambda} )^2}{m^4 a^4} + \frac{Q^4}{30 \pi^2} \frac{(g_{\hat\Lambda} E_{\hat\Lambda})^2}{m^4 a^4}
	 \right] (g_{\hat\Lambda} \dot E_{\hat\Lambda}) + \cdots\,.
	 \label{eq:eom_gauge}
\end{align}
Here we have explicitly indicated bare quantities  which should be renormalized with the subscript $\hat \Lambda$.

One can easily see that the first parenthesis is nothing but the running gauge coupling evaluated at the fermion mass $m$
\begin{align}
	\frac{1}{g^2_m} = \frac{1}{g_{\hat\Lambda}^2} + \frac{Q^2}{12 \pi^2} \ln \frac{\hat \Lambda^2}{m^2}\,,
	\label{eq:running}
\end{align}
and hence it is finite.
Moreover, recalling the Ward-Takahashi identity, one finds
$g_{\hat\Lambda} E_{\hat\Lambda} = g_m E_m$ and $g_{\hat\Lambda} B_{\hat\Lambda} = g_m B_m$ where quantities with a subscript $m$ are the renormalized ones evaluated at the fermion mass scale.
Noticing that the electromagnetic field appears the combinations $g_{\hat \Lambda} E_{\hat \Lambda}$ or $g_{\hat \Lambda} B_{\hat \Lambda}$ in Eq.~\eqref{eq:eom_gauge}, we may rewrite the equation in terms of the renormalized quantities\footnote{
	Note that this property also holds for the induced current \eqref{eq:induced} as expected. Hence one may keep the same expression given in \eqref{eq:induced} by just replacing all the bare quantities with the renormalized ones.
} 
\begin{align}
	0 = \dot E + \frac{(gQ)^4}{36 \pi^2} \frac{B^2 \dot E}{m^4 a^4} + \frac{(g Q)^4}{30 \pi^2} \frac{E^2 \dot E}{m^4 a^4} +\cdots\,,
	\label{eq:eom_gauge_ren}
\end{align}
where we have suppressed the subscript $m$ for notational brevity.

The final task is to identify the origin of the terms suppressed by the power of the fermion mass in Eq.~\eqref{eq:eom_gauge_ren}.
Such higher dimensional operators arise in the low energy effective action when we integrate out a heavy fermion.
Its concrete form is known as the Euler-Heisenberg action.
On top of the kinetic term for the gauge field, this yields the following corrections~\cite{Heisenberg:1935qt,Dunne:2004nc}
\begin{align}
	{\cal L}_\text{EH} = \frac{(g Q)^4}{360 \pi^2} \frac{1}{m^4 a^4} \left[ \left( \bm{E}^2 - \bm{B}^2 \right)^2 + 7 \left( \bm{E} \cdot \bm{B} \right)^2 \right] + \cdots\,,
	\label{eq:EH}
\end{align}
where the ellipses imply terms higher powers in $E/m^2a^2$ and $B/m^2a^2$.
Inserting the specific gauge field configuration \eqref{eq:A} into Eq.~\eqref{eq:EH}, we find the following terms in the equation of motion for the gauge field after variation with respect to $A_z$:
\begin{align}
	\frac{\delta {\cal L}_\text{EH}}{\delta A_z} = 
	\frac{(g Q)^4}{30 \pi^2} \frac{E^2 \dot E}{m^4 a^4} 
	+ \frac{(g Q)^4}{36 \pi^2} \frac{B^2 \dot E}{m^4 a^4} + \cdots\,.
\end{align}
This coincides with the second and third term in Eq.~\eqref{eq:eom_gauge_ren}, implying that the origin of \eqref{eq:eom_gauge_ren} is nothing but the Euler-Heisenberg action.

Before closing this section we would like to emphasize that, even if one takes a large enough fermion mass to exponentially suppress the fermion production, there always exist power-law suppressed terms in the expectation value of the current.
However their origin is merely higher dimensional operators which should not to be confused with the particle production.
This conclusion holds for any particle production in general.

\section{Applications}
\label{sec:applications}

Fermion production in strong helical, abelian gauge fields plays an important role in many different areas of particle physics, solid state physics and cosmology. While the relevance of these questions for the former fields of research has been known for a long time (see \textit{e.g.}, the seminal papers~\cite{Sauter:1931zz} and \cite{Nielsen:1983rb,Heisenberg:1935qt,Schwinger:1951nm}), the importance of chiral fermion production for cosmological processes has only fairly recently gained significant attention. 
{In the remainder of this section, we will focus on the dual production of gauge fields and fermions in cosmic inflation, triggered by coupling the inflaton to the Chern-Simons term of an Abelian gauge theory. However, the results presented here also have important implications for a much broader class of processes in the early Universe. For example, this coupling to the Chern-Simons term can play a crucial role in the phase of preheating after cosmic inflation~\cite{Adshead:2015pva,Cuissa:2018oiw,Adshead:2019lbr,Adshead:2019igv} and has been employed in~\cite{Caprini:2014mja,Adshead:2016iae,Caprini:2017vnn,Choi:2018dqr, Domcke:2019mnd,Sobol:2019xls} to generate large scale magnetic fields which can survive in the intergalactic voids until today. The (spontaneously) $CP$-violating nature of this coupling moreover makes it a promising candidate to implement baryogenesis, either by a direct coupling of the inflaton (or SM Higgs) to the fermion current~\cite{Chiba:2003vp,Kusenko:2014lra,Kusenko:2014uta,Pearce:2015nga,Yang:2015ida}, or by sourcing a baryon asymmetry from the generated $CP$-violating background of gauge fields~\cite{Joyce:1997uy,Bamba:2006km,Kamada:2016eeb,Kamada:2016cnb,Anber:2015yca,Jimenez:2017cdr,Domcke:2019mnd}. At much lower energy scales, the same coupling to the Chern Simons term can play a crucial role in explaining the smallness of the electroweak scale~\cite{Hook:2016mqo,Choi:2016kke,Tangarife:2017vnd,Tangarife:2017rgl,Fonseca:2018xzp}.}

In this section, we will apply the results derived in the previous sections to analyze particle production in quasi de-Sitter space, \textit{i.e.}, during cosmic inflation. The analysis of gauge field production in this context, sourced by a coupling of the particle driving cosmic inflation (the inflaton) to the Chern-Simons density of an abelian gauge group, dates back to Ref.~\cite{Turner:1987bw}. The coupling to a chiral current of (uncharged) fermion current was first studied in \cite{Adshead:2015kza} and recently refined in \cite{Adshead:2018oaa}. The simultaneous coupling to both abelian gauge fields and massless charged fermions, as dictated by the chiral anomaly equation, was first studied in \cite{Domcke:2018eki}. Here we extend these results to include a finite fermion mass. As we will see, the Chern-Simons coupling induces remarkable signatures in the predicted scalar and tensor spectrum at small (sub-CMB) length scales. The presence of fermions, and in particular the presence of a fermion mass term, drastically alters these predictions.

\subsection{Upper bound on gauge field production during cosmic inflation}
\label{subsec:EBbound}

We consider a pseudo-scalar singlet $\phi$ (referred to as `inflaton' or `axion' in the following), coupled to the Chern-Simons term and to the chiral fermion current,
\begin{align}
 S = \int d^4 x   \left\{ \sqrt{-g} \left[\frac{1}{2} g^{\mu \nu} \partial_\mu \phi \partial_\nu \phi - V(\phi) \right] - \frac{1}{4} F_{\mu \nu}F^{\mu \nu} + \overline\psi (i \slashed{D} - ma) \psi  +  \frac{\alpha}{4 \pi f_a} \phi F_{\mu \nu} \widetilde F^{\mu \nu}  
 \right\} \,,
 \label{eq:action-inflation}
\end{align}
see App.~\ref{app:notation} for details on our notation and conventions.\footnote{{Note in particular that here we have taken the $\phi$-dependent? complex phase in the Dirac mass term to be zero. This corresponds to a particular choice of the couplings transporting the spontaneous $CP$-violation. We leave an investigation of the fully general model parameter space for future work.}
}
We will leave the scalar potential $V(\phi)$ unspecified for now, assuming only that it is flat enough to support slow-roll inflation, $\dot \phi/H \ll 1, \ddot \phi/(\dot \phi H) \ll 1$, where $H = \dot a/a$ denotes the Hubble parameter.\footnote{See Ref.~\cite{Domcke:2016bkh} for a detailed discussions on the role of the shape of the scalar potential.}  The resulting equation of motion for the fermion is given in Eq.~\eqref{eq:eom}. The equations of motion for the scalar field and the gauge fields read
\begin{align}
 \ddot \phi + 3 H \dot \phi + V'(\phi) -  \frac{\alpha}{4 \pi f_a a^4} \vev{ F_{\mu \nu} \tilde F^{\mu \nu} }  & = 0 \,,  \label{eq:inflaton_eom}\\
 \partial_\mu F^{\mu\nu} - \partial_\mu \left( \frac{\alpha \phi}{\pi f_a} \, \tilde F^{\mu \nu} \right) - g Q J_\psi^\nu & = 0 \,. \label{eq:eom_A}
\end{align}
Here as usual, we have decomposed the inflaton field into a homogeneous background with (small) perturbations, $\phi(t,\bm x) \mapsto \phi(t) + \delta \phi(t, \bm x)$. The backreaction of the gauge fields on the inflaton background is encoded in the spatial average $\langle F \tilde F \rangle$ in Eq.~\eqref{eq:inflaton_eom}, whereas the fermion backreaction enters through  $J_\psi^\nu = \overline{\psi} \gamma^\mu \psi$. 
Decomposing the vector potential into longitudinal and transverse modes, $\bm{A} = \bm{A}_L + \bm{A}_T$ with $0 = \bm{\nabla} \cdot \bm{A}_T$, one can readily see that $\dot \phi$ never affects $\bm{A}_L$ but leads to an instability for $\bm{A}_T$ as discussed below.

\paragraph{Without fermion backreaction.} Neglecting for a moment the fermion current $J^\nu_\psi$, Eq.~\eqref{eq:eom_A} can be expressed in Fourier space as
\begin{align}
 0 = \left[ \partial_0^2 + k(k \pm 2 \lambda \xi a H) \right] A_\pm(\eta, \bm{k}) \,,
 \label{eq:eom_A_helical}
\end{align}
where $\xi$ parameterizes the inflaton velocity,
\begin{align}
 \xi \equiv \frac{\alpha \lambda \dot \phi}{2 \pi f_a H} > 0 \,,  \qquad \text{with  } \lambda \equiv \text{sgn}(\dot \phi)
 \label{eq:def_xi}
\end{align}
and we have introduced the helicity decomposition for the transverse mode
\begin{align}
	\bm A_T (\eta, \bm x) = \int \frac{\dd^3 k}{(2 \pi)^{3/2}} 
	\sum_{\sigma = \pm} \left[
		\bm \epsilon^{(\sigma)} (\bm{k}) a^{(\sigma)}_{\bm{k}} A_\sigma (\eta, \bm{k}) e^{i \bm{k}\cdot \bm{x}}
		+ \text{H.c.}
	\right],
	\label{eq:Adecom}
\end{align}
with the polarization tensors obeying
$\bm{k} \cdot \bm{\epsilon}^{(\sigma)} (\bm{k}) = 0$,
$\bm{\epsilon}^{(\sigma)} (\bm{k})^\ast \cdot \bm{\epsilon}^{(\sigma')} (\bm{k}) = \delta_{\sigma \sigma'}$, and  
$\bm{k} \times \bm{\epsilon}^{(\sigma)} = - \sigma i k \bm{\epsilon}^{(\sigma)}(\bm{k})$. Assuming $\dot \xi/H \ll 1$, consistent with the slow-roll approximation, Eq.~\eqref{eq:eom_A_helical} is solved by the Whittaker functions $W_{k,m}(z)$. This leads in particular to one exponentially growing mode (in accordance with the negative effective squared mass in Eq.~\eqref{eq:eom_A_helical}). After imposing Bunch-Davies initial conditions in the far past, the solution for this mode reads
\begin{align}
 A_{- \lambda} (\eta, \bm k) \simeq \frac{e^{\pi \xi/2}}{\sqrt{2 k}} W_{- i \xi, 1/2} (2 i k \eta) \,.
\end{align}
From this, we can compute the expectation values of the electric and magnetic fields as
\begin{align}
	\vev{\hat{\bm{E}}^2}
	&=
	\frac{e^{2 \pi \xi}}{\abs{\xi}^3} H^4 \times 
		\left( \frac{\abs{\xi}^3}{4 \pi^2} e^{- \pi \xi}
			\int^{\kappa_\text{UV}}_0 \dd \kappa\, \kappa^3
			\abs{\frac{\partial}{\partial \kappa}W_{- i \lambda \xi,1/2} (-2i \kappa)}^2 \right) 
	 \simeq 2.6 \times 10^{-4} \,
	\frac{e^{2 \pi \xi}}{\abs{\xi}^3} H^4,
	\label{eq:E2}
\end{align}
\begin{align}
	\vev{\hat{\bm{B}}^2}
	&=
	\frac{e^{2 \pi \xi}}{\abs{\xi}^5} H^4 \times 
	\left(
		\frac{\abs{\xi}^5}{4 \pi^2} e^{- \pi \xi}
			\int^{\kappa_\text{UV}}_0 \dd \kappa\, \kappa^3 \abs{W_{- i \lambda \xi,1/2} (-2i \kappa)}^2
	\right) 
	 \simeq
	3.0 \times 10^{-4} \,
	\frac{e^{2 \pi \xi}}{\abs{\xi}^5} H^4,
	\label{eq:B2}
\end{align}
\begin{align}
	\vev{\hat{\bm{E}}\cdot \hat{\bm{B}}}
	&=
	\lambda \frac{e^{2 \pi \xi}}{\xi^4} H^4 \times
	\left( -
		\frac{\xi^4}{8 \pi^2}	e^{- \pi \xi}	\int^{\kappa_\text{UV}}_0 \dd \kappa\, \kappa^3 \frac{\partial}{\partial \kappa} \abs{W_{- \lambda i \xi,1/2} (-2i \kappa)}^2
	\right)
& \simeq 
	2.6 \times 10^{-4} \, \lambda
	\frac{e^{2 \pi \xi}}{\xi^4} H^4,
	\label{eq:EB}
\end{align}
with $\kappa = - k \eta = k/(a H)$ and where we have introduced physical electric and magnetic fields defined by $\hat E = E / a^2$ and $\hat B = B / a^2$, respectively. See \textit{e.g.,}~\cite{Jimenez:2017cdr} for a detailed derivation and discussion of these results.  Note that in accordance with statistical isotropy, we expect $\langle \bm E \rangle = 0 = \langle \bm B \rangle$ when averaging over the entire Universe. However, well within any given Hubble patch, the electric and magnetic fields are homogeneous, (anti-)parallel and thus locally select a preferred direction. On average, the magnitude of these fields in given by Eqs.~\eqref{eq:E2} and \eqref{eq:B2}, respectively. Under the assumption that both $\xi$ and $H$ are approximately constant, the magnitudes of these physical fields are constant. Hence, well within a given Hubble patch, the gauge field configuration sourced by Eq.~\eqref{eq:action-inflation} corresponds precisely to the constant, helical external gauge field configuration studied in Secs.~\ref{sec:eom} and \ref{sec:PP}. We note that the choice of sign of $\dot \phi$ distinguishes between parallel and anti-parallel $\bm E$ and $\bm B$ fields, with $ \lambda > 0$ corresponding to $B > 0$ in Secs.~\ref{sec:eom} and \ref{sec:PP}.

\paragraph{The induced current.} The analysis above neglected the fermion current. As discussed in Sec.~\ref{sec:PP}, the presence of strong electric and magnetic fields leads to the production and acceleration of charged fermions, resulting in the induced current $J_\text{ind}$.
In Secs.~\ref{sec:eom} and \ref{sec:PP} we considered these gauge fields to be classical, external fields. In the context of axion inflation these gauge fields are sourced dynamically as discussed above and hence the generation and acceleration of charged fermions drains energy from these fields. We thus expect the amplitude of the gauge fields to be reduced compared to the expressions~\eqref{eq:E2} - \eqref{eq:EB}. See Ref.~\cite{Domcke:2018eki} for an analogous discussion for the case of massless fermions.

In Section~\ref{subsec:Jind} we computed the different contributions to the induced current assuming homogeneous, constant (anti-)parallel electric and magnetic fields and a static Universe. From the discussion above, we note that approximately homogeneous and constant \textit{physical} (anti-)parallel electric and magnetic fields are sourced during axion inflation. We now turn on the cosmic expansion adiabatically and assume de Sitter spacetime. Note that this adiabatic approximation only holds 
if all relevant microphysical processes are
much faster than the cosmic expansion, see discussion below.
By replacing the time with the conformal time $\eta$, one may rewrite Eq.~\eqref{eq:induced} as
\begin{align}
	\partial_\eta \left( g Q \vev{J_z}_\text{ind} \right) = a^4 \times \frac{(g \abs{Q})^3}{2 \pi^2} \hat E \abs{\hat B}\,
	\coth \left( \frac{\pi \abs{\hat B}}{\hat E} \right)\,
	e^{- \frac{\pi m^2}{g \abs{Q} \hat E}}\,,
\end{align}
Assuming a constant physical electromagnetic field, we can perform the conformal time integral, which gives
\begin{align}
	g Q \vev{\hat J_z}_\text{ind} \equiv \frac{g Q \vev{J_z}_\text{ind}}{a^3} = \frac{(g \abs{Q})^3}{6 \pi^2} \hat E \abs{\hat B}\,
	\coth \left( \frac{\pi \abs{\hat B}}{\hat E} \right)\,
	e^{- \frac{\pi m^2}{g \abs{Q} \hat E}}\times \frac{1}{H}\,,
	\label{eq:Jind}
\end{align}
where we have defined the physical current by $\hat J = J / a^3$. 
Taking the mass to zero in this expression, we can recover the known result for the chiral fermions in Refs.~\cite{Abramchuk:2016afc,Domcke:2018eki}.
One may also take $B \to 0$ instead. Then we recover the known result for the Schwinger effect without the magnetic field~\cite{Kobayashi:2014zza,Hayashinaka:2016qqn}.
This indicates that our understanding smoothly connects massive/massless regimes with/without magnetic field.

\begin{figure}
\center
 \includegraphics[width = 0.48 \textwidth]{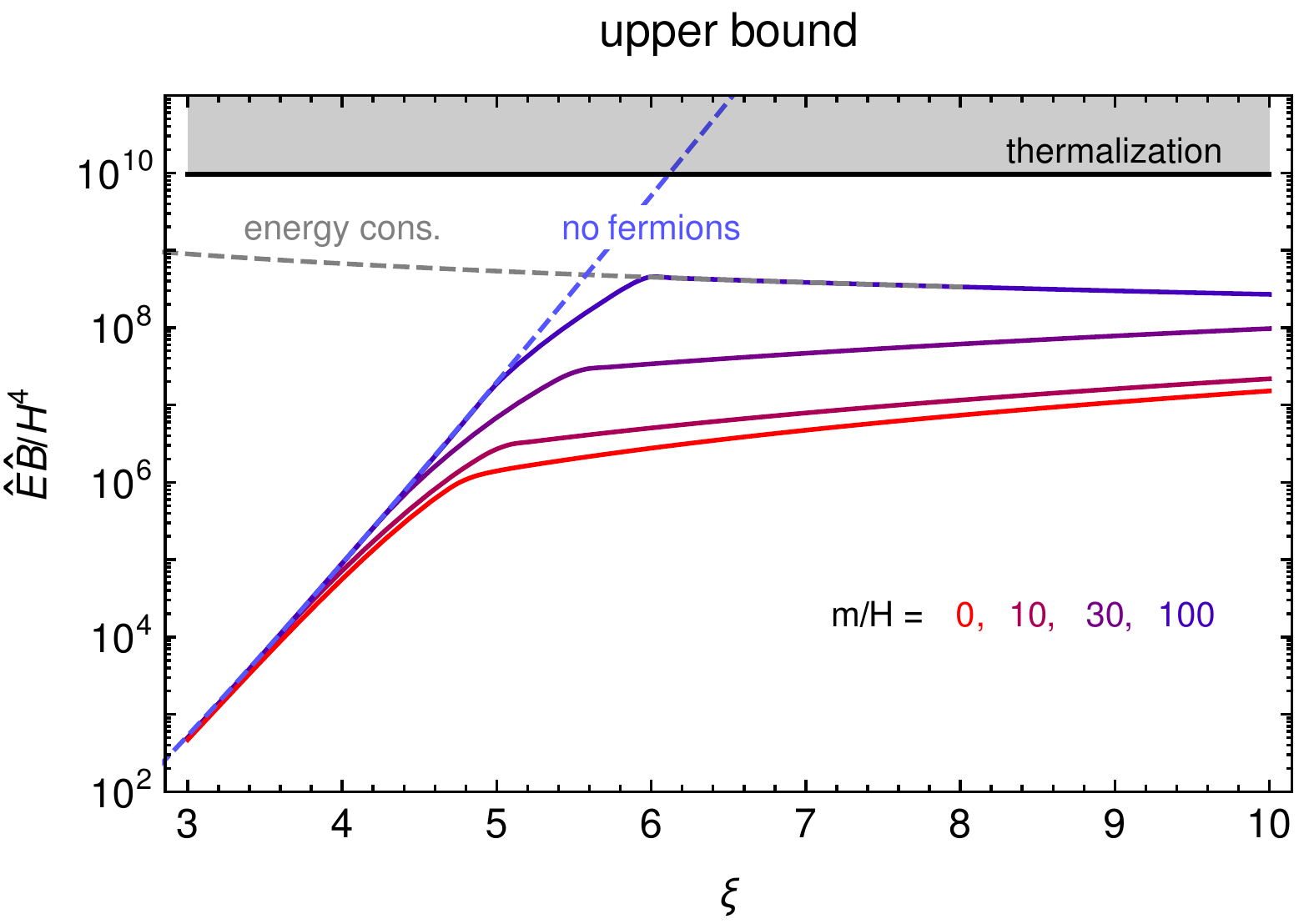} \hfill
 \includegraphics[width = 0.48 \textwidth]{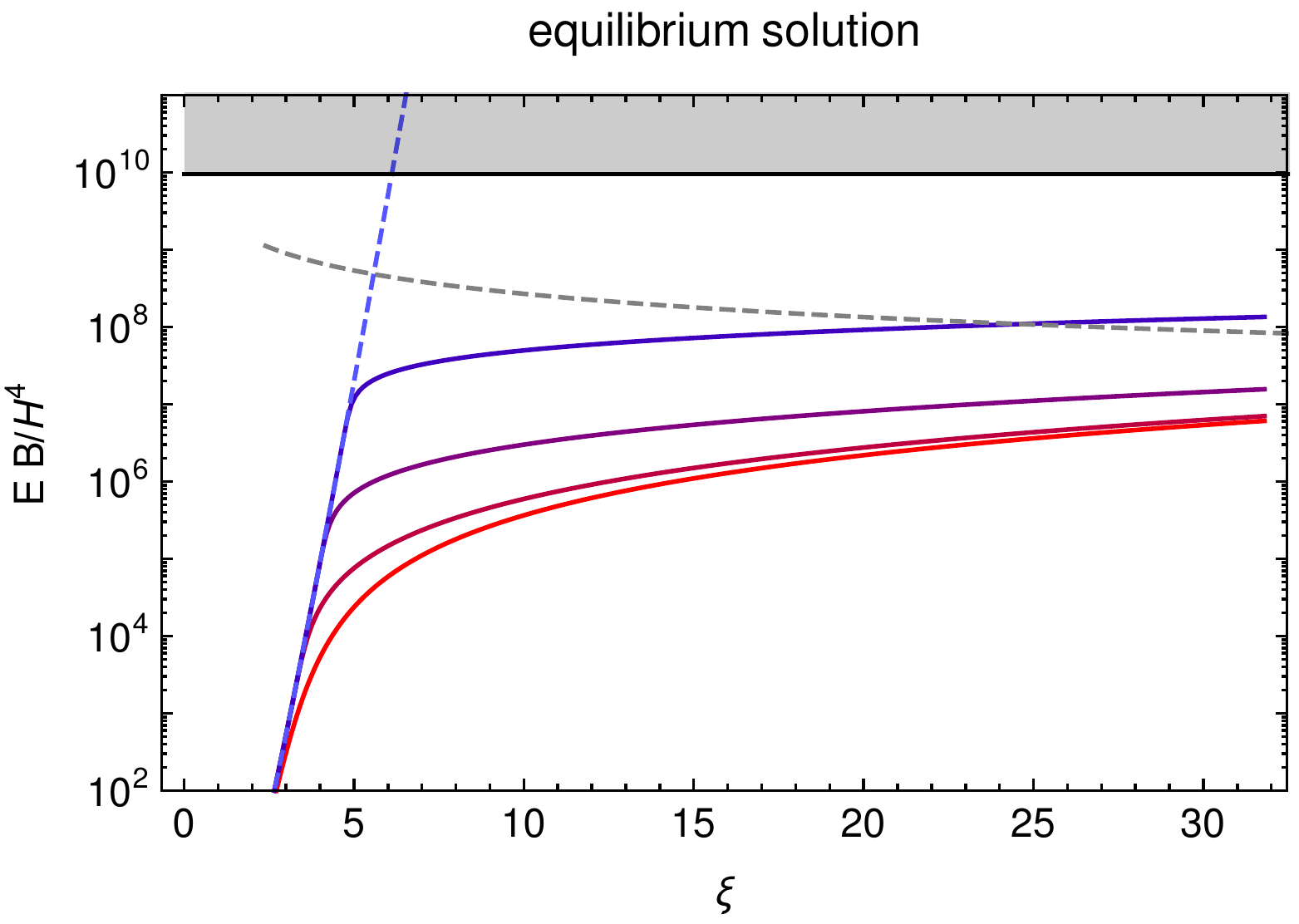}
 \caption{Magnitude of the Chern-Simons term generated in the presence of charged fermions with mass $m/H = \{0 , 10,30, 100 \}$ (red to blue, \textit{i.e.}, bottom to top). Left panel: upper bound obtained by extremizing over Eq.~\eqref{eq:consistency}. Right panel: equilibrium solution obtained by selfconsistently replacing $\xi \mapsto \xi_\text{eff}$ in Eq.~\eqref{eq:EB}.
 }
 \label{fig:EB_bounds}
\end{figure}

{Eq.~\eqref{eq:Jind} is a key result for the study of cosmological applications in the remainder of this section. While Eq.~\eqref{eq:induced} is an exact result, the generalization \eqref{eq:Jind} to de Sitter space requires some additional assumptions. Let us pause here for a moment to clarify these.

\begin{enumerate}
 \item As pointed out in Ref.~\cite{Banyeres:2018aax}, the effect of cosmic expansion becomes relevant for weak electric fields, when the acceleration induced by the electric field is comparable or even smaller than the Hubble expansion. The latter case can even lead to an effective motion of the fermions antiparallel to the electric field lines. The validity of Eq.~\eqref{eq:Jind} thus requires $g Q |\hat E|/(m H) \gg 1$. From Fig.~\ref{fig:EB_bounds}, we see immediately that this condition is fulfilled in the entire parameter space where the induced current is relevant.
    \item
    {
    As we have seen in Sec.~\ref{subsec:Jind}, 
    the induced current contains the Euler-Heisenberg terms
    which we ignored in the discussion above.
    For (mildly) strong field case $g \vert Q \vert \hat{E} \gtrsim m^2$,
    the leading Euler-Heisenberg term sets the renormalization scale for the coupling
    of $\mathcal{O}(g\vert Q \vert \hat{E})$,
    while the real part of the rest is suppressed by $m^2/g \vert Q \vert \hat{E}$
    (see~\cite{Dunne:2004nc} as a review).
    The latter is negligible compared to the particle production
    as long as $g\vert Q \vert \hat{E}/(mH) \gg 1$.\footnote{
    The same condition is obtained in the case without the magnetic field in~\cite{Banyeres:2018aax}.
    }
    It by chance coincides with the previous condition and is satisfied for
    the entire parameter space of our interest. 
    Thus the Euler-Heisenberg terms are important only for $g \vert Q \vert \hat{E} \ll m^2$,
    which is out of our interest since the particle production 
    and hence the fermion backreaction are anyway negligible.
    Here we implicitly assumed that $\hat{E} \sim \vert \hat{B}\vert$.
    It is valid in the case of our interest with of order unity $\xi$ or $\xi_\mathrm{eff}$.
    Note that it is $\xi_\mathrm{eff}$ that determines the hierarchy between
    $\hat{E}$ and $\vert \hat{B} \vert$ for the equilibrium solution.
    }
 \item In the discussion above, we neglected cosmic expansion when computing the fermion production rate. This is valid as long as the relevant microphysical length scale is much smaller than the Hubble horizon, \textit{i.e.}, $m_T \gg H$. From Eq.~\eqref{eq:asymptotic} we note that particle production is most efficient for $m_T^2 \sim g |Q| E / \pi$, and hence the flat space expressions for the production rate can be applied if $E \gg H^2$, which again holds whenever the induced current is relevant. Note that for massless fermions, this condition becomes equivalent to requiring that the particle production is fast compared to the cosmic expansion rate (see Ref.~\cite{Domcke:2018eki}), whereas the additional mass scale of massive fermions relaxes this condition.

\end{enumerate}}

The equation of motion~\eqref{eq:eom_A} implies the energy conservation equation
\begin{align}
 \dot \rho_A = - 4 H \rho_A + 2 \xi \hat E |\hat B| - g Q \hat E \hat J_\text{ind}^z \,,
\end{align}
where $\rho_A =  \frac{1}{2}(\hat E^2 + \hat B^2)$ denotes the physical energy density stored in the gauge fields. 
To obtain this expression, we have assumed that the average of $\vev{\hat{\bm{E}} \cdot \hat{\bm{J}}}$ is approximated with $\hat{E}\hat{J}^z_\text{ind}$.
The second term on the right-hand side describes the increase of the gauge fields sourced by the inflaton motion whereas the third term describes the transfer of gauge field energy into the fermion sector. Assuming dynamical equilibrium between the inflaton, gauge field and fermion sector, $\dot \rho_A = 0$, and 
inserting the explicit expression for the induced current, Eq.~\eqref{eq:Jind}, this leads to an algebraic consistency equation for $\hat E$ and $\hat B$,
\begin{align}
 - 2 H (\hat E^2 + \hat B^2) + 2 \xi_\text{eff} H \hat E |\hat B| = 0 \,,
 \label{eq:consistency}
\end{align}
with
\begin{align}
 \xi_\text{eff} \equiv \xi - \frac{(g |Q|)^3}{12 \pi^2} \text{coth} \left( \frac{\pi |\hat B|}{\hat E} \right) \exp\left(- \frac{\pi m^2}{g |Q| \hat E}\right) \frac{\hat E}{H^2} \,.
 \label{eq:xi_eff}
\end{align}

Neglecting the induced current (i.e the second term in Eq.~\eqref{eq:xi_eff}), Eq.~\eqref{eq:consistency} has two solutions (for any given $\xi$), which can be depicted as two straight lines in the $\hat E$ vs $\hat B$ plane. The analytical solution given by Eqs.~\eqref{eq:E2}, \eqref{eq:B2} corresponds to a particular point on one of these lines. Switching on the induced current, Eq.~\eqref{eq:consistency} describes a closed contour in the $\hat E$ vs $\hat B$ plane. Since the equation of motion is highly non-linear, we are no longer able to solve it analytically. However, we can place an upper bound on $\hat E$, $\hat B$ or $\hat E \hat B$ by extremizing these quantities over the closed contour described by Eq.~\eqref{eq:consistency}.\footnote{{Based on the analytical results above, we  employ the analytical results obtained in the absence of fermion production as upper bounds in the extremization procedure and choose the branch with $\hat E > \hat B$.}}  

This is depicted in the left panel of Fig.~\ref{fig:EB_bounds} for different values of the fermion mass. For simplicity, we have neglected the running of gauge coupling (see Eq.~\eqref{eq:running}), simply setting $Q = 1$ and $g = 1/\sqrt{2}$ for the purpose of this figure. The dashed blue line indicates the solution with negligible backreaction (Eq.~\eqref{eq:EB}), which is applicable for small values of $\xi$.
The dashed gray line indicates a violation of the slow-roll condition,
\begin{align}
 \xi \hat E |\hat B| \lesssim V(\phi) \,,
 \label{eq:slow-roll-bound}
\end{align}
where for the purpose of Fig.~\ref{fig:EB_bounds} we have set $H = 8 \times 10^{13}$~GeV.
If this constraint is violated, the gauge field production consumes the entire energy of the inflaton sector in less than a Hubble time. Approaching this bound the assumption of an approximately constant $\xi$ breaks down, and the analysis we perform here is no longer valid. When treating $\dot \phi$ and hence $\xi$ as dynamical parameters by solving Eq.~\eqref{eq:inflaton_eom}, this constraint is automatically fulfilled since the backreaction of the gauge fields on inflaton equation of motion limits the growth of $\dot \phi$. In practice, this implies that the large values of $\xi$ required for a violation of Eq.~\eqref{eq:slow-roll-bound} are not reached dynamically, see also Sec.~\ref{subsec:axion_inflation}.
Finally, the gray shaded region indicates the regime where thermalization of the produced fermions can no longer be neglected according to the estimate~\eqref{cond:therm}.
Here we only show the condition for the massless case because the bound becomes the most stringent as can be readily seen from Eq.~\eqref{cond:therm}.
We also check that the bound relaxes only by an $\mathcal O(1)$ magnitude for the heaviest mass parameter taken in Fig.~\ref{fig:EB_bounds}.
In summary, as expected, increasing the fermion mass reduces the fermion production and hence the induced current, leading to larger values of the electric and magnetic fields.

The upper bounds depicted in the left panel of Fig.~\ref{fig:EB_bounds} are clearly conservative and here is no compelling reason why they should be saturated. 
We can proceed by solving Eq.~\eqref{eq:consistency} under an additional assumption: if all the equilibration processes between the inflaton, gauge field and fermion sector are sufficiently fast, we expect $\hat E$ and $\hat B$ to be given by Eqs.~\eqref{eq:E2} to \eqref{eq:EB} with the replacement $\xi \mapsto \xi_\text{eff}$. This set of equations can be solved selfconsistently, and we depict the results for this `equilibrium' solution in the right panel of Fig.~\ref{fig:EB_bounds}. Comparing Fig.~\ref{fig:EB_bounds} with the conditions listed below Eq.~\eqref{eq:Jind}, we see that also for the equilibrium solution, these are always fulfilled in the regime of interest. 

We emphasize that both panels of Fig.~\ref{fig:EB_bounds} only provide an estimate for the magnitude of the generated gauge fields. An exact solution of the non-linear equation of motion~\eqref{eq:eom_A} would be highly desirable for a more precise study of the phenomenology of these models, but is unfortunately beyond the scope of this paper. Nevertheless, we very clearly see that the presence of fermions, and in particular their mass, changes the prediction for the gauge field production by several orders of magnitude.

\paragraph{Thermalization and Pauli blocking.} 
The key quantity in the study of backreaction is the induced current.
So far, we have neglected the interaction of fermions in the estimation of induced current.
Here we discuss the condition justifying this approximation.
As one may see from Eqs.~\eqref{eq:induced} and \eqref{eq:Jind}, we have assumed that the acceleration by the electric field continues for $1 / H$.
However, if fermion scatterings are efficient, this approximation is no longer justified.

Let us start our discussion with the ``would-be'' temperature if the particles were thermalized:
\begin{align}
	\hat T_\text{wb} \simeq \left(\frac{30}{\pi^2 g_\ast} n_\psi \omega \right)^\frac{1}{4} \,, \quad
	\omega = \left(g |Q| \hat E \right)^\frac{1}{2} \, 
	\max \left[
		1\,, \left( g |Q| \hat E  \right)^\frac{1}{2} \min [\tau_\text{s}\,, H^{-1}]
	\right]\,,
\end{align}
where $\omega$ denotes the energy of the most relevant Landau level after experiencing acceleration in the electric field for a time period $\min [\tau_\text{s}\,, H^{-1}]$ with $\tau_\text{s}$ being a typical time scale of scatterings.
If we suppose that the plasma were thermalized, the typical interaction rate of particles in the thermal plasma is 
\begin{align}
	\tau_\text{th}^{-1} = \alpha^2 \hat T_\text{wb}\,.
	\label{eq:large_angle}
\end{align}
In most cases of our interest, we find that the typical energy of fermions after production is much larger than the would-be temperature, $\omega \gg \hat T_\text{wb}$.
This leads to an additional suppression of the scattering rate of the generated fermions compared to $\tau_\text{th}$:
\begin{align}
	\tau_s > \tau_\text{th}\,.
\end{align}
For instance, the concrete form including this suppression factor is $\alpha^2 \hat T_\text{wb} (\hat T_\text{wb}/\omega)^{1/2}$~\cite{Gyulassy:1993hr,Arnold:2001ba,Arnold:2002ja,Kurkela:2011ti} if the generated particles are also charged under non-Abelian gauge field as in the SM~\cite{Harigaya:2013vwa,Mukaida:2015ria,Domcke:2018eki}.
Here, to avoid complications and model-dependent discussions, we just conservatively require $1 < \tau_\text{th} H$  as a condition to neglect the interactions of the generated fermions.
One can readily see that, for $1 < \tau_\text{th} H$, even an initially thermal plasma would drop out of thermal equilibrium, indicating 
how conservative this requirement is.
The condition, $1 < \tau_\text{th} H$, gives\footnote{{Alternatively, one may estimate the scattering rate of the initially non-thermal fermions as $\tau_\text{s} =  n_\psi \sigma_\text{sc}$ with $\sigma_\text{sc} \sim \alpha^2/\omega^2$. Numerically, this leads to a very similar result for $\tau_\text{s}$.}}
\begin{align}
	1 < \tau_\text{th} H \lesssim \alpha^{-2} (g |Q|)^{-\frac{3}{4}} \left( \frac{\pi^4 g_\ast}{5} \right)^\frac{1}{4} \left( \frac{\hat B}{\hat E} \right)^\frac{1}{8} \left( \frac{H^4}{\hat E \hat B} \right)^\frac{3}{8} e^\frac{\pi m^2}{4 g |Q| \hat{E}} \,.
	\label{cond:therm}
\end{align}
Here we have assumed $m^2 \lesssim g |Q| \hat E$ and $g |Q| \hat E > H^2$, since otherwise the backreaction from the induced current is anyway negligible.
For the parameter choices in Fig.~\ref{fig:EB_bounds}, this condition is always fulfilled. 
However, we note that for a smaller energy scale of inflation, the upper bound on $\hat E \hat B/H^4$ from Eq.~\eqref{eq:slow-roll-bound} becomes less stringent, potentially opening up some parameter space where thermalization is relevant, in particular for a heavier mass.

It is sometimes argued that the production of fermions is not efficient due to Pauli blocking in the final state. In our computations in Sec.~\ref{sec:PP} this is already intrinsically accounted for, and is reflected \textit{e.g.}, in the condition $|\alpha|^2 + |\beta|^2 = 1$. Nevertheless, we find significant particle production. Here we interpret these results in terms of simple quantum mechanical arguments.

Immediately at their generation, the fermions are characterized by a vanishing velocity in $z$-direction and a finite (for $n > 0$) velocity in the transverse direction, encoded in the transverse mass $m_T$. From the spacing of the energy levels in Eq.~\eqref{eq:mT}, we see that the quantum mechanical uncertainty in position space is given by $\dd x \dd y \sim 2 \pi/(g s Q \hat B)$. With this, we find for the production rate within a box of dimensions $\dd x \,  \dd y \, L$,
\begin{align}
 \Gamma_p^{(n)} = \dot n_\psi \dd x\, \dd y L  \simeq  \frac{g^2 Q^2}{4 \pi^2} \hat E \hat B \exp\left(- \frac{2 \pi n \hat B}{\hat E}\right) \exp\left( - \pi m^2/g |Q| \hat E \right) \times \frac{2 \pi}{ g s Q \hat B} \times L \,,
\end{align}
where we have employed the asymptotic solution~\ref{eq:asymptotic} to obtain $\dot n_\psi$. Consequently the acceleration in $z$-direction accumulated before the creation of the subsequent fermion is given by
\begin{align}
 \Delta p_\text{acc} = g Q \hat E \, \Gamma_p^{-1} \,.
\end{align}
Comparing this to the spacing of the $p_z$ levels, $\Delta p_z = 2 \pi/L$, we conclude that Pauli block is inefficient if
\begin{align}
 1 <  \frac{\Delta p_\text{acc}}{\Delta p_z} =   \exp\left(\frac{2 \pi n \hat B}{\hat E}\right) \exp\left( \pi m^2/g |Q| \hat E \right) \,,
\end{align}
which is clearly intrinsically satisfied. In particular, we note that for massless fermions in the lowest Landau level, this bound is saturated (as expected for fermion production from a gap-less dispersion relation), whereas massive fermions and higher Landau levels feature an occupation number significantly below the Pauli limit. This extends and improves the discussion given on this matter in Ref.~\cite{Domcke:2018eki}.

\subsection{Axion inflation}
\label{subsec:axion_inflation}

In the previous section we discussed the gauge field production induced by the non-vanishing velocity of a pseudo-scalar field. This is a key ingredient in axion inflation (see \textit{e.g.}, \cite{Barnaby:2011qe} for a review). In this context, the gauge field production has shown to lead to striking and observable signatures both in the scalar and tensor power spectrum~\cite{Barnaby:2010vf,Barnaby:2011qe,Barnaby:2011vw,Meerburg:2012id,Linde:2012bt, Garcia-Bellido:2016dkw,Domcke:2017fix,Cook:2011hg,Barnaby:2011qe,Barnaby:2011vw,Anber:2012du,Domcke:2016bkh,Bartolo:2016ami}. In this section, we revisit these predictions accounting for the presence of massive fermions.

For the purpose of this discussion, we will choose a linear scalar potential in Eq.~\eqref{eq:inflaton_eom},
\begin{align}
 V(\phi) = \mu^3 \phi \,, \qquad \mu = 6.1 \cdot 10^{-4} M_P  \,,
\end{align}
motivated by models of axion monodromy~\cite{McAllister:2008hb} with the parameter $\mu$ determined by the observed amplitude of the scalar power spectrum at CMB scales. For the resulting phenomenology it will be mainly relevant that this leads to an inflaton velocity which grows monotonously during inflation, for a discussion of the impact of the choice of scalar potential see \cite{Domcke:2016bkh}. In the figures below, we will further choose $\alpha/(\pi f_a) = (0.02\,  M_P)^{-1}$, which is the maximal value consistent with current constraints on scalar non-gaussianities in the CMB.

\begin{figure}
\centering
 \includegraphics[width = 0.48 \textwidth]{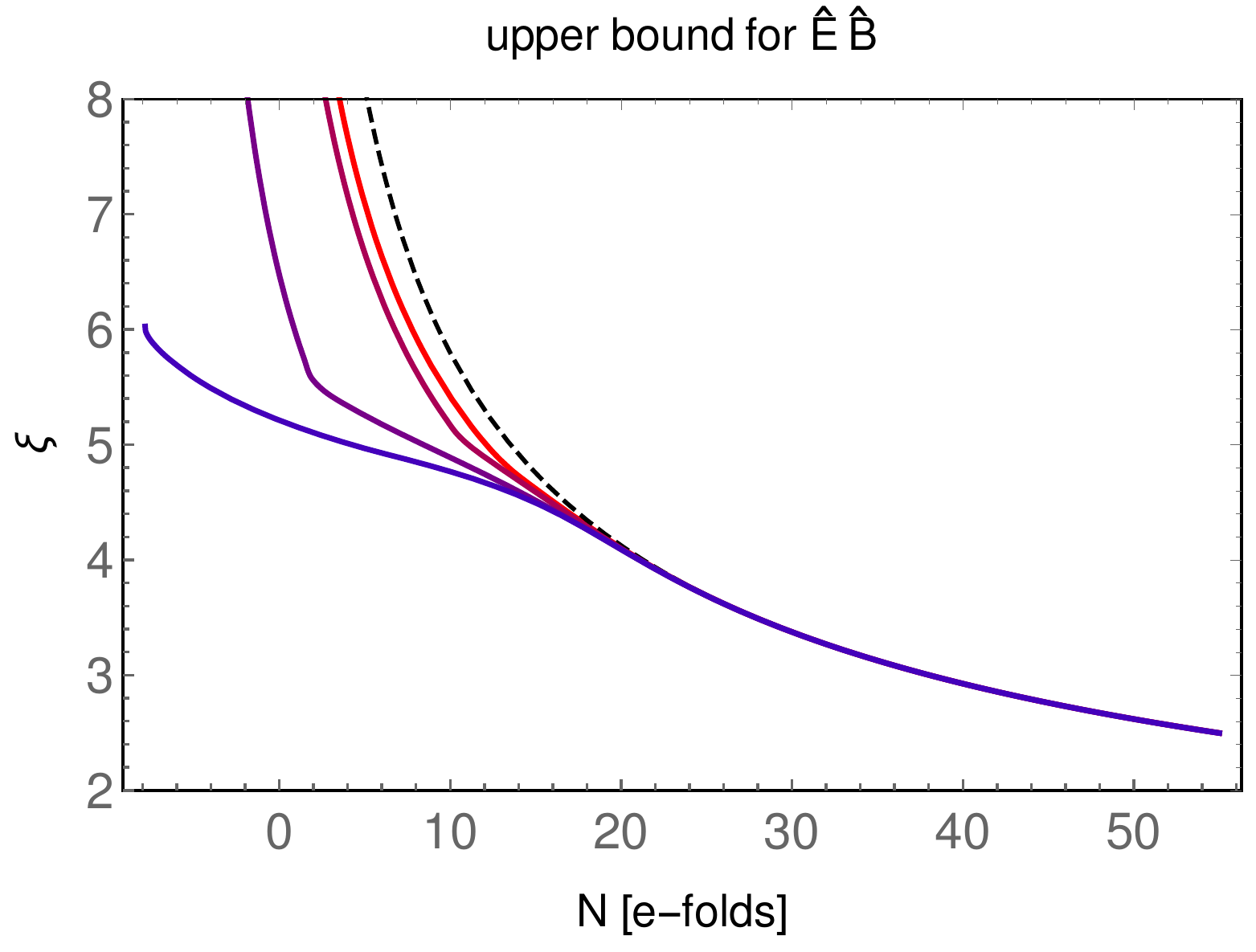} \hfill
 \includegraphics[width = 0.48 \textwidth]{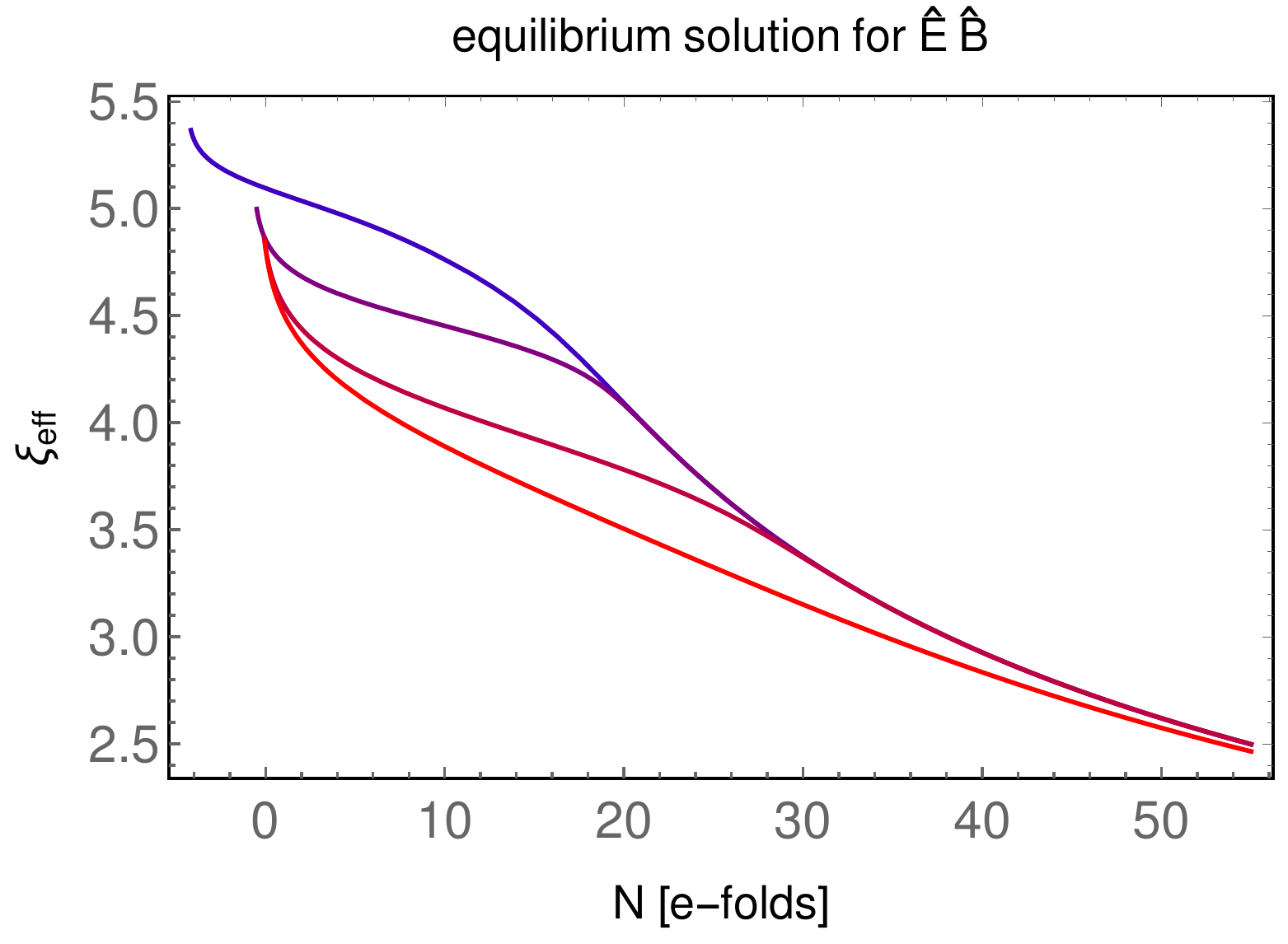}
 \caption{Evolution of the parameter $\xi$ (proportional to the inflaton velocity) for different fermion masses.  Left panel: upper bound for $\hat E \hat B$. Right panel: equilibrium solution for $\hat E \hat B$. Color coding as in Fig.~\ref{fig:EB_bounds}. }
 \label{fig:xi}
\end{figure}

With these choices, we can numerically solve the inflaton equation of motion~\eqref{eq:inflaton_eom} inserting the results for $\hat E \hat B (\xi)$ depicted in Fig.~\ref{fig:EB_bounds}.
Fig.~\ref{fig:xi} illustrates how the fermion masses impact the evolution of inflaton velocity. For massless fermions (red curve) we reproduce the result of Ref.~\cite{Domcke:2018eki}: due to the efficient generation of fermions, the gauge field abundance and hence the backreaction on the inflaton field are strongly suppressed, leading to $\xi \propto \dot \phi/H \propto N^{-1/2}$. This is the usual result for inflation in monomial potentials without a gauge field background. As the fermion mass increases, the energy conversion from gauge fields into fermions becomes less efficient. This results in a stronger gauge field background and hence in a stronger backreaction on the inflaton field, reducing the inflaton velocity towards the end of inflation. Defining the end of inflation as $(\dot \phi)^2/(2 H^2 M_P^2) = 1$, this additional friction prolongs the duration of inflation, explaining why the $x$-axis in Fig.~\ref{fig:xi} extends to negative values (here $N = 0$ is defined as the end of inflation in the absence of gauge fields).
This is particular visible for $m/H = 100$ (purple curve) in the left panel of Fig.~\ref{fig:xi}, which essentially reproduces the result of axion inflation in the absence of fermions, see \textit{e.g.}, Ref.~\cite{Barnaby:2011qe}. In the right panel of Fig.~\ref{fig:xi} we show the corresponding evolution for $\xi_\text{eff}$, defined in Eq.~\eqref{eq:xi_eff}. In this case, the induced current is included in the effective parameter $\xi_\text{eff}$. Lighter fermions correspond to a smaller value of $\xi_\text{eff}$ and hence a less efficient gauge field production.
In summary, light fermions lead to a strong induced current which makes gauge field production more difficult, whereas sufficiently heavy fermions effectively decouple.

\paragraph{Scalar power spectrum}

In slow-roll approximation, the equation of motion for the inhomogeneous perturbations $\delta \phi$ of the inflaton field is given by~\cite{Linde:2012bt}
\begin{align}
 \delta \ddot{\phi} + 3 \beta H \delta \dot{\phi} - \frac{\nabla^2}{a^2} \delta \phi + V_{,\phi \phi}(\phi) \delta \phi = -  \frac{\alpha}{\pi f_a} \left(\hat{\bm E} \cdot \hat{\bm B} -  \langle \hat 
{\bm E} \cdot \hat{\bm B} \rangle \right) \,,
\end{align}
with $\beta \equiv 1 + 2 \xi \alpha \vev{  \hat{\bm E} \cdot \hat{\bm B} } /(3 \pi H \dot \phi f_a)$. From this, one can estimate the scalar power spectrum as~\cite{Linde:2012bt}\footnote{{Here we are assuming that the estimate of the variance of $\hat E \hat B$ is given by the expression found in \cite{Linde:2012bt} in the absence of fermions. In other words, we are assuming that the fermions only change the magnitude of the generated gauge fields but not their spectral shape. It will be crucial to scrutinize this assumption using dedicated numerical simulations in the future.}}
\begin{align}
 \Delta_s^2 = \frac{H^2}{\dot \phi^2} \langle \delta \phi^2 \rangle  \simeq 
 \left( \frac{H^2}{2 \pi \dot \phi} \right)^2 + 
 \left( \frac{\alpha \langle  \hat{\bm E} \cdot \hat{\bm B} \rangle }{3 \pi \beta H \dot \phi f_a} \right)^2 \,,
 \label{eq:Ds}
\end{align}
where the first term is the standard vacuum contribution and the second term in sourced by the Chern-Simons charge. We note that both the homogeneous equation of motion for the inflaton as well as the scalar power spectrum for its fluctuations are sensitive to the gauge field and fermion production only through the magnitude of $\hat E \hat B$, which in turn is a function of $\xi \propto \dot \phi$.
Consequently we can convert the estimates depicted in Fig.~\ref{fig:EB_bounds} into estimates of the scalar power spectrum, taking into account both the backreaction of the gauge fields in Eq.~\eqref{eq:inflaton_eom} as well as the additional source term in Eq.~\eqref{eq:Ds}.

\begin{figure}
\centering
 \includegraphics[width = 0.48 \textwidth]{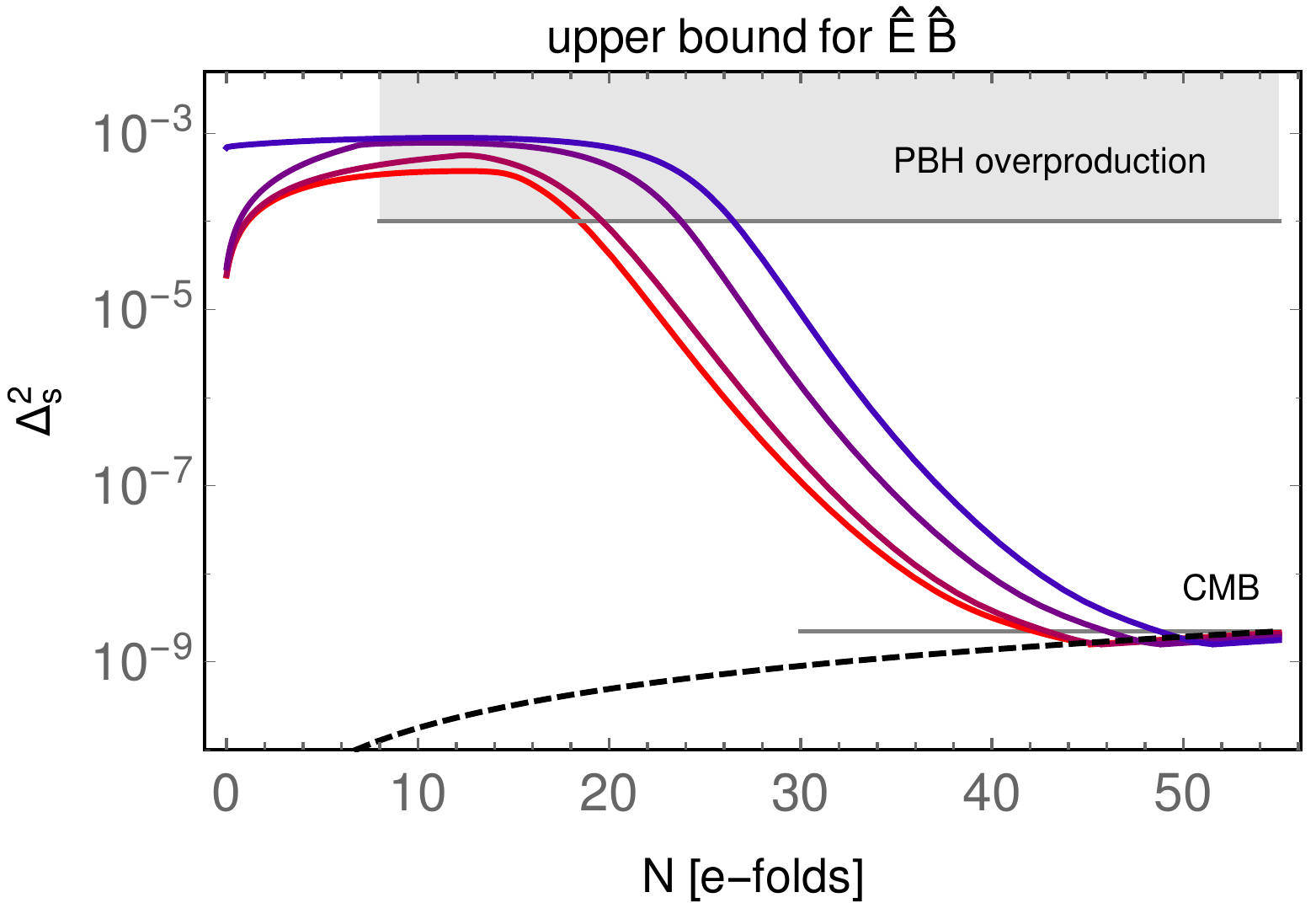} \hfill
 \includegraphics[width = 0.48 \textwidth]{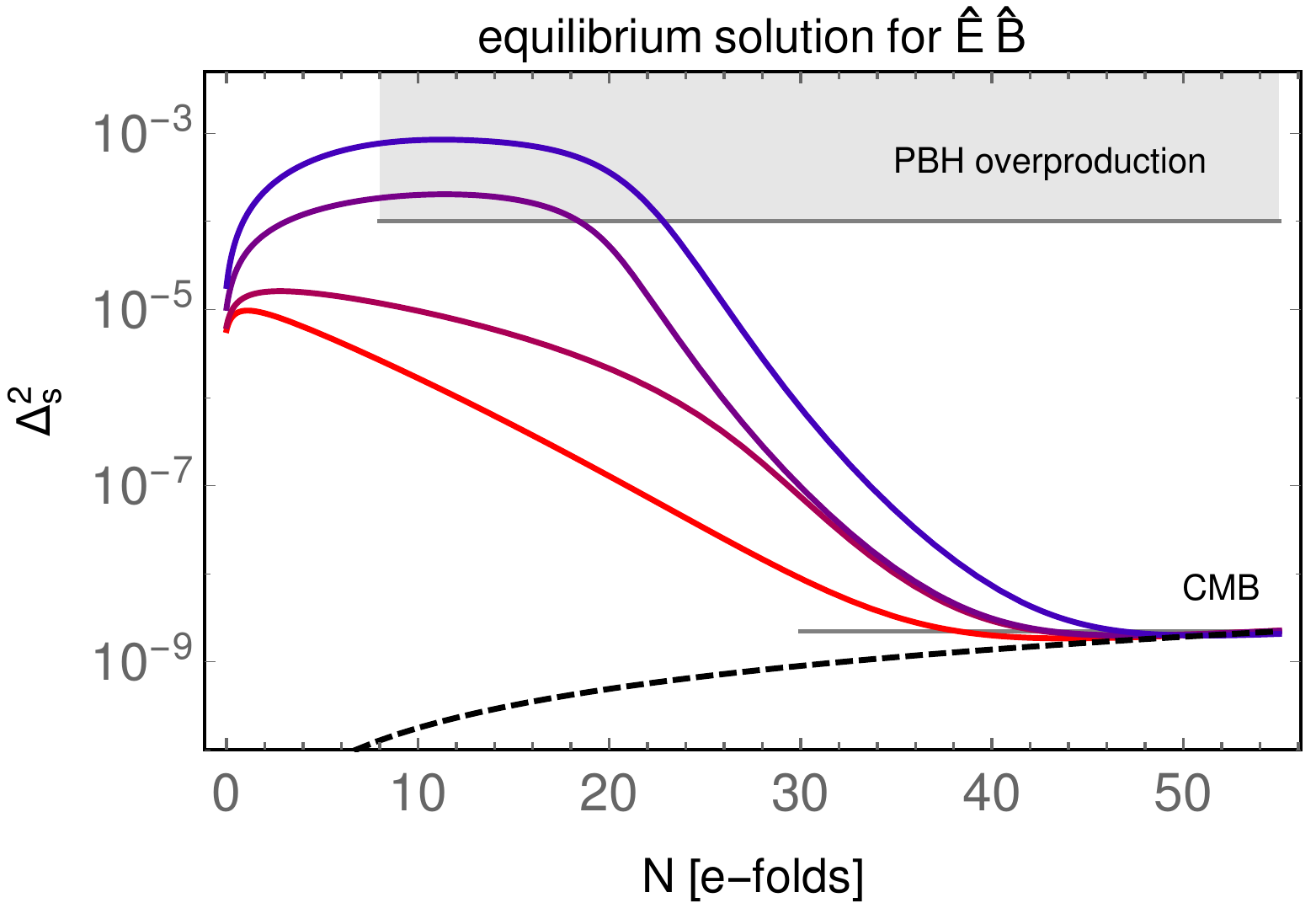}
 \caption{Scalar power spectrum.  Left panel: upper bound for $\hat E \hat B$. Right panel: equilibrium solution for $\hat E \hat B$. Color coding as in Fig.~\ref{fig:EB_bounds}. }
 \label{fig:As}
\end{figure}

The resulting scalar power spectrum for different values of the fermion mass is shown in Fig.~\ref{fig:As}. At CMB scales ($N \simeq 55$), the parameter $\xi$ is small and both gauge field and fermion production are inefficient. At these scales, the model thus closely resembles a standard single-field slow roll inflation model (indicated by the dashed black line). At smaller scales, towards the end of inflation, $\xi$ increases and the scalar power spectrum is strongly enhanced. Lighter fermion masses lead to a smaller scaler power spectrum, since the induced current inhibits the gauge field production. The slightly earlier rise in the scalar power spectrum for heavy fermions is due to the definition of $N = 0$ as $\dot \phi^2/(2 H^2 M_P^2) = 1$, which occurs a few e-folds later when efficient gauge field production induced a strong friction in the inflaton equation of motion which delays the end of inflation. 

For strong gauge fields, $\beta \gg 1$, the scalar power spectrum can be approximated as $\Delta_s^2 \simeq 1/(2 \pi \xi)^2$. An upper bound on the gauge fields implies an upper bound on the amount of gauge friction and thus translates to a lower bound on $\xi$. In this sense, the curves in the left panel of Fig.~\ref{fig:As} can be viewed as upper bounds for the scalar power spectrum for different fermion masses.
The equilibrium solution predicts overall smaller gauge fields compared to the upper bound, resulting in a smaller scalar power spectrum, in particular for light fermions.

The horizontal lines in Fig.~\ref{fig:As} indicate the CMB normalization and the primordial black hole bound \cite{Linde:2012bt}, respectively. Above the latter, the large scalar perturbations lead to a too large probability of producing primordial black holes, in contradiction with observations. This can be remedied by considering couplings to multiple abelian gauge groups~\cite{Domcke:2016bkh}. However, given the significant uncertainties in our estimates of $\hat E$ and $\hat B$ as well as the limits of perturbation theory for large values of $\xi$~\cite{Ferreira:2015omg, Peloso:2016gqs}, we shall for the moment ignore this problem. The main purpose of this section is not to derive a precise upper bound on the fermion mass which would circumvent the primordial black hole bound, but rather to highlight the significant impact of (massive) fermions on the scalar power spectrum of axion inflation.

\paragraph{Gravitational wave spectrum}

The computation of the tensor power spectrum requires the knowledge of the anisotropic contribution to the energy momentum tensor, sourced by vacuum fluctuations as well as by the gauge field and fermion sector. The contribution from the vacuum perturbation is given by the usual expression, $\Omega_\text{GW}^\text{vac} = \Omega_r/12 (H/(\pi M_P))^2$ with $\Omega_r = 9.1 \cdot 10^{-5}$ denoting the fraction of radiation energy density today. In the absence of fermions, the gauge field contribution can be estimated as (see e.g.~\cite{Barnaby:2011qe,Barnaby:2011vw})
\begin{align}
 \Omega_\text{GW}^\text{A} = \frac{\Omega_r}{12} \left(\frac{H}{\pi M_P} \right)^2 \times 4.3 \cdot 10^{-7} \frac{H^2}{M_P^2} \frac{e^{4 \pi \xi}}{\xi^6} \,.
 \label{eq:GWsA}
\end{align}
The contribution from  uncharged chiral fermions, in the absence of gauge fields, was computed in~\cite{Adshead:2019aac} and found small compared to the vacuum contribution. A full computation of the gravitational wave spectrum resulting from the interplay of the fermion and gauge field sector requires the solution of the inhomogeneous non-linear gauge field equations taking into account the fermion backreaction, which is beyond the scope of this paper. However, for the equilibrium solution (requiring a self-consistent solution of Eq.~\eqref{eq:E2} and \eqref{eq:B2} with $\xi \mapsto \xi_\text{eff}$), we can perform an estimate for the gravitational wave spectrum by inserting the resulting $\xi_\text{eff}$ into Eq.~\eqref{eq:GWsA}, adding the vacuum contribution but neglecting the fermion contribution.

\begin{figure}
 \centering
 \includegraphics[width = 0.48 \textwidth]{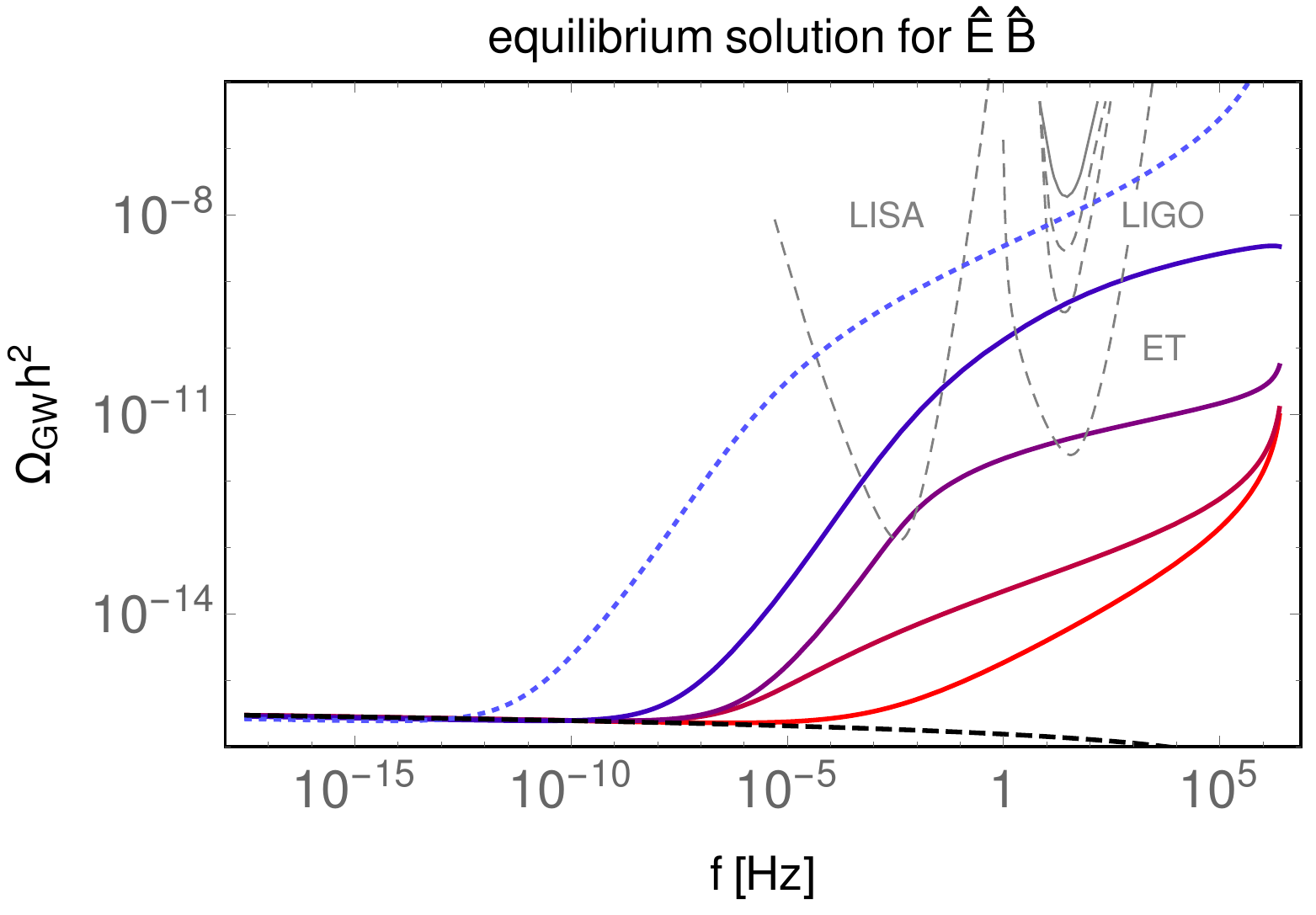}
 \caption{Predicted gravitational wave spectra for different fermion masses, employing the equilibrium solution. The (dashed) gray lines indicate the power-law sensitivity curves of the (future) gravitational wave interferometers LIGO, LISA and the Einstein Telescope. Color coding as in Fig.~\ref{fig:EB_bounds}. For reference, the dashed black line indicates the vacuum contribution and the dotted blue line indicated the predicted spectrum in the absence of fermions.}
 \label{fig:GWs}
\end{figure}

The resulting spectra are shown in Fig.~\ref{fig:GWs}. As expected, the presence of light fermions reduces the gauge field abundance and hence the expected gravitational wave signal. This is a striking example of the ability of gravitational wave detectors to probe the properties of particle physics models. 

Finally, we note that the equilibrium solution assumes a fast backreaction of the generated fermions on the gauge field abundance. If this backreaction is slower than, e.g., the change in $\xi$, we would expect the gauge fields to `overshoot' their equilibrium value. Moreover, the gravitational wave spectrum depicted in Fig.~\ref{fig:GWs} neglects any contribution from the fermion sector. Given that we found the fermion production to be fast compared to the Hubble expansion rate, we expect a larger fermion abundance than found from the gravitational production in Ref.~\cite{Adshead:2019aac}, and hence the resulting contribution to the gravitational wave spectrum may be relevant. In view of this, the results depicted in Fig.~\ref{fig:GWs} should be viewed as \textit{lower bounds} on the gravitational wave production in axion inflation. We leave a more detailed investigation to future work.

\subsection{Baryogenesis from axion inflation}

As pointed out in Refs.~\cite{Anber:2015yca,Jimenez:2017cdr,Domcke:2019mnd}, axion inflation with the abelian gauge group identified as the SM hypercharge $U(1)_Y$ can provide the initial conditions required for successful baryogenesis from decaying helical hypermagnetic fields~\cite{Joyce:1997uy,Bamba:2006km,Kamada:2016eeb,Kamada:2016cnb}. If the SM Higgs expectation value is stabilized at zero during inflation, axion inflation leads to a dual production of helical hyper gauge fields and massless chiral fermions. The latter thermalize once Yukawa interactions and sphaleron processes become relevant in the thermal plasma after inflation. However, under certain conditions, depending also on the chiral charge in the fermion sector, the helical hypermagnetic fields can survive until the electroweak phase transition. During the electroweak phase transition the hypermagnetic fields are converted into electromagnetic fields, thereby generating a baryon asymmetry which can explain the matter-antimatter asymmetry observed in our Universe~\cite{Domcke:2019mnd}. 

If the SM Higgs expectation value is stabilized at non-zero vev during inflation, axion inflation will lead to a dual production of \textit{electromagnetic} gauge fields and massive chiral fermions. Depending on the details of the Higgs potential and the properties of thermal plasma, the electroweak symmetry may be restored after inflation, implying a conversion of the electromagnetic fields to hyper gauge fields. The magnitude of these hyper gauge fields, together with chiral charge stored in the fermion sector at this point in time, determines the efficiency of the subsequent baryogenesis process. We leave a detailed study of this (model-dependent) symmetry restoration process and the subsequently generated baryon asymmetry to future work.

\section{Conclusions}
\label{sec:conclusions}

The production and transport of charged particles in the presence of electromagnetic fields are ubiquitous in nature.
In this paper, we study the production of charged massive Dirac fermions in parallel electric and magnetic fields.
By solving the equation of motion for fermions in a background gauge field, we compute various quantities semi-analytically in terms of Bogoliubov coefficients.
We analytically show how the chiral anomaly arises from the equation of motion [Eqs.~\eqref{eq:q5dot_temp}, \eqref{eq:cs_proof}, and \eqref{eq:mass_proof}], and also numerically evaluate the chiral charge and the pseudo-scalar operator yielding the correct chiral anomaly equation [Eqs.~\eqref{eq:q5_t_int} and \eqref{eq:mass_t_int}],
which guarantees a smooth connection between the case of massless and massive fermions.
In addition, we evaluate the induced current and explicitly show how three contributions appear: (i) the current from fermion production [Eq.~\eqref{eq:induced}], (ii) the running of the gauge coupling constant [Eq.~\eqref{eq:running}], and (iii) the Euler-Heisenberg terms in a $1/m$-expansion [Eq.~\eqref{eq:current_vac}].
This result clarifies that, while particle production is exponentially suppressed in the limit of heavy fermions, the expectation value of any operator contains terms suppressed in powers of $1/m$, which is nothing but higher dimensional operators from integrating out heavy particles and should not be confused with the particle production. We moreover demonstrate explicitly how the Pauli exclusion principle manifests itself, showing that the occupation number per phase space volume is saturated only for massless fermions in the lowest Landau level.

Equipped with these results, we discuss a pseudo-scalar inflaton which couples to a U$(1)$ gauge theory solely via the Chern-Simons coupling.
The velocity of the inflaton increases over the course of inflation, which triggers the efficient production of helical gauge fields.
We demonstrate how the production of charged fermions backreacts on the gauge field equation of motion and suppresses the production of helical gauge fields.
As expected, this backreaction becomes more significant for lighter fermions, resulting in a suppression of the sourced scalar/tensor perturbations. We find that depending on the fermion mass, the stringent constraints from primordial black hole overproduction may be avoided, while still implying a gravitational wave spectrum which is well above the standard vacuum contribution.
Our discussion neglects the interactions among fermions after production. We clarify the condition for this approximation to be valid (see also~\cite{Domcke:2018eki}), \textit{i.e.,} the condition that ensures that thermalization of the fermion population during inflation is inefficient.
To go beyond this limitation, one would need to treat the evolution of {the phase space density of the} fermions dynamically, as recently discussed in Ref.~\cite{Gorbar:2019fpj} in the context of an $f^2(\phi) FF$ inflation model.
Extending their analysis to our case is surely worth pursuing in future.

Although we restrict ourselves to the Chern-Simons coupling as a starting point for cosmological applications,
one may also consider more general shift-symmetric couplings such as $(\partial \phi) \cdot J_5$ and $\overline \psi e^{i \gamma_5 \phi / f_a} \psi$.
Up to a field redefinition associated with a chiral rotation, one of these three couplings is redundant and we are left with two independent couplings, \textit{i.e.,} the Chern-Simons coupling and $\phi$-dependent mass coupling.
This field redefinition never changes the physical result as it should be, which is however has been explicitly shown only in two limited cases:
the case without the gauge coupling (uncharged fermions) in Ref.~\cite{Adshead:2015kza,Adshead:2018oaa} and the case without mass in Ref.~\cite{Domcke:2018eki}.
Including this basis independence, a smooth connection between two regimes is therefore desirable for more complete understanding of the system at hand.

For simplicity we assume a constant mass for the Dirac fermions throughout this paper. However, if we would like to identify this gauge group as part of the SM gauge group, we need to recast the mass in terms of the Higgs field.
As recently pointed out in Ref.~\cite{Hook:2019vcn}, once we treat the Higgs field dynamically, the asymmetric production of fermions backreacts onto the Higgs, destabilizing the Higgs effective potential in contrast to the usual thermal mass,
which can lead to an interesting signature in the bispectrum {of the scalar perturbations generated during inflation}.
Although Ref.~\cite{Hook:2019vcn} focuses on the production via the $\phi$-dependent mass coupling, it would be interesting to extend their analysis to include the Chern-Simons coupling which then involves the production of helical gauge fields, the production of fermions obeying the anomaly equation, and also the Higgs production via the Schwinger effect.
We hope that our work will stimulate further studies on this direction, including the impact of these effects on the dynamics of the Higgs field and their observational signatures.

\paragraph{Acknowledgements}
It is a pleasure to thank Tomohiro Fujita for helpful discussions related to this project.
This work was funded by the Deutsche Forschungsgemeinschaft under Germany's Excellence Strategy - EXC 2121 ``Quantum Universe'' - 390833306.

\appendix

\section{Notation and conventions}
\label{app:notation}

We consider a  pseudo-scalar singlet $\phi$ together with a Dirac fermion $\psi$ of mass $m$ which is charged under an Abelian gauge group. Taking into account all interaction terms (up to dimension 5) which respect the shift-symmetry of $\phi$, the most general action (which does not explicitly break parity) reads
\begin{align}
 S = \int d^4 x  & \left\{ \sqrt{-g} \left[\frac{1}{2} g^{\mu \nu} \partial_\mu \phi \partial_\nu \phi - V(\phi) \right] - \frac{1}{4} F_{\mu \nu}F^{\mu \nu} + \psi i \slashed{D} \psi \right. \nonumber \\
 & \left. + \, c_A \frac{\alpha}{4 \pi f_a} \phi F_{\mu \nu} \widetilde F^{\mu \nu}  + \frac{c_5}{f_a} (\partial_\mu \phi) J_5^\mu - m \, a \,  \bar \psi e^{2 i c_m \gamma_5 \phi/f_a} \psi  \right\} \,,
 \label{eq:action}
\end{align}
with $\slashed D = (\partial_\mu + i g Q A_\mu)\gamma^\mu$, $\alpha = g^2/4 \pi$ and $Q$ denoting the charge of the fermion $\psi$ under the abelian gauge group with vector potential $A_\mu$. Here $c_A$, $c_5$ and $c_m$ are real coupling constants, $f_a$ indicates the cut-off scale of the effective theory and $g^{\mu \nu}$ denotes the FRW metric.
Performing a chiral fermion rotation, we can eliminate (for example) the term proportional to $(\partial_\mu \phi)J^\mu_5$ where $J^\mu_5 = \bar \psi \gamma^\mu \gamma_5 \psi$. In this paper, we will moreover for simplicity drop the $\phi$-dependent phase of the mass term. This leads to the fermion equation of motion~\eqref{eq:eom}.

In Eq.~\eqref{eq:action} we have introduced the comoving quantities $\psi, A_\mu$ and $g^{\mu \nu}$, related to the corresponding physical quantities (indicated by a hat) as
\begin{align}
 \psi & = a^{3/2} \hat \psi \,, \qquad A_\mu =  (A_0, - \bm A) = \hat A_\mu  \,, \quad A^\mu = a^2 (A_0, \bm A)  = a^2 \hat A^\mu \,. 
\end{align}
and correspondingly, with $\bm E = - \partial_0 \bm A$, $\bm B = \nabla \times \bm A$ in the temporal gauge $A_0 = 0$,
\begin{align}
  \hat{\bm E} = \bm E/a^2 \,, \qquad \hat{\bm B} = \bm B/a^2 \,.
\end{align}
The indices of the physical quantities are raised/lowered by the FRW metric $g^{\mu\nu}$ whereas the indices of the comoving quantities are raised/lowered by the flat metric $\eta_{\mu \nu} = \text{diag}(+,-,-,-) = g_{\mu \nu}/a^2$.

Using the chiral representation of the $\gamma$ matrices, $(\gamma^\mu) = (\gamma_0, \bm{\gamma})$ with
\begin{align}
 \{ \gamma^\mu ,\gamma^\nu \} = 2 \eta^{\mu \nu} \,, \quad  \gamma^0 = \begin{pmatrix}
             0 & 1 \\ 1 &  0 
            \end{pmatrix} \,, \quad
 \bm{\gamma} = \begin{pmatrix}
                0 & \bm{\sigma} \\
                - \bm{\sigma} & 0 
               \end{pmatrix} \,, \quad
 \gamma_5 = \begin{pmatrix}
             -1 & 0 \\
             0 & 1
            \end{pmatrix}
\end{align}
the left(right-)handed component of the four-spinor $\psi = (\psi_L, \psi_R)$ is projected out by the projection operator $P_{L/R} = (1 \mp \gamma_5)/2$. The (dual) field strength tensor of the gauge field is given by
\begin{align}
 F_{\mu \nu} = \partial_\mu A_\nu - \partial_\nu A_\mu \,, \quad \widetilde F^{\mu \nu} = \frac{1}{2}\epsilon^{\mu \nu \rho \sigma} F_{\rho \sigma} \,,
\end{align}
with $\epsilon^{0123} = +1$. 

\section{Solving the fermion equation of motion}
\label{app:eom}

\paragraph{Second order equation of motion.} This appendix provides some additional details on the results derived in Sec.~\ref{sec:eom}. We start be deriving the second order equation of motion~\eqref{eq:eom3} starting from the first order equation~\eqref{eq:eom}:
\begin{align}
 0 & = ( i \slashed D + m a) ( i \slashed D -  m a) \psi \nonumber  \\
  & = \left[ - D_\mu D_\nu {\gamma^\mu \gamma^\nu} - m^2 a^2 {- i D_\mu \gamma^\mu m a + i m a D_\nu \gamma^\nu} \right] \psi \nonumber \\
   & = \left[ - D_\mu D_\nu \left( \eta^{\mu \nu} + \frac{1}{2} [\gamma^\mu, \gamma^\nu]\right) - m^2 a^2 - i m \gamma^0 (D_0 a)  \right] \psi \nonumber \\
  & = \left[- \partial_0^2 + (\bm \nabla - i g Q \bm A)^2  - \frac{g Q}{2}  \sigma^{\mu \nu} F_{\mu \nu} -  m^2 a^2 - i m \gamma^0 a' \right] \psi \,.
  \label{eq:eom2}
\end{align}
Here the relative minus sign in the spatial part of the gauge co-variant derivative arises from our convention $\partial_\mu = (\partial_0 , \bm \nabla)$ and $A_\mu = (A_0, - \bm A)$. In the third term we have introduced the anti-symmetric operator
\begin{align}
 \sigma^{\mu \nu} = \frac{i}{2}[\gamma^\mu, \gamma^\nu ] \quad \text{with} \quad \sigma^{\mu \nu} D_{\mu} D_{\nu} = \frac{1}{2} \sigma^{\mu \nu} [D_\mu, D_\nu] = \frac{1}{2} \sigma^{\mu \nu} (i g Q F_{\mu \nu})
\end{align}
In temporal gauge, $A_0 = 0$, we can immediately evaluate explicitly the relevant components of $\sigma^{\mu \nu}$ and $F_{\mu \nu}$, 
\begin{align}
 \sigma^{0 i} = i \begin{pmatrix}
                   - \bm \sigma & 0 \\
                   0  & \bm \sigma
                  \end{pmatrix} \,, \quad
F_{0i} = E^i \,, \quad \frac{1}{2} \epsilon^{ijk} F_{ij} = - B^k \,.
\end{align}
With this, the third term of Eq.~\eqref{eq:eom} can be expressed as
\begin{align}
 - \frac{g Q }{2} \sigma^{\mu \nu} F_{\mu \nu} = g Q \bm \sigma \begin{pmatrix} 
                                                                  \bm B + i \bm E & 0 \\
                                                                  0 & \bm B - i \bm E
                                                                 \end{pmatrix} \,,
\end{align}
and we  obtain
\begin{align}
 0 = \left[ \left(- \partial_0^2 + (\bm \nabla - i g Q \bm A)^2  - m^2 a^2 \right) \mathbb{1}  + g Q \bm \sigma \begin{pmatrix} 
                                                                  \bm B + i \bm E & 0 \\
                                                                  0 & \bm B - i \bm E
                                                                 \end{pmatrix}
        - i m \gamma^0 a' \right] \psi \,.
        \label{eq:eom3_app}
\end{align}

\paragraph{Solutions to the first order equation of motion.}

For the lowest Landau level ($n = 0$), the second order equation~\eqref{eq:eom3_app} admits four solutions,
\begin{align}
 g_0^{H\,(\pm)}(t) = \exp(\mp i \Omega_0 t) \,, \qquad 
	\Omega_0 \equiv \sqrt{\Pi_z^2 + m^2 a^2}\,.
	\label{eq:g0-app}
\end{align}
see Eq.~\eqref{eq:g0}, corresponding to positive and negative energy solutions for $H = \text{L}, \text{R}$. From the original first order equation of motion~\eqref{eq:eom} we however expect a unique positive and negative energy solution, singling out a unique linear combination of $\chi_s^\text{L}$ and $\chi_s^\text{R}$.

To see this, let us insert the most general solution of \eqref{eq:eom3_app} with arbitrary coefficients $c_s^{r,p}$ (with $p = \pm$ indicating positive/negative frequency) 
\begin{align}
 \widetilde \psi^{r,p}_s(t, x) =  \sum_{r,p} c^{r,p}_s h_0(x_{s}) g_0^p(t) \chi_s^r \,,
\end{align}
into \eqref{eq:eom}. This yields
\begin{align}
 \left[ p \,  \omega \begin{pmatrix}
                  0 & 1 \\
                  1 & 0
                 \end{pmatrix}
 - k_z \begin{pmatrix}
        0 & \sigma_z \\
        - \sigma_z & 0
       \end{pmatrix}
 - m a \begin{pmatrix}
        1 & 0 \\
        0 & 1
       \end{pmatrix}               
\right] \begin{pmatrix}
         \theta(+s) c^{1,p}_+ \\
         \theta(-s) c^{1,p}_- \\
         \theta(+s) c^{2,p}_+ \\ 
         \theta(-s) c^{2,p}_-
        \end{pmatrix} = 0 \,.
\end{align}
with $\theta(\pm s)$ denoting the Heaviside function. 
\begin{align}
 c_+^{2,p} =  \frac{c_+^{1,p}}{a m} ( k_z + p \omega ) \,,\quad  c_-^{2,p} = \frac{c_-^{1,p}}{a m} (- k_z +  p \omega ) \,,
\end{align}
and hence the two (positive and negative energy) solutions for fixed $s$ and $n = 0$ are given by
\begin{align}
 \widetilde \psi^p_{0,s} = {\cal N}^p_s g_0^p(t) h_0(x_{s}) \left[ a m \chi^1_s + ( p \omega + s k_z) \chi^2_s \right] \,,
 \label{eq:A0solution}
\end{align}
with the normalization factor
\begin{align}
 {\cal N}^p_s = \left[  (a m )^2 + (p \omega + s k_z)^2 \right]^{-1/2} \,.
\end{align}
With some algebra, this can be expressed as in Eqs.~\eqref{eq:u0} and \eqref{eq:v0},
\begin{align}
	u_0 &= \frac{e^{- i \Omega_0 t}}{\sqrt{2 \Omega_0}} h_0 \left[ \sqrt{\Omega_0 - s \Pi_z}\, \chi_s^1 + \sqrt{\Omega_0 + s \Pi_z}\, \chi_s^2\right]\,, \label{eq:u0-app} \\
	v_0 &= \frac{e^{i \Omega_0 t}}{\sqrt{2 \Omega_0}} h_0 \left[ \sqrt{\Omega_0 + s \Pi_z}\, \chi_s^1 - \sqrt{\Omega_0 - s \Pi_z}\, \chi_s^2\right]\,. \label{eq:v0-app}
\end{align}

For the higher Landau levels, an analogous procedure leads to Eqs.~\eqref{eq:u1} to \eqref{eq:v2}.


\section{Particle production}
\label{app:PP}

\paragraph{Equations of motion for the Bogoliubov coefficients.} Starting from the definition~\eqref{eq:bogo_def} of the Bogoliubov coefficients,
\begin{align}
 \alpha_n^{(r)} = 
 \left( U_n^{(r)}, \tilde u_n^{(r)} \right)\,, \quad
  \beta_n^{(r)} = 
  \left( V_n^{(r)}, \tilde u_n^{(r)} \right)\,,
\end{align}
we compute their time-derivatives as
\begin{align}
  \partial_t \alpha_n^{(r)} & = \left(  \partial_t U_n^{(r)}, \tilde u_n^{(r)} \right) + \left( U_n^{(r)},  \partial_t \tilde u_n^{(r)} \right) \nonumber \\
  & = 
  i \Omega_{n} \alpha_n^{(r)} +
  \int \dd x \, \left\{  e^{i \int^t \dd t' \Omega_n (t')} 
	\begin{pmatrix}
		h_n \chi_s^1 &
		h_{n-1} \chi_{-s}^1 &
		h_n \chi_s^2 &
		h_{n-1} \chi_{-s}^2
	\end{pmatrix}^\ast  \left[\partial_t \bm{g}^{r, (+)}_n \right]^\ast
	\right\} \tilde u_n^{(r)} 
	 -  i \Omega_{n} \alpha_n^{(r)} \nonumber \\
      & = \int \dd x \, \left\{   e^{ i \int^t \dd t' \Omega_n (t')} 
	\begin{pmatrix}
		h_n \chi_s^1 &
		h_{n-1} \chi_{-s}^1 &
		h_n \chi_s^2 &
		h_{n-1} \chi_{-s}^2
	\end{pmatrix}^\ast   
	\left[\partial_t \bm{g}^{r, (+)}_n \right]^\ast
	\right\} \tilde u_n^{(r)}\,,
\end{align}
where for the first term in the first line we have inserted the solution~\eqref{eq:UV} and for the second term we have used the fact that $\tilde u_n^{(r)}$ solves the equation of motion, 
$i \partial_t \tilde u_n^{(r)} = H_\psi \tilde u_n^{(r)} = \Omega_n (\alpha_n^{(r)} U_n^{(r)} - \beta_n^{(r)} V_n^{(r)} )$. 
The equation for $\partial_t \beta_n^{(r)}$ is obtained analogously. 

To proceed, we need to evaluate $\partial_t \bm{g}^{r, (+)}_n$ with $\bm g^{r, (\pm)}_n$ defined in Eqs.~\eqref{eq:gn1} and \eqref{eq:gn2}. It is straight-forward to verify the following relation,
\begin{align}
 \partial_t \left( \frac{\Omega_{n}(t) \pm s \Pi_z(t)}{2 \Omega_{n}(t) } \right)^{1/2} = \pm \frac{s a m_T \dot \Pi_z(t)}{2 \Omega_n^2(t)}  \left( \frac{\Omega_{n}(t) \mp s \Pi_z(t)}{2 \Omega_{n}(t) } \right)^{1/2} \,.
\end{align}
With this, we see that
\begin{align}
 \partial_t \bm{g}^{r, (\pm)}_n = \mp \frac{s a m_T \dot \Pi_z(t)}{2 \Omega_n^2(t)} \bm{g}^{r, (\mp)}_n \,,
\end{align}
implying the time derivative of the positive frequency eigenvector  acts as a projector on the negative frequency solution and vice versa.
Finally, recalling $\tilde u_n^{(r)} = \alpha_n^{(r)} u_n^{(r)} + \beta_n^{(r)} v_n^{(r)}$, we obtain
\begin{align}
	\partial_t \alpha_n^{(r)} = - \beta_n^{(r)} \times  \frac{s \dot \Pi_z(t)}{2 \Omega_n^2}  e^{2 i \int^t \dd t' \Omega_n (t')} m_T a \,, 
	\qquad
	\partial_t \beta_n^{(r)} = \alpha_n^{(r)} \times  \frac{s \dot \Pi_z(t)}{2 \Omega_n^2}  e^{- 2 i \int^t \dd t' \Omega_n (t')} m_T a\,.
	\label{eq:bogo_evo_temp-app}
\end{align}

\paragraph{Chiral charge.}  Starting from the definition of the chiral charge in Eq.~\eqref{eq:chiral_charge},
\begin{align}
 q_5
	= \frac{1}{\operatorname{vol} (\mathbb{R}^3)} \int \dd^3 x\, \frac{1}{2}\vev{ \left[ \psi^\dag, \gamma_5 \psi \right] } \,,
	\label{eq:chiral_charge-app}
\end{align}
our goal here is to evaluate this expression, deriving Eq.~\eqref{eq:chiral_charge_bogo} of the main text. Since the higher Landau levels are symmetric under parity transformations (exchange of left- and right-handed particles), is suffices to consider the contribution of the lowest Landau level to $\psi$.
Also one can explicitly check that the contributions from the higher Landau levels cancel out between $r=1$ and $r=2$.
Inserting Eq.~\eqref{eq:FT} and more specifically expanding in terms of the creation and annihilation operators defined at some time $t > 0$ yields
\begin{align}
 q_5 & = \frac{1}{\operatorname{vol}( \mathbb{R}^3)} \int \dd x \int\frac{\dd k_y \dd k_z}{(2 \pi)^2} \int\frac{\dd k'_y \dd k'_z}{(2 \pi)^2} {\int \dd y \, \dd z \, e^{iy(k_y - k_y')} e^{iz(k_z - k_z')}}  \frac{1}{2} \, \vev{  \left[ \tilde \psi^\dagger, \gamma_5 \tilde \psi \right] } \nonumber \\
 & = \frac{1}{\operatorname{vol} ( \mathbb{R}^3) } \int \dd x \int\frac{\dd k_y \dd k_z}{(2 \pi)^2} \int{\dd k'_y \dd k'_z} { \delta(k_y - k_y') \delta(k_z - k_z')}  \frac{1}{2} \, \vev{  \left[ \tilde \psi^\dagger, \gamma_5 \tilde \psi \right] }
\nonumber \\
 & = \frac{1}{\operatorname{vol} (\mathbb{R}^3)} \int \dd x \int\frac{\dd k_y \dd k_z}{(2 \pi)^2} \frac{1}{2}  \left[  U_0^\dagger \gamma_5 U_0 \vev{ \hat B_0^\dagger \hat B_0 - \hat B_0 \hat B_0^\dagger } + V_0^\dagger \gamma_5 V_0 \vev{ \hat D_0 \hat D_0^\dagger - \hat D_0^\dagger \hat D_0 } \right. \nonumber \\
 & \qquad \qquad \qquad \qquad  \left.  +   \left( U_0^\dagger \gamma_5 V_0 \vev{ \hat B_0^\dagger \hat D_0^\dagger - \hat D_0^\dagger \hat B_0^\dagger } + \text{H.c.} \right)
\right] \,.
 \label{eq:q5_1}
\end{align}
Using the explicit expression for the eigenvectors $U_0$ and $V_0$ in Eq.~\eqref{eq:UV0} we find
\begin{align}
  U_0^\dagger \gamma_5 U_0 = -  V_0^\dagger \gamma_5 V_0 = s \, h_0^2 \frac{\Pi_z}{\Omega_0} \,, \quad  U_0^\dagger \gamma_5 V_0 =  (V_0^\dagger \gamma_5 U_0)^* = - h_0^2  e^{2 i \int^t \dd t' \Omega_0(t')} \frac{m a}{\Omega_0} \,.
\end{align}
Expanding  $\hat B_0$ and $\hat D_0$ in terms of the creation and annihilation operators of the original vacuum at $t < 0$ yields
\begin{align}
\vev{ \hat B_0^\dagger \hat B_0 - \hat B_0 \hat B_0^\dagger } & =  |\beta_0|^2 \vev{ \hat d_0 \hat d_0^\dagger } -  |\alpha_0|^2 \vev{ \hat b_0 \hat b_0^\dagger } = (|\beta_0|^2 - |\alpha_0|^2) \delta^{(2)} (0) = (2 |\beta_0|^2 - 1) \delta^{(2)}(0) \,, \nonumber \\
 \vev{ \hat D_0 \hat D_0^\dagger - \hat D_0^\dagger \hat D_0 } & =  |\alpha_0|^2 \vev{ \hat d_0 \hat d_0^\dagger } -  |\beta_0|^2 \vev{ \hat b_0 \hat b_0^\dagger } = (|\alpha_0|^2 - |\beta_0|^2) \delta^{(2)} (0) = (- 2 |\beta_0|^2 + 1) \, \delta^{(2)}(0) \,, \nonumber \\
 \vev{ \hat B_0^\dagger \hat D_0^\dagger - \hat D_0^\dagger \hat B_0^\dagger } & =  -  \alpha_0^* \beta_0 \vev{ \hat d_0 \hat d_0^\dagger } -  \alpha_0^* \beta_0 \vev{ \hat b_0 \hat b_0^\dagger }  = - 2 \alpha_0^* \beta_0 \, \delta^{(2)}(0) \,. 
\label{eq:BD}
\end{align}
The factor $\delta^{(2)}(0)$ arises since the creation and annihilation operators contain the same wave vector $k_y$ and $k_z$, as enforced by the delta function in the second line of Eq.~\eqref{eq:q5_1}.

Plugging these results back into Eq.~\eqref{eq:q5_1}, we obtain
\begin{align}
 q_5 & =  \frac{1}{\operatorname{vol} (\mathbb{R}^3)}  {\int \dd x} \int {\frac{\dd k_z}{2 \pi} \delta(0)} \int  {\frac{\dd k_y }{2 \pi}  \,  h_0^2 \delta(0)}  \left[ \frac{ s \Pi_z}{\Omega_0} (2 |\beta_0|^2 - 1) +  \frac{m a }{\Omega_0} \left( e^{ 2 i \int^t \dd t' \Omega_0(t')} \alpha_0^* \beta_0 + \text{H.c.} \right) \right] \nonumber \\
 & =  \lim_{\hat\Lambda \rightarrow \infty} \frac{L_x L_y L_z}{\operatorname{vol} (\mathbb{R}^3)}  \;  \frac{g Q B }{2 \pi} \int \frac{\dd k_z}{2 \pi} R_{\hat \Lambda}(|\Omega_0|)  \left[ \frac{ \Pi_z}{\Omega_0} \left( 2 |\beta_0|^2 - 1 \right) +  s \frac{m a}{\Omega_0} \left( e^{ 2 i \int^t \dd t' \Omega_0(t')} \alpha_0^* \beta_0 + \text{H.c.} \right) \right] \,.
 \label{eq:q5}
\end{align}
To treat the vacuum contribution unambiguously, we have explicitly inserted the regulator $R_{\hat\Lambda}$ with $\hat\Lambda$ denoting a UV-cutoff.
By changing the integration variable $k_z \mapsto \Pi_z$, one can see that the vacuum contribution proportional to ``$-1$'' in the first bracket vanishes because the integrand including the regulator is an odd function of $\Pi_z$.
This confirms that, in the case at hand, the vacuum contribution originating from the $\eta$-invariant vanishes.
In contrast, the $\eta$-invariant plays a crucial role in the case discussed in Ref.~\cite{Domcke:2018gfr}. 
In summary, we obtain the following expression for the chiral charge:
\begin{align}
	q_5 (t) = \frac{gQB}{2 \pi} \int \frac{\dd k_z}{2 \pi} \left[  
		\frac{\Pi_z}{\Omega_0}\, 2 \abs{\beta_0}^2 
		+ s \frac{m a}{\Omega_0} \left( e^{2 i \int^t \dd t' \Omega_0 (t')} \alpha_0^\ast \beta_0 + \text{H.c.} \right)
	\right]\,.
	\label{eq:chiral_charge_bogo-app}
\end{align}

\paragraph{Induced current.}
Here we explicitly give the derivation of Eq.~\eqref{eq:current} starting from the definition \eqref{eq:current_def}:
\begin{align}
	\vev{J_z} = \frac{1}{\operatorname{vol}(\mathbb{R}^3)} \int \dd^3 x\, \frac{1}{2} \vev{ \left[ \overline{\psi},  \gamma^3 \psi \right] } \,.
\end{align}
Inserting Eqs.~\eqref{eq:FT} and \eqref{eq:expand_t} into Eq.~\eqref{eq:current}, we get
\begin{align}
	\vev{J_z}  &=  \frac{1}{\operatorname{vol} ( \mathbb{R}^3) } \int \dd x \int\frac{\dd k_y \dd k_z}{(2 \pi)^2} \int{\dd k'_y \dd k'_z} { \delta(k_y - k_y') \delta(k_z - k_z')}  \frac{1}{2} \, \vev{  \left[ \tilde \psi^\dagger, \gamma^0 \gamma^3 \tilde \psi \right] } \nonumber \\
	&=  \frac{1}{\operatorname{vol} ( \mathbb{R}^3) }  \int \dd x \int\frac{\dd k_y \dd k_z}{(2 \pi)^2} \frac{1}{2} \sum_{n,r}
	\left[  \overline U_n^{(r)} \gamma^3 U_n^{(r)} \vev{ \hat B_n^{(r)\dagger} \hat B_n^{(r)} - \hat B_n^{(r)} \hat B_n^{(r) \dagger} } 
	+ \overline V_0^{(r)}  \gamma^3 V_n^{(r)} \vev{ \hat D_n \hat D_n^{(r) \dag} - \hat D_n^{(r)\dagger} \hat D_n^{(r)} } 
	\right. \nonumber \\
	& \qquad \qquad \qquad \qquad  \left. 
	  +   \left( \overline U_n^{(r)} \gamma^3 V_n^{(r)} \vev{ \hat B_n^{(r)\dagger} \hat D_n^{(r)\dagger} - \hat D_n^{(r)\dagger} \hat B_n^{(r)\dagger} } + \text{H.c.} \right)
\right]\,.
\end{align}
Using the following properties of the eigenvectors $U_n^{(r)}$ and $V_n^{(r)}$,
\begin{align}
	\int \dd k_y\, \overline U_n^{(r)} \gamma^3 U_n^{(r)} =
	- \int \dd k_y\, \overline V_n^{(r)} \gamma^3 V_n^{(r)} = \frac{s g Q B}{2 \pi} \frac{\Pi_z}{\Omega_n}\,, \quad
	\int \dd k_y\, \overline U_n^{(r)} \gamma^3 V_n^{(r)}
	= - \frac{g Q B}{2 \pi} \frac{m_T a}{\Omega_n} e^{2 i \int \dd t' \Omega_n (t')}\,,
\end{align}
and expectation values of creation and annihilation operators by the original vacuum,
\begin{align}
	\vev{ \hat B_n^{(r)\dagger} \hat B_n^{(r)} - \hat B_n^{(r)} \hat B_n^{(r)\dagger} } &= \vev{ \hat D_n^{(r)} \hat D_n^{(r)\dagger} - \hat D_n^{(r)\dagger} \hat D_n^{(r)}} = \left( 2 |\beta_n^{(r)}|^2 - 1 \right) \delta^{(2)}(0)\,, \nonumber \\
	\vev{ \hat B_n^{(r)\dagger} \hat D_n^{(r)\dagger} - \hat D_n^{(r)\dagger} \hat B_n^{(r)\dagger} }  &= - 2 \alpha_n^{(r) \ast} \beta_n \, \delta^{(2)}(0)\,,
\end{align}
we arrive at
\begin{align}
	\vev{ J_z } = \lim_{\hat \Lambda \to \infty} 
	\frac{g Q B}{2 \pi} \int \frac{\dd k_z}{2 \pi} \sum_{n,r} R_{\hat \Lambda}(|\Omega_n|) \left[
		\frac{2 \Pi_z}{\Omega_n} \left( 2 \abs{\beta_n^{(r)}}^2 - 1 \right) + \frac{m_T a}{\Omega_n} \left( \alpha_n^{(r) \ast} \beta_n^{(r)} e^{2 i \int^t \dd t' \Omega_n} + \text{H.c.} \right) 
	\right]\,.
\end{align}
As in the estimation of $q_5$, the term proportional to ``$-1$'' in the first parenthesis vanishes because the integrand including the regulator is an odd function of $\Pi_z$.
Therefore we finally obtain
\begin{align}
	\vev{ J_z } = \lim_{\hat \Lambda \to \infty}
	\frac{g Q B}{2 \pi} \int \frac{\dd k_z}{2 \pi} \sum_{n,r}  R_{\hat \Lambda}(|\Omega_n|) \left[
		\frac{2 \Pi_z}{\Omega_n} 2 \abs{\beta_n^{(r)}}^2 + \frac{m_T a}{\Omega_n} \left( \alpha_n^{(r) \ast} \beta_n^{(r)} e^{2 i \int^t \dd t' \Omega_n} + \text{H.c.} \right) 
	\right]\,.
\end{align}
We keep the regulator to make it clear that it contains the divergence.


\section{Weak field expansion}
\label{sec:WFE}

In this appendix, we derive expressions for the Bogoliubov coefficients
in the weak field approximation, \text{i.e.} in the limit $E/m^2a^2 \ll 1$.
We also assume that the time dependence of $E$ is weak, or $\abs{\dot{E}}/E m a \ll 1$ and so on.
The time evolution of the Bogoliubov coefficients is governed by Eq.~\eqref{eq:bogo_evo_temp-app}.
For our purpose, however,
it is more useful to deal with $\alpha_n^{(r)\ast} \beta_n^{(r)}$ and $\abs{\beta_n^{(r)}}^2$ 
instead of $\alpha_n^{(r)}$ and $\beta_n^{(r)}$.
Their equations of motion are given by (see also Eq.~\eqref{eq:bogo_evo}),
\begin{align}
	\frac{\dd}{\dd t}\left(\alpha_n^{\ast} \beta_n\right) 
	&= s  m_T a \frac{g Q E}{2\Omega_n^2}
	\left(1 - 2\abs{\beta_n}^2\right)e^{-2i\int^t \dd t' \Omega_n}\,, 
	\label{eq:ab} \\
	\frac{\dd}{\dd t}\abs{\beta_n}^2
	&=
	s  m_T a \frac{g Q E}{\Omega_n^2}
	\Re\left(\alpha_n^\ast \beta_n e^{2i\int^t \dd t' \Omega_n}
	\right)\,.
	\label{eq:b2}
\end{align}
Since the equations of motion are the same for $r = 1, 2$, 
we omit the index $r$ here and hereafter for notational simplicity.
We solve these equations in the weak field expansion in the following.

In the weak field limit, one can clearly separate terms that contribute
to the Bogoliubov coefficients into two classes.
The first are those that accompany the exponential suppression factor
$e^{- \frac{\pi m^2 a^2}{g \abs{Q} E}}$,
which corresponds to the particle production due to the electric field.
We call it a ``non-perturbative" contribution
since it is not obtained from the weak field expansion.
The second are those that do not accompany such a suppression factor,
which we call a ``perturbative" contribution.
It is not understood as particle production, but it nevertheless
has effects on, \textit{e.g.}, the induced current of the fermion
and hence the equation of motion of the gauge field.
Indeed, the gauge coupling renormalization and the Euler-Heisenberg Lagrangian 
originate from the latter class of terms, as shown in Sec.~\ref{subsec:Jind}.
For this reason, we sometimes refer to the latter as a vacuum contribution. 

In the following, we focus on the perturbative contribution.
We first describe how the weak field expansion proceeds 
with the help of integration by parts.
We then compute the perturbative contribution
up to third order in the weak field expansion..
The results are used in Secs.~\ref{subsec:anomaly} and~\ref{subsec:Jind}.

\paragraph{Weak field expansion as integration by parts.}
The weak field expansion can be systematically performed by integration by parts.
Eq.~\eqref{eq:ab} is formally solved as
\begin{align}
	\alpha_n^{\ast} \beta_n (t) 
	&= s  m_T a \int^t \dd t' \frac{g Q E}{2\Omega_n^2}
	\left(1 - 2\abs{\beta_n}^2\right) e^{-2i\int^{t'} \dd t'' \Omega_n} \nonumber \\
	&= i s m_T a \frac{g Q E}{4\Omega_n^3} \left(1 - 2\abs{\beta_n}^2\right) e^{-2i\int^t \dd t' \Omega_n}
	- i s m_T a\int^t \dd t' \frac{\dd}{\dd t'}
	\left[\frac{gQE}{4\Omega_n^3} \left(1 - 2\abs{\beta_n}^2\right) \right] 
	e^{-2i\int^{t'} \dd t'' \Omega_n}\,,
	\label{eq:int_by_parts}
\end{align}
where we have performed the integration by parts to get the second line.
We ignore the time dependence of the scale factor here and henceforth.
Note that $E$ and its derivatives have to vanish at the initial time
for the initial state to be defined unambiguously.
Since the time derivatives of $\Omega_n$ and $\abs{\beta_n}^2$ contain additional $E$
or $\dot{E}$, the second term in Eq.~\eqref{eq:int_by_parts}
is sub-leading compared to the first term for the perturbative contribution.
In this sense, the above integration by parts is equivalent to the weak field expansion.
By repeating the same procedure, 
we obtain the following formula:
\begin{align}
	\left(\alpha_n^{\ast} \beta_n (t) e^{2i\int^{t} \dd t' \Omega_n}\right)_\mathrm{P}
	= i s m_T a\left[\sum_{j=0}^{\infty}
	\left(\frac{-i}{2\Omega_n}\frac{\dd}{\dd t}\right)^j 
	\frac{gQE}{4\Omega_n^3}\left(1 - 2 \abs{\beta_n}^2_\mathrm{P}\right)
	\right]\,,
	\label{eq:weak_field_expansion}
\end{align}
where the subscript ``P" stands for ``perturbative".
The corresponding $\abs{\beta_n}^2_\mathrm{P}$ is obtained 
by integrating Eq.~\eqref{eq:b2}.
In the following, we will recursively solve these equations
to derive the explicit forms of $\left(\alpha_n^\ast \beta_n\right)_\mathrm{P}$ 
and $\abs{\beta_n}^2_\mathrm{P}$ in the weak field expansion.

\paragraph{Leading order contribution.}
It is easy to compute the leading order contribution
to $\alpha_n^\ast \beta_n$.
From Eq.~\eqref{eq:weak_field_expansion}, it is given by
\begin{align}
	\left(\alpha_n^{\ast} \beta_n e^{2 i \int^t \dd t' \Omega_n}\right)_\mathrm{P}
	\simeq i s m_T a \frac{g Q E}{4 \Omega_n^3}\,,
	\label{eq:wfe_leading_app}
\end{align}
at the leading order.
The result with $n = 0$ is used to check the anomaly equation
in Sec.~\ref{subsec:anomaly}.

\paragraph{Higher order perturbative contribution.}
Now we compute higher order perturbative terms in the weak field expansion.
In particular, we focus on $\Re[\alpha_n^\ast \beta_n e^{2i\int^t \dd t' \Omega_n}]_\mathrm{P}$
and $\abs{\beta_n}^2_\mathrm{P}$ 
that are necessary to derive the Euler-Heisenberg term.
We compute terms up to third order in the weak field expansion
that correspond to the lowest order terms of the Euler-Heisenberg Lagrangian.
If one goes beyond the third order terms, 
one obtains higher order Euler-Heisenberg terms.

Both $\Re[\alpha_n^\ast \beta_n e^{2i\int^t \dd t' \Omega_n}]_\mathrm{P}$
and $\abs{\beta_n}^2_\mathrm{P}$ vanishes at the zero-th order 
since $\left(\alpha_n^\ast \beta_n e^{2i\int^t \dd t' \Omega_n}\right)_\mathrm{P}$
is purely imaginary at this order.
At the first order, we obtain from Eq.~\eqref{eq:weak_field_expansion}
\begin{align}
	\Re\left(\alpha_n^\ast \beta_n e^{2i\int^t \dd t' \Omega_n}\right)_{\mathrm{P}}
	&=
	s\frac{m_T a}{8\Omega_n} \frac{\dd}{\dd t}\left[\frac{gQ E}{\Omega_n^3}\left(1-2\abs{\beta_n}^2_\mathrm{P}
	\right)\right]
	+ \cdots \nonumber \\
	&=
	s\frac{m_T a}{8\Omega_n} \frac{\dd}{\dd t}\left[\frac{gQ E}{\Omega_n^3}\right]
	+ \cdots\,,
\end{align}
where we have ignored $\abs{\beta_n}^2_\mathrm{P}$ 
in the second line since it vanishes at the zero-th order.
Plugging it into the equation for $\abs{\beta_n}^2$, we obtain at the first order
\begin{align}
	\abs{\beta_n}^2_\mathrm{P} 
	&= 
	\int^t \dd t' \frac{m_T^2 a^2 gQ E}{8\Omega_n^3}\frac{\dd}{\dd t'}\left[\frac{gQ E}{\Omega_n^3}\right]
	+ \cdots \nonumber \\
	&=
	m_T^2 a^2\frac{\left(gQ\right)^2}{16\Omega_n^6}E^2 + \cdots\,.
	\label{eq:b2_1st}
\end{align}
Now we move to the higher order terms.
The second order terms vanish since the second order term of 
$\left(\alpha_n^\ast \beta_n e^{2i\int^t \dd t' \Omega_n}\right)_\mathrm{P}$ is again purely imaginary.
Up to the third order, we obtain
\begin{align}
	\Re\left(\alpha_n^\ast \beta_n e^{2i\int^t \dd t' \Omega_n}\right)_{\mathrm{P}}
	&=
	s m_T a \left[
	\frac{1}{8\Omega_n}\frac{\dd}{\dd t} - \frac{1}{32}\left(\frac{1}{\Omega_n}\frac{\dd}{\dd t}\right)^3
	\right]
	\left[\frac{gQE}{\Omega_n^3}\left(1 - 2\abs{\beta_n}^2_\mathrm{P}\right)\right]
	\nonumber \\
	&=
	s \frac{m_T a}{8 \Omega_n} \left[ 
		\frac{g Q \dot E}{\Omega_n^3} - \frac{(g Q)^3}{8 \Omega_n^9} E^2 \dot E \left( 3 m_T^2 a^2 - 38 \Omega_n^2 + 248 \Pi_z^2 \right) 
	\right] + \text{(Odd in $\Pi_z$)} + \cdots\,,
\end{align}
where we have inserted Eq.~\eqref{eq:b2_1st} in the second line,
and ignored terms higher than the third order.
We have also ignored terms of $\mathcal{O}\left(\dddot{E}\right)$
that correspond to higher derivative terms in the effective action
for the gauge field.
Correspondingly $\abs{\beta_n}_\mathrm{P}^2$ is given by
\begin{align}
	\abs{\beta_n}^2_\mathrm{P}
	&= s m_T a \int^t \dd t' \frac{gQE}{\Omega_n^2}
	\Re{\left(\alpha_n^\ast \beta_n e^{2i\int^t \dd t' \Omega_n}\right)}_\mathrm{P} \nonumber \\
	& = 
	\frac{7}{32} m_T^2 a^2 \frac{(g Q)^3}{\Omega_n^{10}} E^2 \dot E \, \Pi_z + \text{(Even in $\Pi_z$)} + \cdots\,.
\end{align}
These expressions are used in Sec.~\ref{subsec:Jind} to derive the Euler-Heisenberg term.
Note that only lower order terms
are necessary to obtain terms at a specific order (the third order in the above case),
and hence this procedure can be systematically extended to higher order terms. 

\small
\bibliographystyle{utphys}
\bibliography{refs}
  
\end{document}